\documentclass[a4paper,11pt]{article}
\pdfoutput=1 

\usepackage{jheppub} 
\usepackage{bm,amssymb,slashed,graphicx,multirow,soul,mathtools,xspace,array}                     
\allowdisplaybreaks
\usepackage{ bbold }
\usepackage{subfigure}
\newcommand{\op}{{\cal O}}

\newcommand{\todo}[1]{{\color{red} \ifmmode\else[todo]\fi #1}}
\usepackage[usenames,dvipsnames]{xcolor}
     \definecolor{hgreen}{rgb}{0,.3,0}
      \definecolor{darkgreen}{rgb}{0.3,.8,0.2}
     \definecolor{hred}{rgb}{.3,0,0}
     \definecolor{hblue}{rgb}{0,0,.3}
     \definecolor{LightGray}{gray}{0.95}

\newcommand{\ov}{\overline}

\newcommand{\dds}{D^{(*)}}

\newcommand{\Bbar}{\,\overline{\!B}{}}

\newcommand{\leff}{\ensuremath{\Lambda_{\rm eff}}\xspace}

\newcommand{\nt}{{\nu_\tau}}
\newcommand{\ntb}{{\bar\nu_\tau}}

\newcommand{\bddstn}{{\Bbar \to D^{(*)}\tau \bar\nu}}

\newcommand{\bddsln}{{\Bbar \to D^{(*)} l \bar\nu}}

\newcommand{\bddstaun}{{\Bbar \to D^{(*)} \tau \bar\nu}}

\newcommand{\nn}{\nonumber}
\newcommand{\TeV}{\text{TeV}}
\newcommand{\GeV}{\text{GeV}}
\newcommand{\Br}{\text{Br}}

\newcommand{\beq}{\begin{equation} }
\newcommand{\eeq}{\end{equation}} 
\newcommand{\bi}{\begin{itemize} }
\newcommand{\ei}{\end{itemize} }

\definecolor{Red}{rgb}{1.,0.,0.}
\definecolor{Grn}{rgb}{0.,0.75,0.}
\definecolor{Blu}{rgb}{0.,0.,1.}

\usepackage[T1]{fontenc} 

\usepackage{amsmath,amssymb,epsfig,ulem,color,slashed}
\allowdisplaybreaks  

\title{\boldmath Right-handed Neutrinos and $R(D^{(*)})$ }

\author[1]{Dean Robinson,}
\author[1,2]{Bibhushan Shakya,}
\author[1]{Jure Zupan}

\affiliation[1]{Department of Physics, University of Cincinnati, Cincinnati, Ohio 45221,USA}
\affiliation[2]{Leinweber Center for Theoretical Physics (LCTP),
University of Michigan, Ann Arbor, Michigan 48109, USA}

\emailAdd{dean.robinson@uc.edu}
\emailAdd{shakyabn@ucmail.uc.edu}
\emailAdd{zupanje@ucmail.uc.edu}

\abstract{We explore scenarios where the $R(D^{(*)})$ anomalies arise from semitauonic decays to a right-handed sterile neutrino. We perform an EFT study of all five simplified models capable of generating at tree-level the lowest dimension electroweak operators that give rise to this decay. We analyze their compatibility with current $R(D^{(*)})$ data and other relevant hadronic branching ratios, and show that one simplified model is excluded by this analysis. The remainder are compatible with collider constraints on the mediator semileptonic branching ratios, provided the mediator mass is of order TeV. We also discuss the phenomenology of the sterile neutrino itself, which includes possibilities for displaced decays at colliders and direct searches, measurable dark radiation, and gamma ray signals.}

\preprint{LCTP-18-19}

\begin{document} 

\maketitle

\flushbottom

\section{Introduction}
Measurements of the semitauonic to light semileptonic ratios at multiple experiments~\cite{Lees:2012xj, Lees:2013uzd,Huschle:2015rga, Abdesselam:2016cgx, Abdesselam:2016xqt,Aaij:2015yra},
\begin{equation}\label{eq:ratio}
	R(\dds) = \frac{\Br[\bddstn]}{\Br[\bddsln]}\,,\qquad l = e,\mu\,,
\end{equation}
exhibit a $4\sigma$ tension with respect to the Standard Model (SM) predictions, once both $D$ and $D^*$ measurements are combined~\cite{HFAG} 
(see also Refs.~\cite{Bernlochner:2017jka,Bigi:2017jbd,Jaiswal:2017rve,Alok:2016qyh,Bhattacharya:2016zcw}). Beyond the Standard Model (BSM) explanations of this anomaly typically require new physics (NP) close to the TeV scale. Since the SM neutrino is part of an electroweak doublet, corresponding constraints necessarily arise from high-$p_T$ measurements of $p p \to \tau^+ \tau^-$ at the LHC \cite{Faroughy:2016osc}, $Z$ and $\tau$ decays \cite{Feruglio:2016gvd,Feruglio:2017rjo}, and contributions to flavor changing neutral currents (FCNCs), that can be severe.

As discussed in Refs.~\cite{Asadi:2018wea,Greljo:2018ogz} (see also Refs.~\cite{He:2012zp,He:2017bft}), the observed enhancements of $R(\dds)$ can be achieved not only through 
NP contributions to the $b \to c \tau \ntb$ decay, where $\nu_\tau$ is the SM left-handed $\tau$ neutrino, but also via a new decay channel, 
$b \to c \tau \bar N_R$, where $N_R$ is a sterile right-handed neutrino. The $b \to c \tau \bar\nu$ decay becomes an incoherent sum of two contributions:
To streamline notation we denote $\nu = N_R$ or $\nu_\tau$, so that  $\Br[b \to c \tau \bar \nu]=\Br[b \to c \tau \bar \nu_\tau]+ \Br[b \to c \tau \bar N_R]$.
Since the NP couples to right-handed neutrinos, this can relax many of the  electroweak constraints from the $\tau$ processes mentioned above.

In the specific context of Refs.~\cite{Asadi:2018wea,Greljo:2018ogz}, the $b \to c \tau \bar N_R$ decay is mediated by an $SU(2)_L$ singlet $W'$, which can be UV completed in a `3221' model. In this paper we generalize the EFT studies of Refs.~\cite{Asadi:2018wea,Greljo:2018ogz} to the full set of dimension-six operators involving $N_R$  (for earlier partial studies see \cite{Fajfer:2012jt,Becirevic:2016yqi,Cvetic:2017gkt}). 
Assuming that the NP corrections are due to a tree level exchange of a new mediator, there are five possible simplified models for $b \to c \tau \bar N_R$, whose mediators are: the $SU(2)_L$-singlet vector boson -- the $W'$; a scalar electroweak doublet; and three leptoquarks.

For each simplified model we identify which regions of parameter space are consistent with the $R(\dds)$ anomaly, subject to exclusions from the $B_c \to \tau \nu$ branching ratio~\cite{Li:2016vvp,Alonso:2016oyd,Celis:2016azn}. We further examine the variation in the signal differential distributions expected for each simplified model. While some electroweak constraints are relaxed, these simplified models nonetheless typically imply various sizeable semileptonic branching ratios for the tree-level mediators, for which moderately stringent collider bounds already exist. We show that, depending on the ratios of NP couplings in the simplified model, these in turn set lower bounds of $\mathcal{O}(\TeV)$ on the mediator masses.
We then proceed to examine the implications for neutrino phenomenology, such as bounds from radiative contributions to the SM neutrino masses, astrophysical constraints from sterile neutrino electromagnetic decays, plausible cosmological histories that admit these sterile neutrinos, and displaced decays at colliders and direct searches.
In our analysis, we will require the $N_R$ to be light --  $m_{N_R} \lesssim \mathcal{O}(100)$\,MeV -- in order not to disrupt the measured missing invariant mass spectrum in the full $\bddstaun$ decay chain. Whether heavier sterile neutrinos are compatible with data requires a dedicated forward-folded study, performed by the experimental collaborations.

The paper is structured as follows. Section \ref{sec:EFT} contains the EFT analysis of the $R(\dds)$ data for the case of the right-handed neutrino and introduces the five possible tree-level mediators.   Collider constraints on these simplified models are studied in Section \ref{sec:simplified}, while Section \ref{sec:sterile:neutrino} contains the related sterile neutrino phenomenology. Our conclusions follow in Section \ref{sec:conclusions}.  Appendix \ref{app:dd} examines the structure of the $b \to c \tau \bar\nu$ differential distributions for the simplified models.

\section{EFT analysis}
\label{sec:EFT}
\subsection{EFTs and simplified models}
We consider the extension of the SM field content by a single new state, a right handed, sterile neutrino transforming as $N_R \sim (\bm{1}, \bm{1},0)$ under $SU(3)_c \times SU(2)_L \times U(1)_Y$. 
This state may couple to the SM quarks via higher dimensional operators. Above the electroweak scale, one therefore adds to the renormalizable SM Lagrangian the following effective interactions,
\begin{equation}
\label{eq:LeffEW}
	{\cal L}_{\rm eff}^{\rm EW}=\sum_{a,d} \frac{C_{ad}}{\leff^{d-4}} Q_a+\cdots,
\end{equation}
where $Q_a$ are dimension-$d$ operators, $C_{ad}$ are the corresponding dimensionless Wilson coefficients (WCs), and $\leff$ is the effective scale defined to be
\begin{equation}
\label{eq:leff}
	\leff=\big(2\sqrt 2 G_F V_{cb}\big)^{-1/2}\simeq 0.87 \, \bigg[\frac{40 \times 10^{-3}}{V_{cb}}\bigg]^{1/2}\,\TeV\,.
\end{equation}
The most general basis of dimension-6 operators that can generate the charged current  $b \to c \tau \bar N_R$ decay is given by 
\begin{subequations}
\label{eq:Q6}
\begin{align}
	Q_{\rm SR}&=\epsilon_{ab}\big(\bar Q_L^a  d_R\big)\big(\bar L_L^b N_R \big), 
	 &Q_{\rm SL}&=\big(\bar u_R  Q_L^a \big) \big(\bar L_L^a  N_R \big),
	\\
	Q_{\rm T}&=\epsilon_{ab} \big(\bar Q_L^a \sigma^{\mu\nu} d_R\big)\big(\bar L_L^b\sigma_{\mu\nu} N_R \big), 
	& Q_{\rm VR}&=\big(\bar u_R \gamma^\mu  d_R\big)\big(\bar \ell_R \gamma_\mu N_R \big).
\end{align}
\end{subequations}
Here $a,b$ are $SU(2)_L$ indices, $\epsilon_{ab}$ is an antisymmetric tensor with $\epsilon_{12}=-\epsilon_{21}=1$, and we use the four-component notation, with $Q_L$ the SM quark doublet, $u_R$ and $d_R$ the up- and down-quark singlets, and $L_L$ the SM lepton doublet.  
(As usual, there is only one non-vanishing tensor operator, since $\sigma_{\mu\nu}P_L \otimes \sigma^{\mu\nu}P_R=0$, which immediately follows from the relation $\sigma_{\mu\nu}\otimes \sigma^{\mu\nu}\gamma_5= \sigma_{\mu\nu}\gamma_5 \otimes \sigma^{\mu\nu}$.) 
One may also include the dimension-8 operator
\begin{equation}
	\label{eqn:QVL}
	Q_{\rm VL}=\big(\bar Q_L \tilde H \gamma^\mu H^\dagger Q_L \big)\big(\bar \ell_R \gamma_\mu N_R \big),
\end{equation}
where $\tilde H=\epsilon H^*$, as well as the operators with the left-handed sterile neutrino field, $N_R^c$, that start at dimension-7,
\begin{subequations}
\begin{align}
	Q_{\rm SR}'&=\big(\bar Q_L \tilde H  d_R\big)\big(\bar \ell_R N_R^c \big), 
	 &Q_{\rm SL}'&=\big(\bar u_R  H^\dagger Q_L \big) \big(\bar \ell_R  N_R^c \big),
	\\
	Q_{\rm T}'&= \big(\bar u_R \sigma^{\mu\nu} H^\dagger Q_L \big)\big(\bar \ell_R\sigma_{\mu\nu} N_R^c \big)\,, 
	&Q_{\rm VR}'&=\big(\bar u_R \gamma^\mu  d_R\big)\big(\bar L_L H \gamma_\mu N_R^c \big)\,,
\end{align}
\end{subequations}
and the dimension-9 equivalent of $Q_{\rm VL}$,
\begin{equation}
	\label{eqn:QVLp}
	Q_{\rm VL}'=\big(\bar Q_L \tilde H \gamma^\mu H^\dagger Q_L \big)\big(\bar L_L H \gamma_\mu N_R^c \big).
\end{equation}
Each of the SM fields also carries a family index, i.e., $Q_L^i$, $u_R^i$, $d_R^i$, $L_L^i$, $i=1,2,3$, and similarly for the Wilson coefficients, $C_{ad}^{ijk}$, and the operators, $Q_{ad}^{ijk}$, in Eq.~\eqref{eq:LeffEW}, which we have omitted for the sake of simplicity. Since we focus exclusively 
on the generation of $b \to c \tau \bar \nu$ decays below, we drop the family indices hereafter, unless otherwise stated.
Consistency with bounds from direct searches requires that the Wilson coefficients in Eq.~\eqref{eq:LeffEW} be at most $\mathcal{O}(1)$.

Below the electroweak scale, the top quark, the Higgs, and the $W$ and $Z$ bosons are integrated out.  At the scale $\mu \sim m_{c,b}$, the effective Lagrangian, including SM terms (see, e.g., \cite{Buchalla:1995vs}), can be written 
\begin{equation}
\label{eq:Leff:low}
{\cal L}_{\rm eff}={\cal L}_{\rm eff}^{\rm SM}+\frac{1}{\leff^2} \sum_i c_i \op_i\,,
\end{equation}
in which the NP contributions to $b \to c \tau \bar\nu$, induced by the dimension-6 operators in~\eqref{eq:Q6}, are described by the following four-fermion operators, 
\begin{subequations}
\label{eqn:O}
\begin{align}
\label{eq:QS}
\op_{\rm SR}&=\big(\bar c_L  b_R\big)\big(\bar \tau_L N_R \big), 
&\op_{\rm SL}&=\big(\bar c_R b_L\big)\big(\bar \tau_L N_R\big),
\\
\label{eq:QV}
\op_{\rm VR}&=\big(\bar c_R \gamma^\mu  b_R\big)\big(\bar \tau_R \gamma_\mu N_R \big), 
& \op_{\rm T}&=\big(\bar c_L \sigma^{\mu\nu} b_R \big)\big(\bar \tau_L \sigma_{\mu\nu} N_R\big).
\end{align}
\end{subequations}
The scalar and tensor operators run under the Renormalization Group. The RG evolution from $M > m_t$ to $\mu < m_b$ gives at one-loop order in the leading log approximation for the Wilson coefficients at the low scale \cite{Freytsis:2015qca,Dorsner:2013tla}, for $X = \text{SR},\text{SL},\text{T}$,
\beq
\begin{split}
	c_X(\mu)  & = \bigg[\frac{\alpha(m_b)}{\alpha(\mu)}\bigg]^{\gamma_X/2\beta_0^{(4)}}\bigg[\frac{\alpha(m_t)}{\alpha(m_b)}\bigg]^{\gamma_X/2\beta_0^{(5)}}\bigg[\frac{\alpha(M)}{\alpha(m_t)}\bigg]^{\gamma_X/2\beta_0^{(6)}}c_X(M)\,\\
	& \equiv \rho_X(\mu; M) c_X(M)\,,
\end{split}
\eeq
with anomalous dimensions $\gamma_{\text{SR},\text{SL}} = -8$, $\gamma_{\text{T}} = 8/3$ and the one loop $\beta$-function coefficient $\beta^{(n)}_0 = 11 - 2 n /3$. The running of $c_{\text{SR},\text{SL},\text{T}}$ depends only weakly on the high scale $M$, and hereafter we set $M = \leff$. Fixing the scale low scale to $\mu = \sqrt{m_c m_b}$ --  anticipating the chosen matching scale of QCD onto HQET for the $B \to \dds$ form factor parametrization -- one finds 
\begin{equation}
	\label{eqn:RGErhos}
	\rho_{\text{SR},\text{SL}} \simeq 1.7\,, \qquad \rho_{\text{T}} \simeq 0.84\,.
\end{equation}
Assuming the flavor indices are given in the mass eigenstate basis,
the NP operators~\eqref{eq:LeffEW} can be matched onto the operators~\eqref{eq:Q6} as $c_X{(\leff)} = C_{X}^{233}$,
neglecting the tiny mixing of active neutrinos into $N_R$. 
Note that the operators $\op_{\rm SR,T, SL}$ are accompanied by the $SU(2)_L$ related 
operators 
\begin{equation}
	\label{eqn:Os}
	\op_{\rm SR}^{s}=\big(\bar s_L  b_R\big)\big(\bar \nu_\tau N_R \big), \qquad\quad \op_{\rm T}^{s}= \big(\bar s_L \sigma^{\mu\nu} b_R \big)\big(\bar \nu_\tau \sigma_{\mu\nu} N_R\big),  
\end{equation}
and $\big(\bar c_R t_L\big)\big(\bar \nu_\tau N_R\big)$. The Wilson coefficients of these operators, $c^s_{\text{SR}, \text{T}, \text{SL}}$, correspond to $c_{\text{SR}, \text{T}, \text{SL}}$, respectively, up to 
one-loop or higher-order corrections.

Each of the dimension-six operators in Eq. \eqref{eq:Q6} can arise from the tree level exchange of a new state, either a scalar or a vector. 
The possible mediators, together with the Wilson coefficients $c_X$ they can contribute to, are listed in Table~\ref{tab:mediators}.
Two of these mediators are color singlets: the charged vector resonance $W_\mu'$, discussed extensively in Refs.~\cite{Asadi:2018wea,Greljo:2018ogz}, and the weak doublet 
scalar $\Phi$. The remaining mediators are leptoquarks, for which  we use the notation from Ref.~\cite{Dorsner:2016wpm}. 
In some cases the structure of the mediator Lagrangian, $\delta \mathcal{L}_{\text{int}}$, implies relations between the various Wilson coefficients, 
denoted by equalities in Table~\ref{tab:mediators}. In particular, for the $\tilde R_2$ and $S_1$ models, $c_{\text{SR}}{(\leff)} = \pm 4 c_{\text{T}}{(\leff)}$, which evolves to
\begin{equation}
	c_{\text{SR}}{(\mu)} = \pm 4 r\, c_{\text{T}}{(\mu)}\,, \qquad r \equiv \rho_{\text{SR}}/\rho_{\text{T}} \simeq 2.0\,,
\end{equation}
at the $B$ meson scale. 

\begin{table}[t] 
\begin{center}
\renewcommand{\arraystretch}{1.4}
\newcolumntype{C}{ >{\centering\arraybackslash } m{2cm} <{}}
\newcolumntype{D}{ >{\centering\arraybackslash } m{6cm} <{}}
\newcolumntype{E}{ >{\centering\arraybackslash } m{3cm} <{}}
\begin{tabular}{CCDE}
	\hline\hline
	mediator & irrep & $\delta \mathcal{L}_{\text{int}}$ & WCs
	 \\
	\hline
	$W_\mu'$ & $(1,1)_1$ 	&$g' \big( c_q  \bar u_R\gamma_\mu d_R  + c_N  \bar \ell_R\gamma_\mu N_R\big) W'^{\mu}$&$c_{\rm VR}$		
	\\
	$\Phi$ & $(1,2)_{1/2}$ 	& $y_u \bar u_R Q_L \epsilon \Phi+y_d \bar d_R Q_L  \Phi^\dagger+y_N \bar N_R L_L \epsilon \Phi$&$c_{\rm SL}{(\mu)},\quad c_{\rm SR}{(\mu)}$	
	\\
	$U_1^\mu$ & $(3,1)_{2/3}$ 	& $\big(\alpha_{LQ} \bar L_L\gamma_\mu Q_L +  \alpha_{\ell d}\bar \ell_R\gamma_\mu d_R\big) U_1^{\mu\dagger}+  \alpha_{uN} \big(\bar u_R\gamma_\mu N_R\big) 	U_1^{\mu}$ &  $c_{\rm SL}{(\mu)},\quad c_{\rm VR}$
	\\
	$\tilde R_2$ & $(3,2)_{1/6}$ 	& $\alpha_{Ld} \big(\bar L_L d_R\big) \epsilon \tilde R_2^\dagger+ \alpha_{QN} \big(\bar Q_L N_R\big) \tilde R_2$ & $c_{\rm SR}{(\mu)}=4r c_{\rm T}{(\mu)}$
	\\
	$S_1$ & $(\bar 3,1)_{1/3}$& $z_u (\bar U_R^c \ell_R)S_1+z_d (\bar d^c_R N_R)S_1+z_Q (\bar Q_L^c \epsilon L_L)S_1$ & $c_{\rm VR},$ $c_{\rm SR}{(\mu)}=-4r c_{\rm T}{(\mu)}$
	 \\
\hline
\hline
\end{tabular}
\end{center}
\caption{ \label{tab:mediators} The tree-level mediators that can generate the four-fermion operators with right-handed neutrino, $N_R$, in Eqs. \eqref{eqn:O}. The relevant Wilson coefficients are shown in the final column, explicitly defined at scale $\mu$ where relevant, and including the factor $r \equiv \rho_{\text{SR}}/\rho_{\text{T}} \simeq 2.0$.
}
\end{table}

For completeness, we list the remaining $b\to c \tau \bar N_R$ dimension-6 operators at $\mu\sim m_{c,b}$, 
\begin{subequations}
\label{eqn:Op}
\begin{align}
\label{eq:QSc}
\op_{\rm SR}'&=\big(\bar c_L  b_R\big)\big(\bar \tau_R N_R^c \big), 
&\op_{\rm SL}'&=\big(\bar c_R b_L\big)\big(\bar \tau_R N_R^c\big),
\\
\label{eq:QVc}
\op_{\rm VR}'&=\big(\bar c_R \gamma^\mu  b_R\big)\big(\bar \tau_L \gamma_\mu N_R^c \big), 
&\op_{\rm VL}'&=\big(\bar c_L \gamma^\mu b_L \big)\big(\bar \tau_L\gamma_\mu N_R^c \big),
\\
\label{eq:QTc}
\op_{\rm T}'&=\big(\bar c_R \sigma^{\mu\nu} b_L \big)\big(\bar \tau_R \sigma_{\mu\nu} N_R^c\big), & \op_{\rm VL}&=\big(\bar c_L \gamma^\mu b_L \big)\big(\bar \tau_R\gamma_\mu N_R \big).
\end{align}
\end{subequations}
The generation of these operators from the electroweak scale four-Fermi operators \eqref{eqn:QVL}--\eqref{eqn:QVLp} requires additional insertions of the Higgs vev, $v_{\text{EW}}$, and, apart from $\op_{\rm VL}$, also the left-handed sterile neutrino $N_R^c$. These $\op_a'$ operators are the same as those in Ref.~\cite{Freytsis:2015qca}, but with $N_R^c$ 
replacing the SM neutrino $\nt$. Eqs.~\eqref{eqn:O} and~\eqref{eqn:Op} together form a complete basis of $b\to c \tau \bar N_R$ dimension-six four-fermion operators. 
Since the Wilson coefficients of the operators in Eq.~\eqref{eqn:Op} are suppressed by additional powers of $v_{\text{EW}}/\leff$, we will only focus on the dimension-6 operators listed in Eq.~\eqref{eq:Q6} and~\eqref{eqn:O} in the remainder of this paper.

\subsection{Fits to $R(\dds)$ data}
\label{sec:Fits}
The present experimental world-averages for $R(\dds)$ are~\cite{HFAG}
\begin{equation}
	\label{eqn:RDDsdata}
	R(D)\big|_{\rm exp} = 0.407 \pm 0.046\,, \qquad R(D^*)\big|_{\rm exp} = 0.304\pm 0.015\,, \qquad \text{corr.} = -0.20\,.
\end{equation}
The SM predictions, e.g. making use of the model-independent form factor fit `$L_{w\ge1}\text{+SR}$' of Ref.~\cite{Bernlochner:2017jka}  (see also Refs.~\cite{Jaiswal:2017rve,Bigi:2017jbd}), are
\begin{equation}
	R(D)\big|_{\rm th} = 0.299 \pm 0.003, \qquad R(D^*)\big|_{\rm th}  = 0.257 \pm 0.003,\qquad \text{corr.} = +0.44\,.
\end{equation} 
With the addition of a right-handed neutrino decay mode, the $\bddstn$ decays become an incoherent sum of two contributions: the SM 
decay $b\to c\tau \ntb$ and the new mode $b\to c\tau \bar N_R$. The $N_R$ contributions therefore increase both of the $\bddstn$ branching ratios above the SM predictions, as would be required to explain the experimental measurements of $R(\dds)$.

\begin{figure}[t]
\begin{center}
\includegraphics[width=7.5cm]{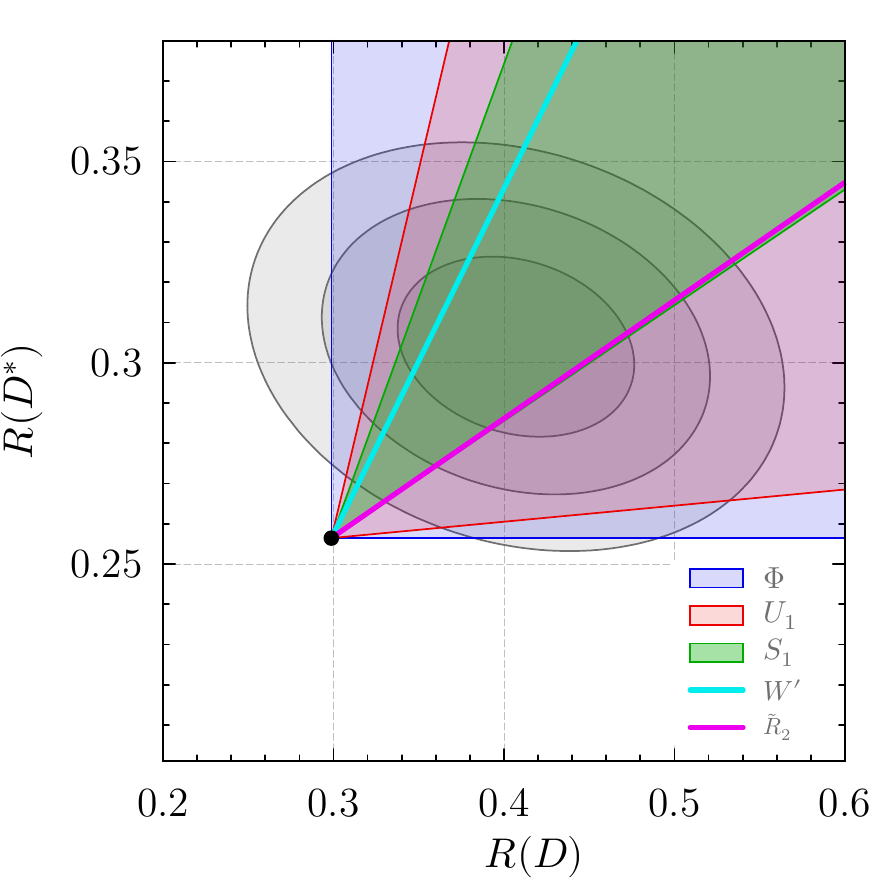}
\end{center}
\caption{The enhancements of $R(\dds)$ from $b \to c \tau \bar N_R$ decays for various simplified models. The world average experimental $1\sigma$, $2\sigma$, and $3\sigma$ fit regions are shown in decreasing shade of gray. 
The SM point is denoted by a black dot. }
\label{fig:RDDs}
\end{figure}

In Fig.~\ref{fig:RDDs}, we show for each simplified model of Table \ref{tab:mediators} the accessible contours or regions in the $R(D)-R(D^*)$ plane, compared to the experimental data. 
The predictions for NP corrections to $R(\dds)$ are obtained from the expressions in Ref.~\cite{Ligeti:2016npd}, making use of the form factor fit `$L_{w\ge1}\text{+SR}$' 
of Ref.~\cite{Bernlochner:2017jka}. This fit was  performed at next-to-leading order in the heavy quark expansion, with matching scale $\mu = \sqrt{m_b m_c}$ and quark masses defined 
in the $\Upsilon(1S)$ scheme, relevant for a self-consistent treatment of the $B_c \to \tau \nu$ constraints below. 
The $W'$ and $\tilde{R}_2$ simplified models have only a single free Wilson coefficient and are constrained to a contour:
Since the $N_R$ contributions add incoherently to the SM, the phase of each Wilson coefficient is unphysical.
By contrast, $\Phi$, $U_1$, and $S_1$ have two free Wilson coefficients, corresponding to two free magnitudes and a physical relative phase, permitting them to span a region. 

\begin{figure}[t]
\begin{center}
\includegraphics[width=0.31\linewidth]{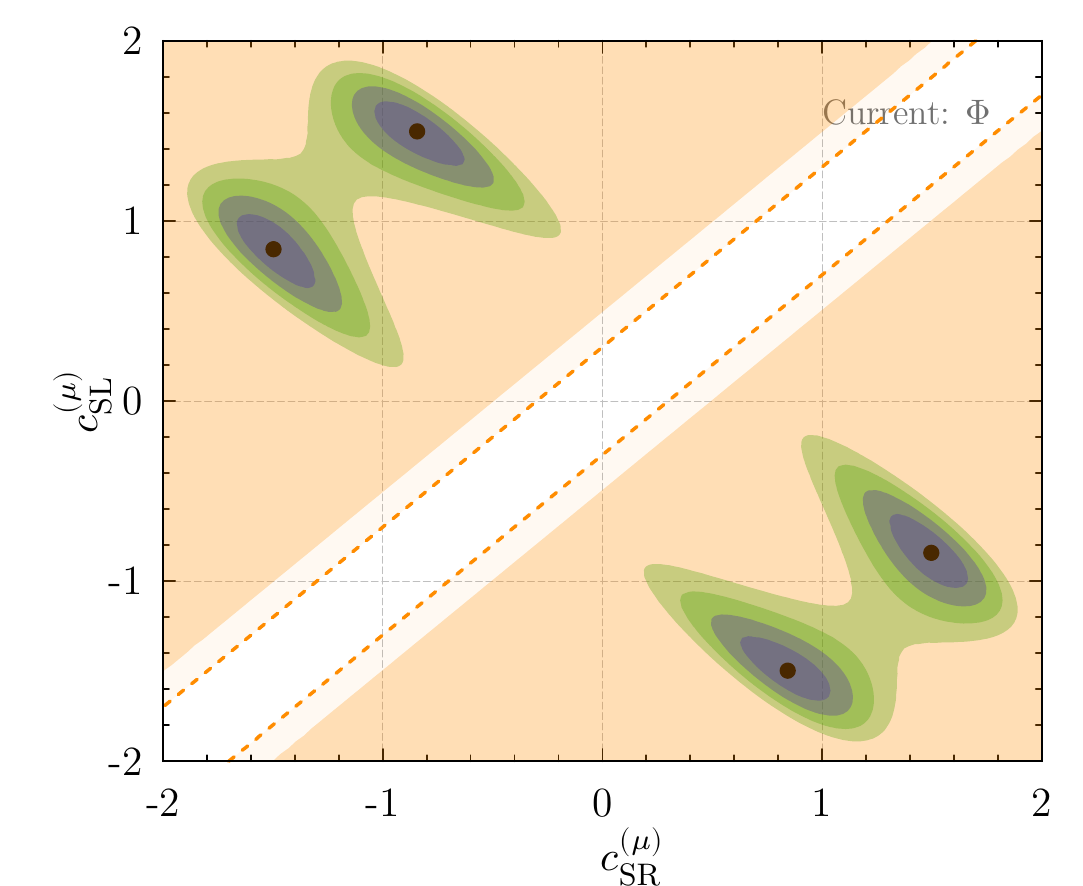}
\includegraphics[width=0.31\linewidth]{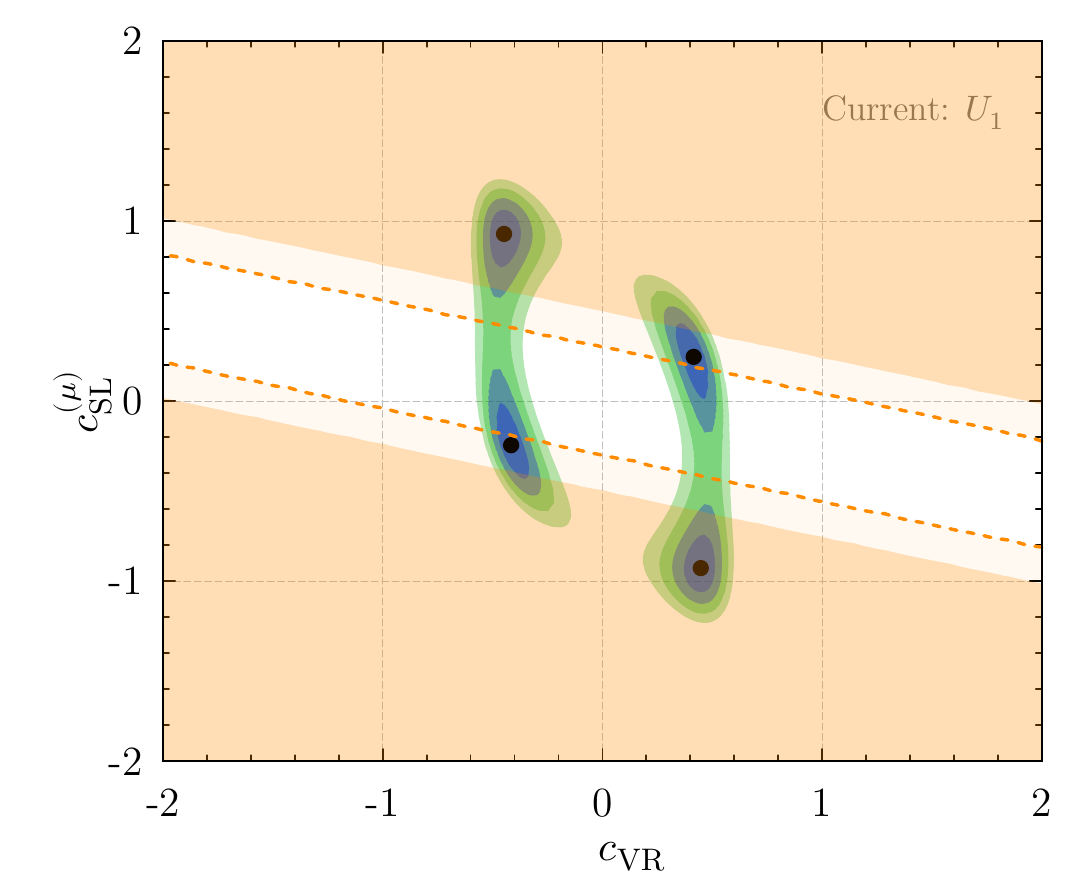}
\includegraphics[width=0.31\linewidth]{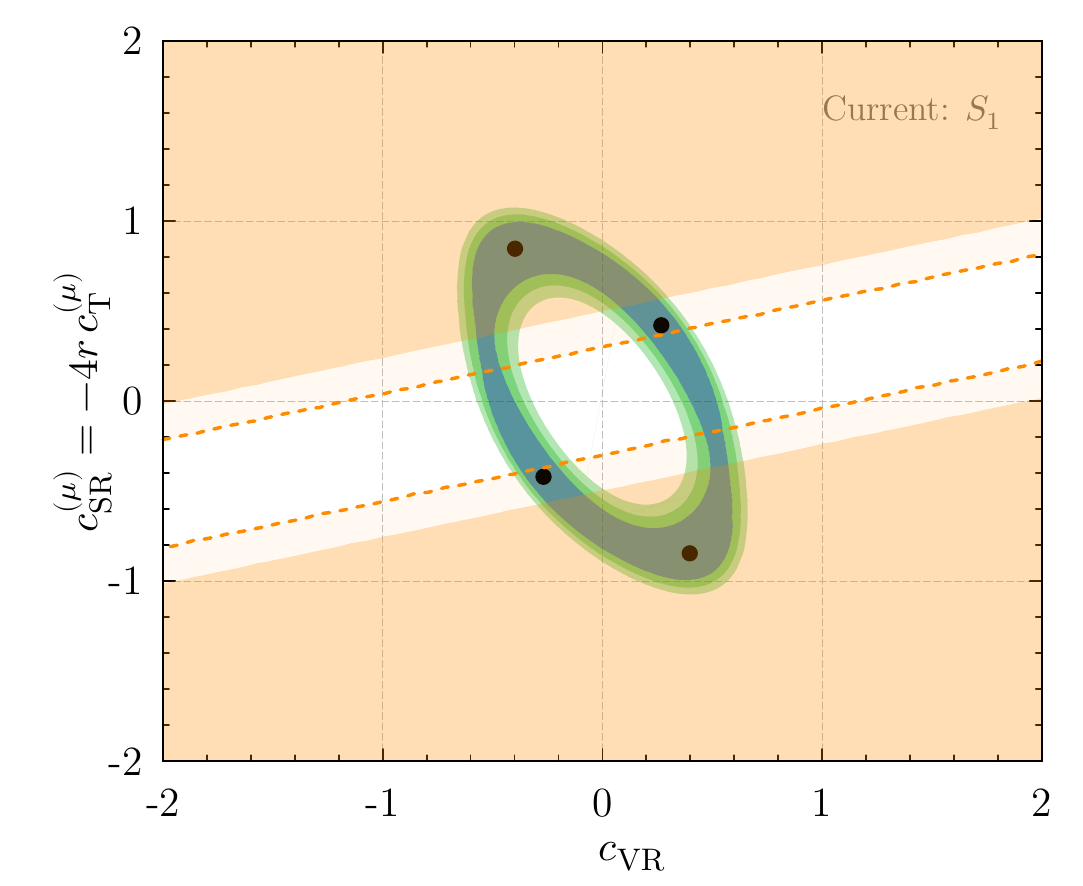}\\
\includegraphics[width=0.31\linewidth]{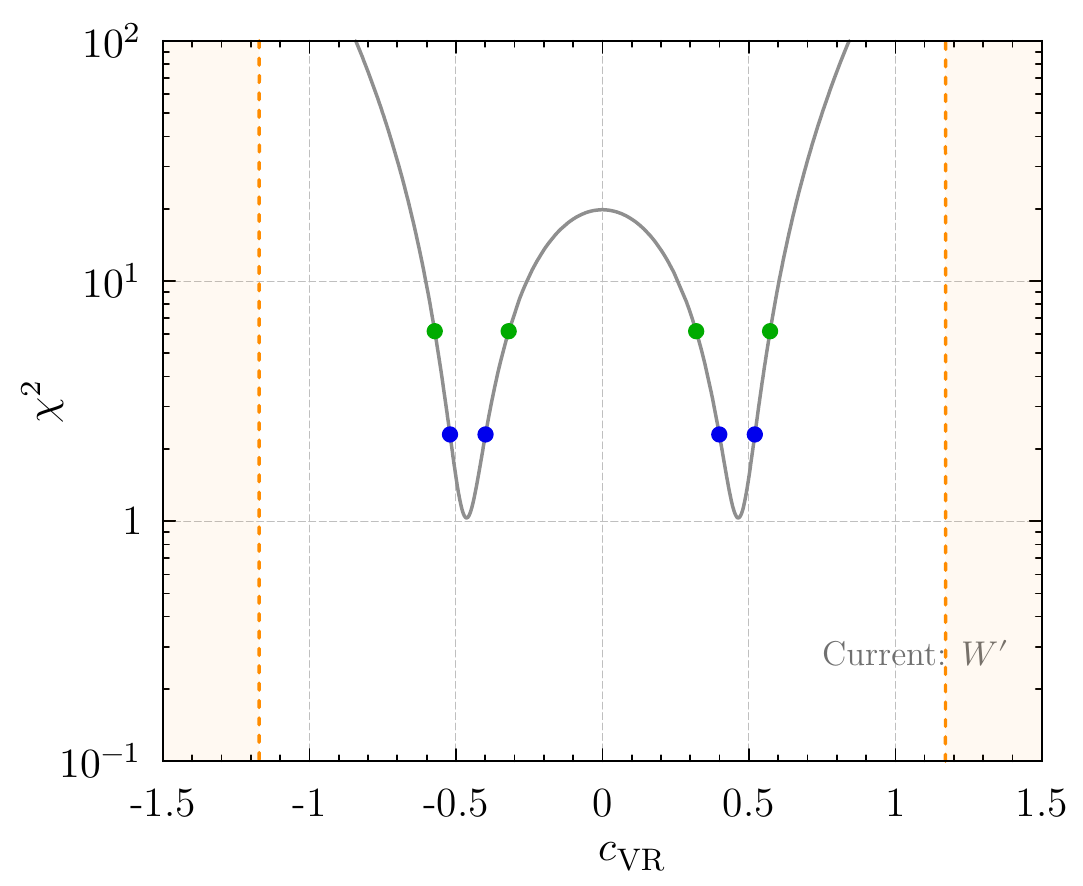}
\includegraphics[width=0.31\linewidth]{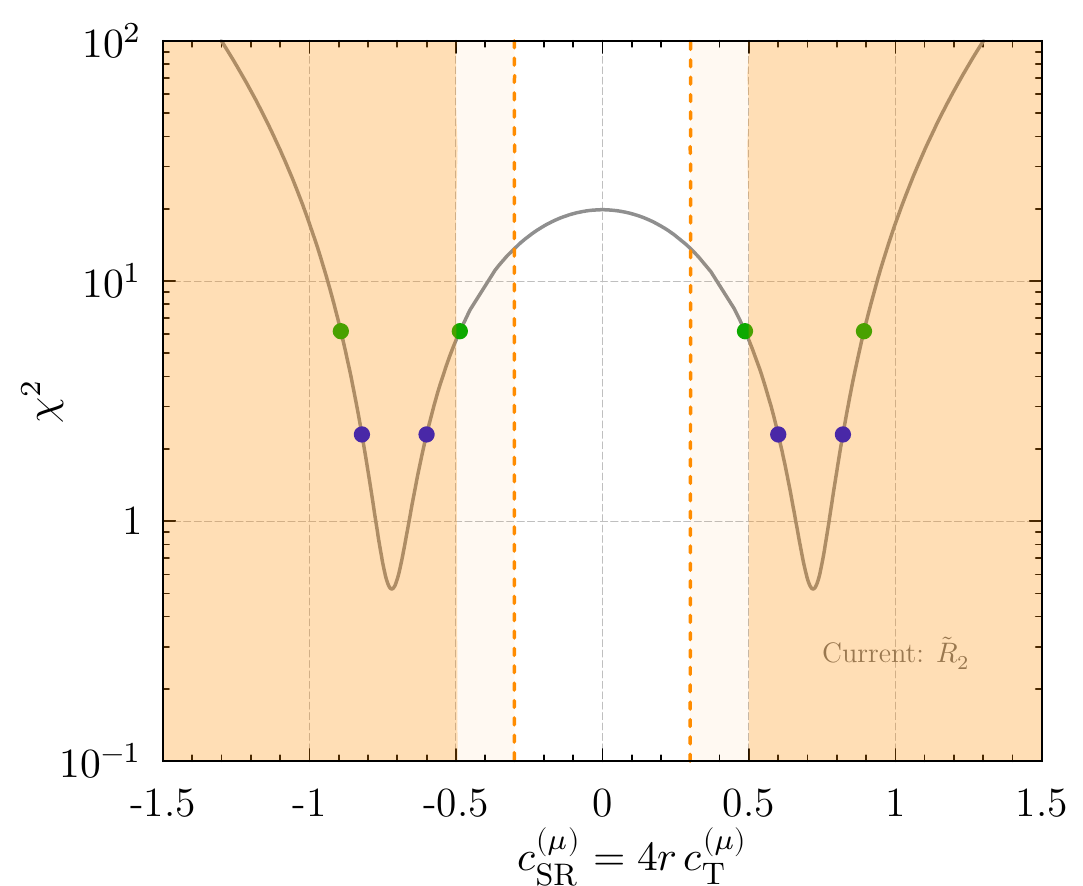}
\end{center}
\caption{{\it Top:} The fit regions for $\Phi$, $U_1$, and $S_1$ models with respect to the $R(\dds)$ results~\eqref{eqn:RDDsdata} in the relevant Wilson coefficient spaces, assuming that all Wilson coefficients are real. Shown are $0.5\sigma$, $1\sigma$ CLs (dark, light blue) and $1.5\sigma$, $2\sigma$ CLs (dark, light green). Best fit points are shown by black dots. {\it Bottom:} The $\chi^2$ ($\text{dof} = 2$) for the $W'$ and $\tilde{R}_2$ models
 in the relevant Wilson coefficient space. The $1\sigma$ and $2\sigma$ CLs are shown by blue and green dots, respectively. Also shown are $B_c \to \tau \nu$ exclusion regions
 requiring $\text{Br}[B_c \to \tau \nu] <10\%$ (dark orange). For a sense of scaling, a more aggressive $\text{Br}[B_c \to \tau \nu] < 5\%$ exclusion region is demarcated by a dashed orange line.}
\label{fig:fits}
\end{figure}

Assuming first that all Wilson coefficients are real, we show in Fig.~\ref{fig:fits} the $0.5\sigma$, $1\sigma$ CLs (dark, light blue)
and $1.5\sigma$, $2\sigma$ CLs (dark, light green) in the relevant Wilson coefficient spaces for each simplified model.
These CLs are generated by the $\chi^2$ defined with respect to the $R(\dds)$ experimental data and correlations~\eqref{eqn:RDDsdata}, not including the possible effects of NP errors. That is, 
\begin{equation}
\label{eqn:chi2}
	\chi^2 = \bm{v}^T \sigma_{R(\dds)}^{-1} \bm{v}\,, \qquad \bm{v} = \big(R(D)_{\text{th}} - R(D)_{\text{exp}}\,, R(D^*)_{\text{th}} - R(D^*)_{\text{exp}}\big)\,,
\end{equation}	
The $\chi^2$ CLs (dof =2) in Fig~\ref{fig:fits} then correspond simply to projections of the CL ellipses in Fig.~{\ref{fig:RDDs}. 
We will hereafter refer to the minimal $\chi^2$ points in the WC space for each simplified model 
as the model's `best fit' points with respect to the $R(\dds)$ results~\eqref{eqn:RDDsdata}, though it should be emphasized that this is not the same as a NP WC fit to the 
experimental data, which would require inclusion of the NP errors in the underlying experimental fits. 
In Fig.~\ref{fig:fits} the best fit points are shown by black dots, with explicit values provided in Table~\ref{tab:bfp}. For the $W'$ and $\tilde{R}_2$ models, we show the explicit $\chi^2$, as well as the intervals corresponding to $1\sigma$ and $2\sigma$ CLs ($\text{dof} =2$). 

The additional NP currents from the operators~\eqref{eqn:O} also incoherently modify the $B_c \to \tau \nu$ decay rate 
with respect to the SM contribution (cf. Refs.~\cite{Li:2016vvp,Alonso:2016oyd}), such that
\begin{equation}
\label{eq:BcTauNu}
	\Br(B_c \to \tau \nu) = \frac{\tau_{B_c} f_{B_c}^2m_{B_c} m_\tau^2}{64 \pi \leff^4}\big(1 - m_\tau^2/m_{B_c}^2\big)^2\bigg[1 + \bigg| c_{\text{VR}} + \frac{m_{B_c}^2(c_{\text{SL}}^{(\mu)} - c_{\text{SR}}^{(\mu)})}{m_\tau(\ov m_b + \ov m_c)} \bigg|^2\bigg]\,,
\end{equation}
in which $f_{B_c} \simeq 0.43$\,GeV~\cite{Colquhoun:2015oha} and $\tau_{B_c} \simeq 0.507$\,ps~\cite{PDG}, and $\ov m_{c,b}$ are the $\ov{\text{MS}}$ quark masses,
 obeying $m_Q  \simeq \ov m_Q (1+ \alpha_s/\pi[4/3 - \ln(m_Q^2/\mu^2)])$. Self-consistency with the form factor treatment 
of Ref.~\cite{Bernlochner:2017jka} requires these masses to be evaluated at $\mu = \sqrt{m_b m_c}$ in the $\Upsilon(1S)$ quark mass scheme. 
In Fig.~\ref{fig:fits} we show the corresponding 
exclusion regions for the relevant Wilson coefficient spaces (shaded orange), requiring $\Br(B_c \to \tau\bar\nu) < 10\%$~\cite{Li:2016vvp,Alonso:2016oyd}. 
For a sense of scaling, we also include a more aggressive $\Br(B_c \to \tau\bar\nu) < 5\%$ exclusion demarcated by a dashed orange line.
One sees that the $\Phi$ simplified model is excluded, while the $\tilde{R}_2$ $2\sigma$ CL is not quite excluded by the $\Br(B_c \to \tau\bar\nu) < 10\%$ constraint.
The $U_1$ and $S_1$ best fit points are in mild tension with the aggressive $\Br(B_c \to \tau\bar\nu) < 5\%$ exclusion, but also exhibit allowed regions for their $1\sigma$ CLs.

\begin{table}[t]
\begin{center}
\renewcommand{\arraystretch}{1.25}
\newcolumntype{C}{ >{\centering\arraybackslash $} m{1.25cm} <{$}}
\newcolumntype{E}{ >{\centering\arraybackslash $} m{2.75cm} <{$}}
\newcolumntype{D}{ >{\centering\arraybackslash $} m{3cm} <{$}}
\begin{tabular}{CE|DC|DC}
	\hline\hline
	 &  & \multicolumn{2}{c}{Real} & \multicolumn{2}{c}{Phase-optimized} \\  \cline{3-6}
	\text{Model} & \text{WCs} & \text{Best fit} & \chi^2 & \text{Best fit} & \chi^2 \\
	\hline 
	W' 			&  c_{\text{VR}} 				& \pm0.46 	& 1.0  		& \text{--} 		& \text{--} 	\\
	\hline
	\tilde{R}_2 	& c_{\text{SR}}^{(\mu)}=4r\, c_{\text{T}}^{(\mu)} 		& \pm0.72 	& 0.5  		& \text{--} 		& \text{--} 	\\
	\hline
	\multirow{3}{*}{$\Phi$} & \multirow{3}{*}{$\{c_{\text{SR}}^{(\mu)}, c_{\text{SL}}^{(\mu)}\}$}		& \{\pm1.50, \mp0.84\}	& 0.	 	& \{1.50, -0.84\} 				& 0. \\ 
	& & & 																						& \{1.21,\pm1.21 e^{\pm i0.17\pi}\} 	& 0. \\
	& 															& \{\pm0.84, \mp1.50\}	  	& 0. 		& \{0.84, -1.50\} 				& 0. \\ 
	\hline
	\multirow{2}{*}{$U_1$} & \multirow{2}{*}{$\{c_{\text{VR}}, c_{\text{SL}}^{(\mu)}\}$}		& \{\pm0.45,\mp0.93\}	&  0. 		& \{0.45,-0.93\} 	& 0. \\
	& 																	& \{\pm0.42,\pm0.24\}	&  0.		& \{0.42, 0.24\}	& 0. \\
	\hline
	\multirow{2}{*}{$S_1$} &  	 \{c_{\text{VR}}, 							& \{\pm0.40,\mp0.85\} 	& 0.		& \{0.40, -0.85\} & 0. \\
	&				c_{\text{SR}}^{(\mu)}=-4r\,c_{\text{T}}^{(\mu)}\}			& \{\pm0.27,\pm0.42\} 	& 0.		& \{0.27, 0.42\} & 0.\\
	\hline\hline
 \end{tabular}
\end{center}
\caption{ \label{tab:bfp} Best fit points for each model with respect to the $R(\dds)$ results~\eqref{eqn:RDDsdata}, for real and phase-optimized Wilson coefficients. In the phase-optimized case, we show best fits up to an overall phase, by choosing the first WC to be real and positive definite.}
\end{table}

Lifting the requirement of real Wilson coefficients, the $\Phi$, $U_1$, and $S_1$ models now have a physical phase and inhabit a three dimensional parameter space: 
two Wilson coefficient magnitudes, schematically denoted $|c_{1,2}|$, and a relative phase $\varphi$. For the basis of Wilson coefficients defined by the $N_R$ operators~\eqref{eqn:O}, however, the amplitudes for the $\bddsln$ decay alone have no physical relative phases. (Physical phases do exist once the $D^*$ and $\tau$ decay amplitudes are included.) Consequently, 
for a given choice of $|c_{1,2}|$, there may exist a nontrivial value for $\cos\varphi$ that minimizes the $\chi^2$ for $R(\dds)$ in Eq.~\eqref{eqn:chi2}. We refer to this scenario as the `phase optimized' case, 
denoted $\varphi = \varphi_0(|c_1|, |c_2|)$. In explicit numerical terms, for the form factor and $R(\dds)$ inputs described above, the $\Phi$, $U_1$, and $S_1$ models have non-trivial solutions
\begin{align}
	\cos(\varphi_0) & = \begin{cases} 
		\dfrac{0.24- 0.51 |c_{\text{SR}}|^2 - 0.51 |c_{\text{SL}}|^2}{|c_{\text{SR}}||c_{\text{SL}}|}\,, &\qquad \Phi\,, \\[5pt]
		\dfrac{0.38 - 1.38 |c_{\text{VR}}|^2 - 0.60 |c_{\text{SL}}|^2}{|c_{\text{VR}}| |c_{\text{SL}}|}\,, & \qquad U_1\,,\\[5pt]
		\dfrac{0.32 - 1.40 |c_{\text{VR}}|^2 - 0.61|c_{\text{SR}}|^2}{|c_{\text{VR}}||c_{\text{SR}}|}\,, & \qquad S_1\,,
	\end{cases}
\end{align}
valid only on the domain $|\cos(\varphi_0)| < 1$, and otherwise $\cos(\varphi_0) = \pm1$. These phase-optimized CLs for the $\Phi$, $U_1$, and $S_1$ models are shown in Fig.~\ref{fig:fitsph}, with the explicit best fit points listed in Table~\ref{tab:bfp}. The best fit points for $U_1$ and $S_1$ remain the same, and one sees that these models continue to have non-excluded $1\sigma$ CLs. An additional best fit point emerges for the $\Phi$ simplified model; however, this model remains excluded, and we therefore do not consider it further in this paper.

\begin{figure}[t]
\begin{center}
\includegraphics[width=0.31\linewidth]{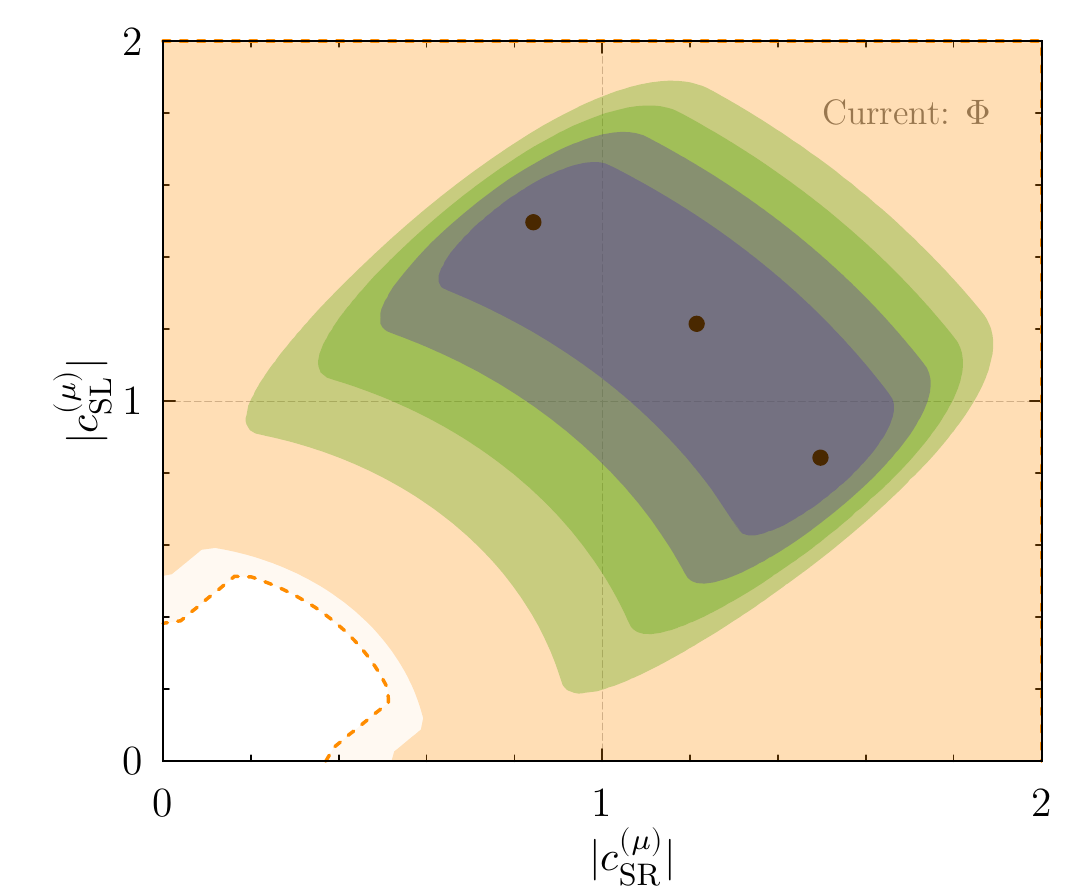}
\includegraphics[width=0.31\linewidth]{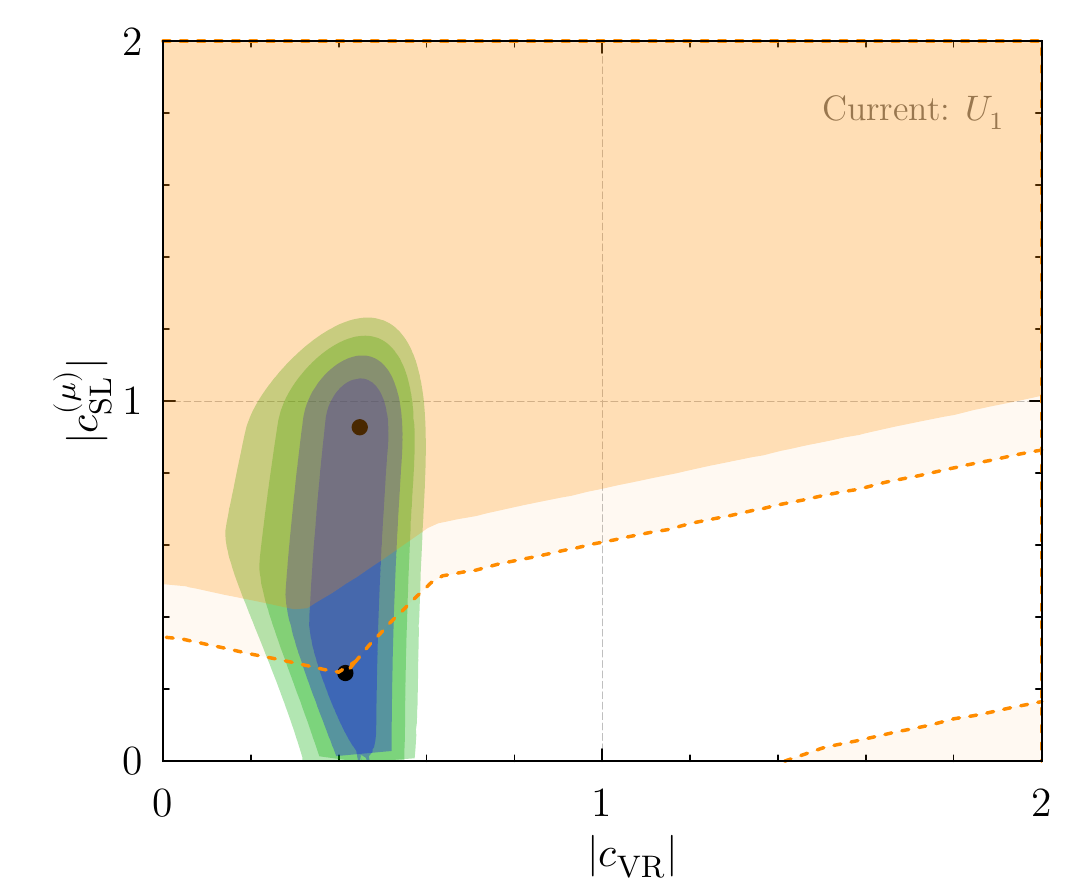}
\includegraphics[width=0.31\linewidth]{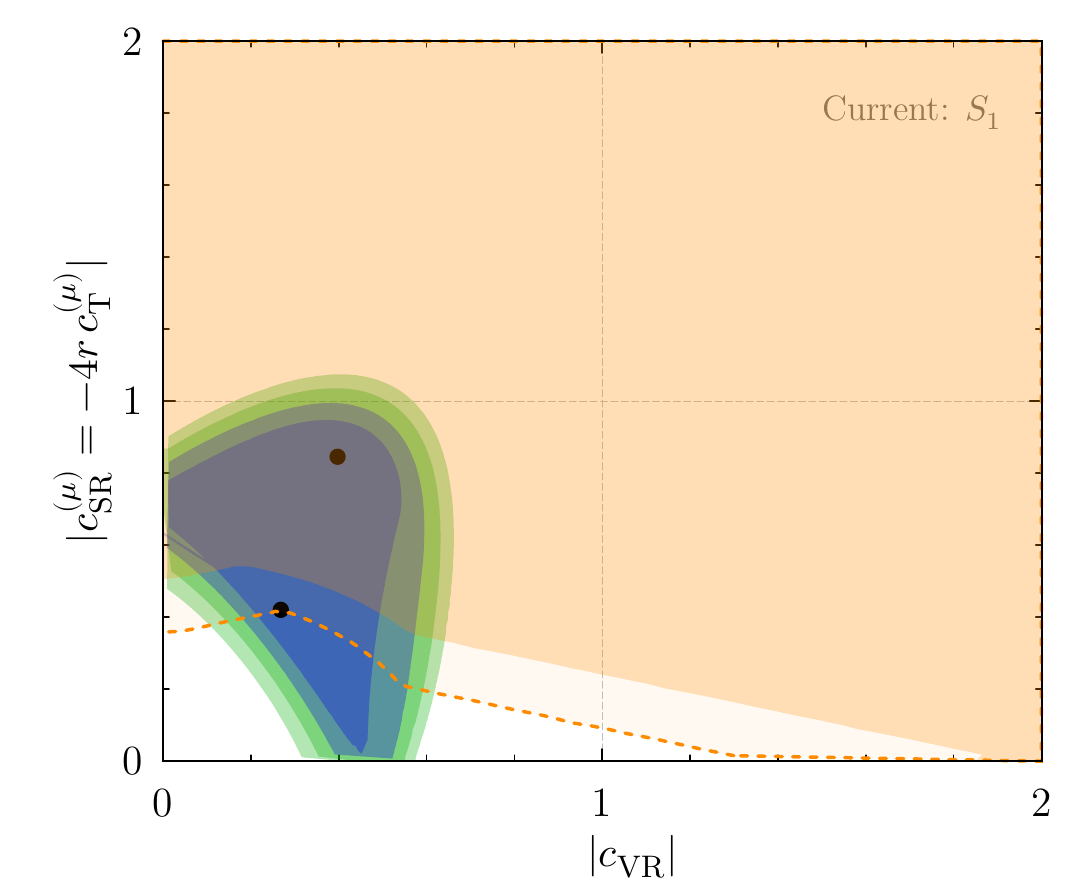}
\end{center}
\caption{The phase-optimized CLs with respect to the $R(\dds)$ results~\eqref{eqn:RDDsdata} for $\Phi$, $U_1$, and $S_1$ models in the relevant Wilson coefficient spaces, 
imposing the condition $\varphi = \varphi_0(|c_1|, |c_2|)$. 
Shown are $0.5\sigma$, $1\sigma$ CLs (dark, light blue) and $1.5\sigma$, $2\sigma$ CLs (dark, light green). Also shown are $B_c \to \tau \nu$ exclusion regions
 requiring $\text{Br}[B_c \to \tau \nu] <10\%$ (dark orange). For a sense of scaling, a more aggressive $\text{Br}[B_c \to \tau \nu] < 5\%$ exclusion region is demarcated by a dashed orange line. Best fit points are shown by black dots.}
\label{fig:fitsph}
\end{figure}

Finally, the exchange of mediators that generates the $c_{\rm SR,T}$ Wilson coefficients also results in $c_{\rm SR,T}^{s}$ of similar size (see Eq.~\eqref{eqn:Os}). The two operators in Eq.~\eqref{eqn:Os} contribute to  $b\to s \nu \bar \nu$ rates. This gives, for instance, for the $B\to K \nu \bar \nu$ decay rate (far enough from the kinematic threshold so that we can neglect all the final state masses)~\cite{Kamenik:2009kc, Kamenik:2011vy}
\begin{align}
	\frac{d\Gamma_{B\to K\nu\bar\nu}}{dz}\Big/ \frac{d \Gamma_{B\to K\nu\bar\nu}}{dz}\biggr|_{\rm SM}&=1+ z \frac{32\pi^2}{3 \alpha^2}\bigg|\frac{V_{cb}}{C_{\nu\nu}^{\rm SM} V_{tb} V_{ts}^*}\bigg|^2  \bigg[\frac{3}{8} \frac{\big(c_{\rm SR}^{s}\big)^2}{(1-z)^2} \frac{f_0^2}{f_+^2} + \big(c_{\rm T}^{s}\big)^2 \frac{f_T^2}{f_+^2}\bigg] \nn 
	\\
	&	\simeq 1+ 5 \times 10^{4}\, z  \bigg[\frac{3}{8}\frac{\big(c_{\rm SR}^{s}\big)^2}{(1-z)^2}  \frac{f_0^2}{f_+^2} + \big(c_{\rm T}^{s}\big)^2 \frac{f_T^2}{f_+^2} \bigg], \label{eq:BKnunu}
\end{align}
with the three $B\to K$ form factors, $f_0(q^2)$, $f_+(q^2)$, $f_T(q^2)$, functions of $q^2$, the invariant mass squared of the neutrino pair, and $z=q^2/m_B^2$. The present experimental bound, $Br(B^+\to K^+\nu\bar\nu)<1.6 \times 10^{-5}$ \cite{Tanabashi:2018oca}, is only a factor of a few above the SM prediction, $Br(B^+\to K^+\nu\bar\nu)|_{\rm SM}\simeq 4 \times 10^{-6}$ \cite{Buras:2014fpa}. This implies that $c_{\rm SR}^{s}$ and $c_{\rm T}^{s}$ are highly suppressed, to the level of ${\mathcal O}(10^{-2})$, introducing tensions with the required size of $c_{\rm SR},c_{\rm T}$ to explain the $R(D^{(*)})$ anomaly. 
In the single mediator exchange models in Table \ref{tab:mediators}, this means that the product $\alpha_{Ld}^3 \alpha_{QN}^2$ for $\tilde R_2$ and the product $z_d^3 z_{Q}^2$ for $S_1$ (and $y_d^{32}$ for $\Phi$) need to be much smaller than what is required to explain $R(D^{(*)})$. This excludes the $\tilde R_2$ as a simple one mediator solution to $R(D^{(*)})$: Additional operators coupling to the second generation of quark doublets must be introduced, whose couplings are tuned appropriately to suppress the contributions to $b\to s\nu\bar \nu$. However, this approach would in turn induce large radiative contributions to the neutrino masses, which would also need to be tuned away (see Sec.\,\ref{sec:sterile:neutrino}). The $S_1$ model also generates too large a $b\to s\nu\bar\nu$ transition rate at the (non-excluded) best fit point, where $c_{\rm SR}$ and $c_{\rm T}$ are nonzero. The dangerous $b\to s\nu\bar\nu$ contribution can be suppressed by taking $z_Q^{23} \to 0$ (see Table\,\ref{tab:mediators}), which forces $c_{\rm SR}=c_{\rm T}\to0$. This $c_{\rm SR}=c_{\rm T}=0$ point leads to only a small change in $\chi^2$, corresponding to a less than 0.5\,$\sigma$ shift in significance, see Fig. \ref{fig:fits}. 

\subsection{Differential distributions}
\label{sec:eftdd}
The reliability of the above $R(\dds)$ fit results turns upon the underlying assumption that the differential distributions, and hence experimental acceptances, of the $\bddstn$ decays are not 
significantly modified in the presence of the NP currents. The $\bddstn$ branching ratios are extracted from a simultaneous float of background and signal data, so that significant modification 
of the acceptances versus the SM template may alter the extracted values. 

To estimate the size of these potential effects, we examine the cascades $\Bbar \to (D^* \to D\pi)(\tau \to \ell \bar\nu_\ell \nt)\bar\nu$ and $\Bbar \to D(\tau \to \ell \nu \nu)\nu$, comparing the purely SM predictions with the predictions for the $2\sigma$ fit regions of the simplified models. We take $N_R$ to be massless, and include the phase space cuts,
\begin{equation}
	\label{eqn:PScut}
	q^2 = (p_B - p_{\dds})^2 > 4~\text{GeV}^2\,, \qquad E_{\ell} > 400~\text{MeV}\,,\qquad m^2_{\text{miss}} > 1.5~\text{GeV}^2\,,
\end{equation}
as an approximate simulation of the BaBar and Belle measurements performed in Refs.~\cite{Lees:2013uzd,Huschle:2015rga}. These distributions are generated as in Ref.~\cite{Ligeti:2016npd}, using a 
preliminary version of the \texttt{Hammer} library~\cite{Hammer_paper}. In Appendix~\ref{app:dd} we show the variation of the normalized differential distributions over the $2\sigma$ fit regions 
in Fig.~\ref{fig:fits} -- i.e. assuming real couplings, for simplicity -- for the detector observables $E_D$, $E_\ell$, $m^2_{\text{miss}}$, $\cos\theta_{D\ell}$ and $q^2$ compared to the SM distributions. 

As already found in Ref.~\cite{Greljo:2018ogz}, the variation of the $W'$ model with respect to the SM is negligible. However, the $\tilde{R}_2$, $U_1$ and $S_1$ theories, since they include interfering scalar and/or tensor currents, 
may significantly modify the spectra, as seen also in Ref.~\cite{Ligeti:2016npd} for the NP tensor current coupling to a SM neutrino. Thus, a fully self-consistent $R(\dds)$ fit for these models will require a forward-folded analysis by the experimental collaborations: Our analysis above and CLs should be taken only as an approximate guide, within likely $1\sigma$ variations in the values of $R(\dds)$.

\section{Collider constraints on simplified models}
\label{sec:simplified}

The simplified models are subject to low energy flavor constraints as well as bounds from collider searches.  
These depend crucially on the assumed flavor structure of the couplings in  Table \ref{tab:mediators}. Furthermore, the sensitivity of the collider searches depend on other open decay channels of the mediators. In this section, we discuss these constraints for the simplified models.

For the $S_1$ and $\tilde{R}_2$ models, the best fit points are naively excluded by bounds on $b\to s \nu\bar \nu$ transitions. These can be avoided by including higher dimensional operators, due to a new set of heavy states, inevitably introducing greater model dependence for LHC studies. To remain as model independent as possible, we study the collider signatures for these models using their ($B_c \to \tau\nu$ consistent) best fit points for $R(D^{(*)})$ as a benchmark, assuming that any new fields required to ameliorate large $b\to s \nu\bar \nu$ (and/or large neutrino mass contributions) are sufficiently heavy that they do not affect mediator production or decay.

\subsection{$W'$ coupling to right-handed SM fermions}
The charged vector boson $W_\mu'$ couples to $SU(2)_L$ singlets only, and transforms as $W'_\mu\sim (1,1)_1$, with
\beq
\label{eq:W':Lagr}
{\cal L}= \frac{g_V}{\sqrt2}  c_q^{ij} \bar u_R^i\slashed W' d_R^j  + \frac{g_V}{\sqrt2} c_N^{i}  \bar \ell_R^i\slashed W' N_R +{\rm h.c.},
\eeq
where $i,j=1,2,3$ are generational indices. As in Table \ref{tab:mediators}, the coefficients $c_q^{ij}$ and $c_N^i$ encode the flavor structure of the interactions, while $g_V$ is the overall coupling strength (in simple gauge models for $W'$ it can be identified with the gauge coupling constant \cite{Asadi:2018wea,Greljo:2018ogz}). A tree level exchange of $W'$ generates the operator $\op_{\rm VR}$, cf. eqs.~\eqref{eq:QV} and \eqref{eq:Leff:low}, with
\beq
\frac{c_{\rm VR}}{\Lambda_{\rm eff}^2}= - \frac{g_V^{2}c_q^{23}c_N^3}{2 m_{W'}^2}.
\eeq
The best fit values for $c_{\text{VR}}$ in Table~\ref{tab:bfp} then imply  \cite{Greljo:2018ogz}
\begin{equation}
	m_{W'} \simeq 540  \big|c_q^{23}c_N^{3}\big|^{1/2} \bigg[\frac{g_V}{0.6}\bigg] \bigg[\frac{40 \times 10^{-3}}{V_{cb}}\bigg]^{1/2}\,\GeV\,. \label{eq:fit}
\end{equation}

\begin{figure}[t]
\begin{center}
\includegraphics[width=7.5cm]{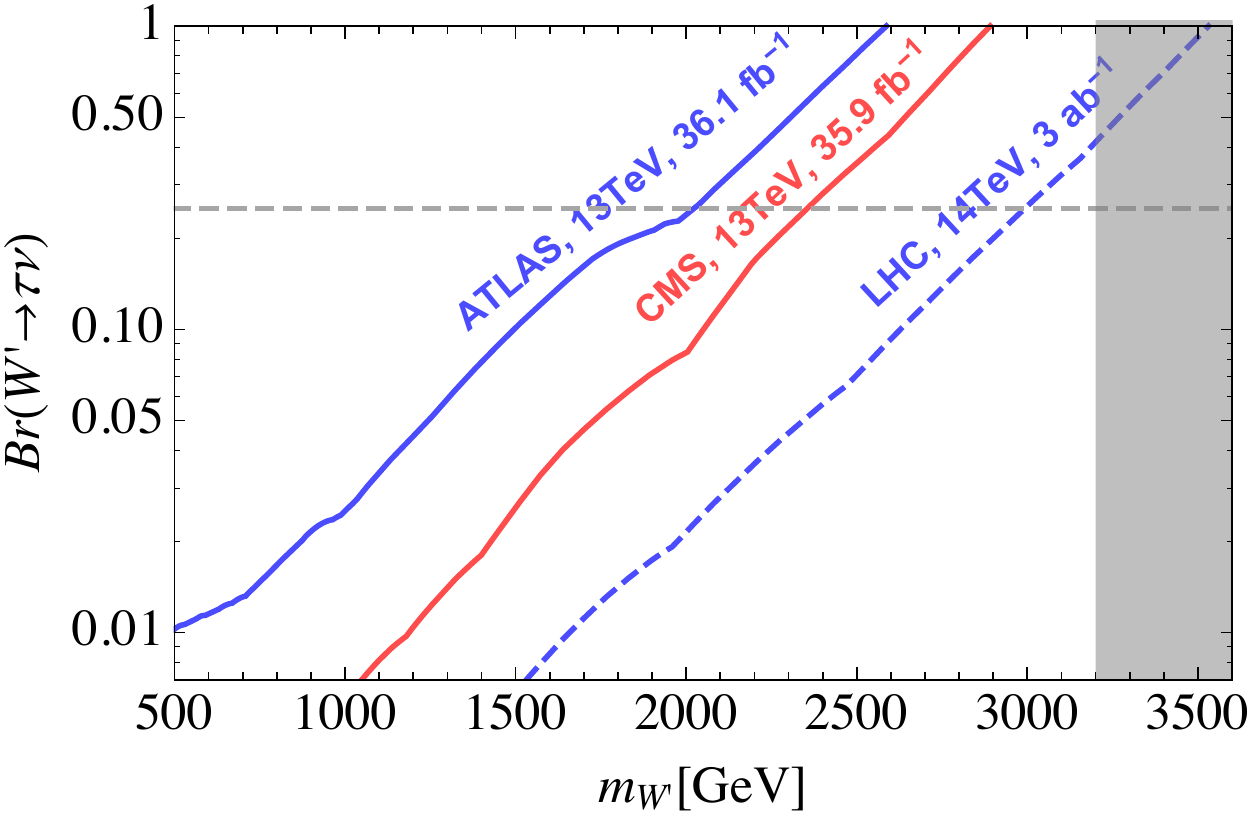}
\end{center}
\caption{The bound on $Br(W'\to \tau \nu)$ as a function of $W'$ mass from the 13 TeV ATLAS \cite{Aaboud:2018vgh} (solid blue) and CMS \cite{Sirunyan:2018lbg} (solid red) searches, as well as the projected reach at the end of the high-luminosity LHC run (dashed blue), for the case $c_q^{23}=c^3_N$, ${W'}$ mass given by Eq.\,\eqref{eq:fit} to fit to $R(\dds)$ data, and the $W'$ couplings to all the other SM quarks set to zero. In this case $Br(W'\to \tau \nu)=0.25$ (dashed grey line) if no other $W'$ decay channels are open. All the bounds assume narrow width for $W'$. The region excluded by unitarity is shaded in grey.}
\label{fig:W'bound}
\end{figure}

In Fig.~\ref{fig:W'bound} we show the minimal set of experimental constraints on such models, applicable to the simplified $W'$ model. 
For this plot we set $c_q^{23}=c^3_N$, take Eq.~\eqref{eq:fit} to provide the $W'$ mass that fits the $R(\dds)$ data, and 
set the $W'$ couplings to all other SM quarks to zero. 
For this scenario, the ATLAS search at 13 TeV with 36.1 fb${}^{-1}$ luminosity \cite{Aaboud:2018vgh} and the CMS search with 35.9 fb${}^{-1}$ \cite{Sirunyan:2018lbg} convert to a 95\,\% CL bounds on $\Br(W'\to \tau\nu)$ shown in Fig.\,\ref{fig:W'bound} (blue and red lines, respectively), see also Refs.~\cite{Khachatryan:2016jww,CMS:2016ppa} for previous bounds. The dashed blue line denotes a naive extrapolation of the expected bound from Ref.~\cite{Aaboud:2018vgh} to the end of the high-luminosity LHC Run 5, assuming 3000 fb${}^{-1}$ integrated luminosity at 14 TeV. For $c_q^{23}=c_N^3$ the two branching ratios of $W'$ are $\Br(W'\to \tau \nu):\Br(W'\to 2j) \simeq1:3$; 
the former is denoted by the horizontal grey dashed line in Fig.\,\ref{fig:W'bound}. The two branching ratios can be correspondingly smaller if other decay channels are open (for instance, to extra vector-like fermions, as contemplated in Refs.~\cite{Greljo:2018ogz,Asadi:2018wea}). The grey shaded region is excluded by unitarity,  which constrains $3 (c_q^{23})^2+(c_N^3)^2< 16 \pi /g_V^2$ \cite{DiLuzio:2017chi}. 
The experimental bounds shown in Fig.~\ref{fig:W'bound} assume  that the $W'$ has a narrow width. This assumption fails for heavy $W'$ with a mass in the few TeV range. According to the results of a recast of the CMS search \cite{Sirunyan:2018lbg} performed for a wide $W'$ \cite{Greljo:2018tzh}, the entire perturbative parameter space of the $W'$ model is excluded, except potentially for the very light $W'$, with masses below $500$ GeV, where a reanalysis of older experiments would need to be carefully performed. Bounds on $W'$ from di-jet production~\cite{Sirunyan:2017nvi,Khachatryan:2016ecr,Sirunyan:2016iap,Aad:2011aj,Abe:1997hm} 
are less stringent and are not relevant for this simplified model.

Since the $W_\mu'$ couples to right-handed quarks, there is significant freedom in terms of the flavor structure of the $c_{q}^{ij}$ and $c_N^i$ couplings. We have limited the discussion to the minimal case, taking only $c_q^{23}, c_N^3\ne 0$, which is non-generic but possible, for instance, in flavor-locked models \cite{Knapen:2015hia, Greljo:2018ogz}. In most flavor models all the $c_q^{ij}, c_N^i$ are non-zero, leading to constraints from precision measurements. In UV completions (see Refs.~\cite{Asadi:2018wea,Greljo:2018ogz}), the $W'$ boson is expected to be accompanied by a $Z'$ state.  
The $Z'$ can, however, be parametrically heavier than the $W'$, in particular if additional sources of symmetry breaking are present.
The collider constraints on $W'$ and $Z'$ are often comparable, while the flavor constraints from FCNCs are far more stringent for $Z'$ in the presence of any appreciable off-diagonal couplings \cite{Greljo:2018ogz}: Contributions from $W'$ exchange to flavor changing neutral currents only arise at one-loop and are significantly less constraining. 

\subsection{Vector leptoquark $U_1^\mu$}
The interaction Lagrangian for the $U_1^\mu\sim(3,1)_{2/3}$ vector leptoquark is
\beq\label{eq:interation:U1}
{\cal L}\supset \alpha_{LQ}^{ij} \big(\bar L_L^i\gamma_\mu Q_L^j\big) U_1^{\mu\dagger}+  \alpha_{\ell d}^{ij} \big(\bar \ell_R^i\gamma_\mu d_R^j\big) U_1^{\mu\dagger}+  \alpha_{uN}^i \big(\bar u_R^i\gamma_\mu N_R\big) U_1^{\mu} +{\rm h.c.},
\eeq
while the kinetic term, following the notation in \cite{Dorsner:2018ynv}, is
\beq
\label{eq:ULagr}
	{\cal L}\supset -\frac{1}{2}U_{\mu\nu}^\dagger U^{\mu\nu}+m_{U_1}^2 U_{1\mu}^\dagger U_1^\mu-i g_s \kappa U_{1\mu}^\dagger T^a U_{1\nu} G^{a\mu\nu},
\eeq
with $U_{\mu\nu}=D_\mu U_{1\nu}-D_\nu U_{1\mu}$ the field strength tensor, and $\kappa$ a dimensionless coupling. 

When the leptoquark is integrated out, eq.~\eqref{eq:interation:U1} gives two four-fermion operators, relevant for $R(\dds)$ anomalies, with the Wilson coefficients
\beq
	\label{eqn:U1:Cs}
	\frac{c_{\rm SL}^{(\mu)}}{\rho_{\text{SL}}\leff^2}=2 \frac{\alpha_{LQ}^{33}\alpha_{uN}^2}{m_{U_1}^2}, \qquad \frac{c_{\rm VR}}{\leff^2}=- \frac{\alpha_{\ell d}^{33}\alpha_{uN}^2}{m_{U_1}^2}.
\eeq
The best fit values for the $U_1$ WCs in Table~\ref{tab:bfp} then imply
\beq
m_{U_1}\simeq 3.2 \big|\alpha_{LQ}^{33}\alpha_{uN}^{2}\big|^{1/2}  \bigg[\frac{40 \times 10^{-3}}{V_{cb}}\bigg]^{1/2}\,\TeV\,, 
 \label{eq:fit:U1:sol1}
\eeq
with
\beq\label{eq:U1:num:alphas}
	\alpha_{\ell d}^{33} \simeq  - 5.8\, \alpha_{LQ}^{33},
\eeq
where we used the lower set of best fits for $U_1$  in Table \ref{tab:bfp} (the upper set is excluded by $B_c\to \tau \nu$, see Fig \ref{fig:fits}). 
If one instead sets $c_{\text{SL}} = 0$, the best fit simply maps onto the $W'$ result (since both models then have the same non-zero coupling $c_{\text{VR}}$): $|c_{\text{VR}}| \simeq 0.46$, and
\begin{equation}
	m_{U_1} \simeq 1.3\big|\alpha_{\ell d}^{33}\alpha_{uN}^{2}\big|^{1/2}  \bigg[\frac{40 \times 10^{-3}}{V_{cb}}\bigg]^{1/2}\,\TeV\,.
\end{equation}

\begin{figure}[t]
\begin{center}
\includegraphics[width=8.0cm]{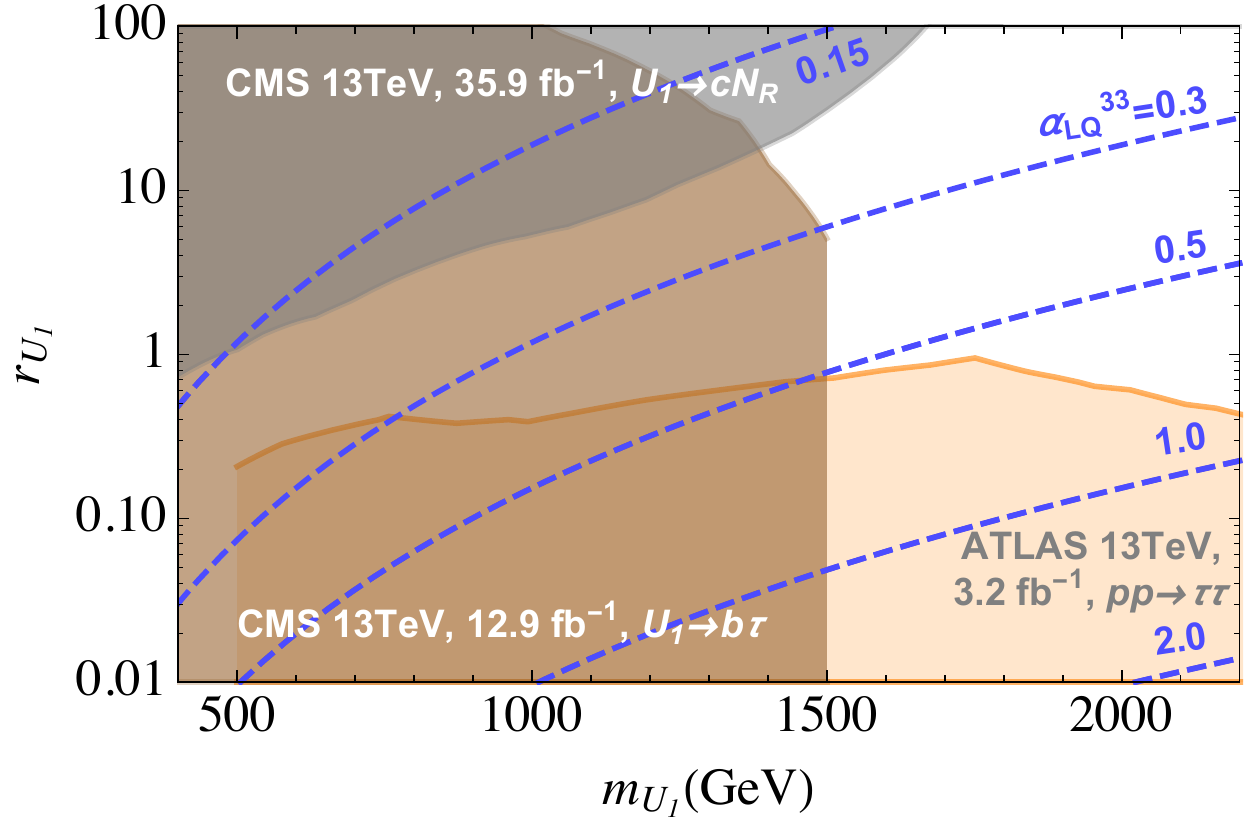}
\end{center}
\caption{The LHC bounds from \cite{CMS:2018bhq} (grey),  \cite{Sirunyan:2017yrk} (brown), and \cite{Faroughy:2016osc,Aaboud:2016cre} (orange) on the $U_1^\mu$ vector leptoquark mass, assuming the relation $\alpha_{\ell d}^{33} \simeq  -5.8\, \alpha_{LQ}^{33}$, arising from the $U_1$ best fit WCs to the $R(\dds)$ data.  Branching ratios for $U_1\to c\nu$, $b\tau$, $t\nu$ decays are fixed by the remaining ratio of coupling constants $r_{U_1}=(\alpha_{uN}^2/\alpha_{LQ}^{33})^2$, assuming no other channels are open. Blue dashed lines denote contours satisfying the $U_1$ best fit mass relation~\eqref{eq:fit:U1:sol1} for $\alpha_{LQ}^{33}=0.15,0.3, 0.5, 1.0$, and $2.0$. }
\label{fig:U1bound}
\end{figure}

 At the LHC, the $U_1$ leptoquark can be singly or pair produced. The pair production, $pp\to U_1 U_1^\dagger$, proceeds through gluon fusion, via the color octet term in \eqref{eq:ULagr}, for which we take $\kappa=1$ following Ref. \cite{CMS:2018bhq}.
The collider signatures of $U_1$ pair production depend on the $U_1$ decay channels. In the minimal set-up we switch on only three couplings, $\alpha_{LQ}^{33}, \alpha_{ld}^{33}$ and $\alpha_{uN}^2$, where $\alpha_{LQ}^{33}$ and $\alpha_{ld}^{33}$ are related through Eq.\,\eqref{eq:U1:num:alphas}, resulting in the branching ratios
\begin{align}
	\Br[U_1\to t \bar \nu_\tau]:\Br[U_1\to b \tau]:\Br[U_1\to &c \bar N_R]
	 =|\alpha_{LQ}^{33}|^2:\big(|\alpha_{LQ}^{33}|^2 + |\alpha_{ld}^{33}|^2\big):|\alpha_{uN}^2|^2 \label{eq:U1:Brs}\\	
	& = \frac{0.03}{1+0.03 r_{U_1}}:\frac{0.97}{1+0.03 r_{U_1}}:\frac{0.03 r_{U_1}}{1+0.03 r_{U_1}}\,, \nn
\end{align}
where 
\beq
\label{eq:def:r}
r_{U_1}=\bigg(\frac{\alpha_{uN}^2}{\alpha_{LQ}^{33}}\bigg)^2.
\eeq
Here, for simplicity, we have neglected the final state masses and the small corrections due to the off-diagonal CKM matrix elements in the $\alpha_{LQ}^{ij} \big(\bar L_L^i\gamma_\mu Q_L^j\big) U_1^{\mu\dagger}$. The presence of left-handed quark doublets also inevitably leads to CKM suppressed transitions $U_1\to c \bar \nu_\tau, u \bar \nu_\tau, s \tau, d \tau$.

The corresponding LHC bounds for $U_1$ are shown in Fig.\,\ref{fig:U1bound}, assuming no other decay channels are open. The most stringent bounds come from $pp\to U_1 U_1$ pair production, with both leptoquarks decaying either as $U_1\to cN_R$ \cite{CMS:2018bhq} (grey region) or $U_1\to b\tau$  \cite{Sirunyan:2017yrk} (brown region). Ref.~\cite{CMS:2018bhq} also gives bounds for the decay channel $U_1\to t\nu_\tau$, which are not shown in Fig.\,\ref{fig:U1bound} as they are always weaker in our setup.
We see that direct searches still allow for $m_{U_1}\geq1.5$ TeV, where the parameters of the model are still perturbative, as an explanation for the $R(\dds)$ anomalies. It is worth noting that a simultaneous fit to all three decay channels by the experiments would improve the sensitivity to $U_1$; such an analysis is likely the most optimal strategy for discovering a $U_1$ state responsible for the $R(D^{(*)})$ anomalies.  

Fig.~\ref{fig:U1bound} also shows the constraint on the $U_1$ model parameter space from the  CMS $pp\to \tau\tau$ search \cite{Aaboud:2016cre} (see also ATLAS search \cite{Aaboud:2017sjh}). In orange is shown the constraint on $r_{U_1}$, as a function of $m_{U_1}$, that is obtained from Fig. 6 of Ref. \cite{Faroughy:2016osc} with the replacement $g_U\to \big[(\alpha_{LQ}^{33})^2+(\alpha_{ld}^{33})^2]^{1/2}$. Assuming the relation $\alpha_{\ell d}^{33} \simeq  -5.8\, \alpha_{LQ}^{33}$, arising from the $U_1$ best fit WCs to the $R(\dds)$ data, the bound on $g_U$ in \cite{Faroughy:2016osc} translates to the excluded region in Fig. \ref{fig:U1bound}.

\subsection{Scalar leptoquark $S_1$}
The scalar leptoquark $S_1\sim (\bar 3,1)_{1/3}$ has the following interaction Lagrangian, 
\beq
{\cal L}\supset z_u (\bar U_R^c \ell_R)S_1+z_d (\bar d^c_R N_R)S_1+z_Q (\bar Q_L^c \epsilon L_L)S_1.
\eeq
Integrating out the leptoquark generates the following interaction Lagrangian above the electroweak scale 
\beq
\begin{split}\label{eq:LeffS1}
{\cal L}_{\rm eff}^{S_1}= &-\frac{z_d z_u^*}{2 m_{S_1}^2} Q_{\rm VR}-\frac{z_d z_Q^*}{2 m_{S_1}^2}\Big(Q_{\rm SR}-\frac{1}{4}Q_{\rm T}\Big)
\\
&+\frac{z_u z_Q^*}{2 m_{S_1}^2} \Big[ \epsilon_{ab} (\bar  \ell_R L_L^a) (\bar u_R Q_L^b)-\frac{1}{4}
 \epsilon_{ab} (\bar \ell_R  \sigma_{\mu\nu} L_L^a)(\bar u_R \sigma^{\mu\nu}  Q_L^b)\Big] +{\rm h.c.},
\end{split} 
\eeq
where the operators $Q_{\rm VR}$, $Q_{\rm SR}$, $Q_{\rm T}$ are defined in \eqref{eq:Q6}. The  $b\to c\tau \bar N_R$ decay is generated if $z_u^{23} z_d^3\ne0$ or $z_Q^{23} z_d^3\ne0$. The two operators in the second line give rise to the $b\to c\tau \nu_i$ decay for $z_{Q}^{3i}z_u^{23} \not = 0$, where $\nu_i$ are the SM neutrinos, which interfere with the SM contribution; for simplicity, we therefore only consider the $b\to c\tau \bar N_R$ decay, setting $z_{Q}^{3i}=0$, so that only the operators in the first line in \eqref{eq:LeffS1} are generated (alternatively, one may consider the regime $z_u$, $z_Q \ll z_d$, so that the contribution from the second line is negligible).

In the analysis of collider constraints, we conservatively keep only the minimal set of $S_1$ couplings required for the $R(\dds)$ anomaly nonzero: $z_u^{23}, z_d^3, z_Q^{23} \ne0$.
The Wilson coefficients of the $b\to c\tau \bar N_R$ operators $\op_{\rm VR}$, $\op_{\rm SR}$, $\op_{\rm T}$ are given by, 
\beq
	\label{eqn:S1:Cs}
\frac{c_{\rm VR}}{\Lambda_{\rm eff}^2}=-\frac{z_u^{23*} z_d^{3}}{2 m_{S_1}^2}, \qquad  \frac{c_{\rm SR}^{(\mu)}}{\rho_{\text{SR}}\leff^2}=-4 \frac{c_{\rm T}^{(\mu)}}{\rho_{\text{T}}\leff^2}=-\frac{z_Q^{23*} z_d^{3}}{2 m_{S_1}^2}.
\eeq
The best fit values for the $S_1$ WCs in Table~\ref{tab:bfp} then imply
\beq
m_{S_1}\simeq 1.2 \big|z_{u}^{23}z_{d}^{3}\big|^{1/2}  \bigg[\frac{40 \times 10^{-3}}{V_{cb}}\bigg]^{1/2}\,\TeV\,, 
 \label{eq:fit:S1:sol1}
\eeq
with 
\beq
\label{eq:zu23}
z_u^{23}\simeq 1.1 z_Q^{23}.
\eeq
using the lower set of best fits for $S_1$  in Table \ref{tab:bfp} (the upper set is excluded by $B_c\to \tau \nu$, see Fig \ref{fig:fits}). 
The branching ratios for $S_1$ decays are thus
\beq
\begin{split}
	\Br[S_1\to c \tau]:\Br[S_1\to b N_R]:\Br[S_1& \to s \nu_\tau] 
		 =\big(|z_u^{23}|^2+|z_Q^{23}|^2\big):|z_d^3|^2:|z_Q^{23}|^2 \label{eq:S1:Brs}\\
		&=\frac{0.69}{1+0.37 r_{S_1}}:\frac{0.37r_{S_1}}{1+0.37 r_{S_1}}:\frac{0.31}{1+0.37 r_{S_1}}\,,
\end{split}		
\eeq
where we have defined
\beq
	\label{eq:rdu}
	r_{S_1}=\bigg(\frac{z_d^3}{z_u^{23}}\bigg)^2.
\eeq

\begin{figure}[t]
\begin{center}
\includegraphics[width=8.0cm]{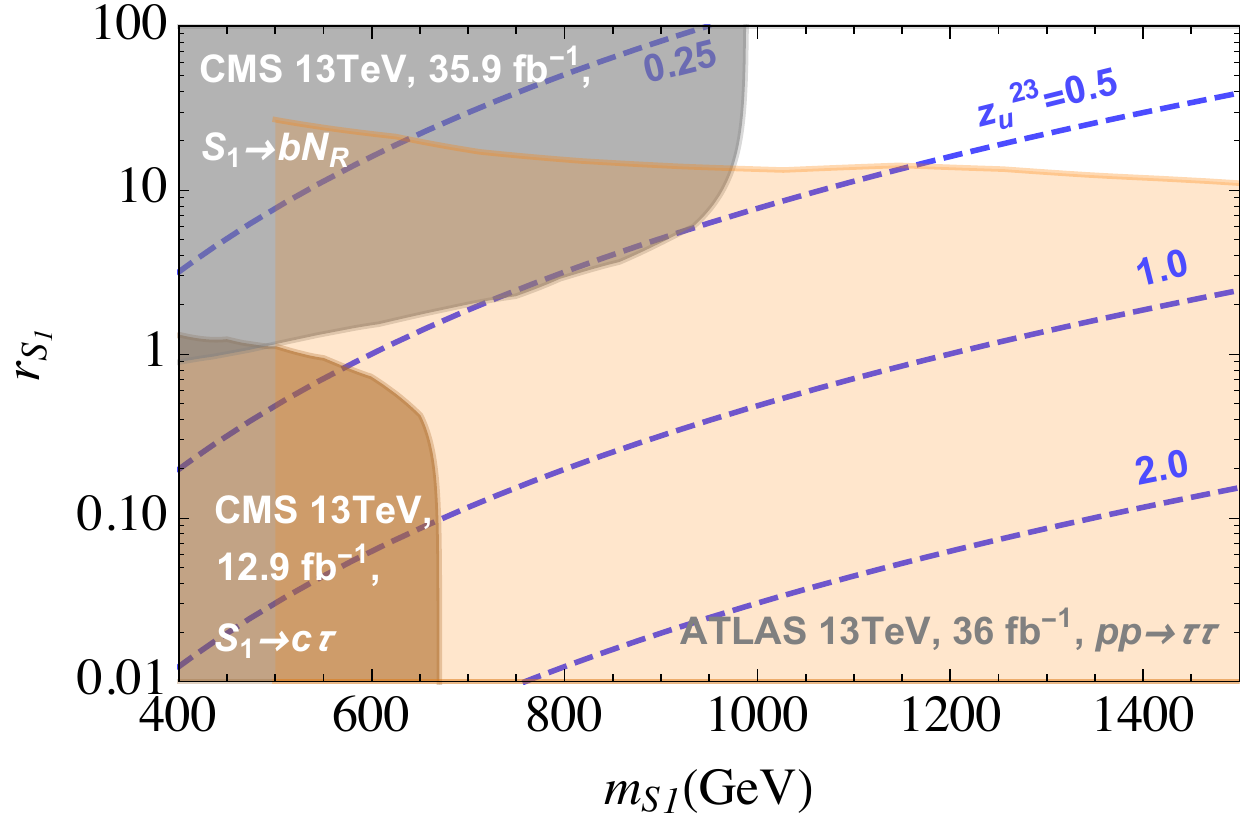}
\end{center}
\caption{The LHC bounds from pair production of $S_1$ leptoquarks followed by  $S_1\to b N_R$ decays \cite{CMS:2018bhq} (grey region) and $S_1\to c \tau$ \cite{Sirunyan:2017yrk} (brown region), and from a recast of the ATLAS $pp\to \tau\tau$ search \cite{Aaboud:2017sjh,Mandal:2018kau} (orange region), as a function of $m_{S_1}$ and the ratio $r_{S_1}=(z_d^3/z_u^{23})^2$~\eqref{eq:rdu}. The remaining ratio of coupling constants is fixed by the relation $z_u^{23}\simeq1.1 z_Q^{23}$, arising from the $S_1$ best fit WCs to the $R(\dds)$ data~\eqref{eqn:RDDsdata}. Contours satisfying the $S_1$ best fit mass relation~\eqref{eq:fit:S1:sol1} are shown by blue dashed lines for $z_{u}^{23}=0.25, 0.5, 1.0$, and $2.0$. }
\label{fig:S1bound}
\end{figure}

The resulting bounds from $pp\to S_1 S_1$ pair production at the 13 TeV LHC are shown in Fig.~\ref{fig:S1bound}. The grey shaded region is excluded by the CMS search \cite{CMS:2018bhq} with 35.9 fb${}^{-1}$ integrated luminosity, assuming both $S_1$ decay as $S_1\to b N_R$ with the branching ratio in \eqref{eq:S1:Brs}. The brown shaded region is excluded by the CMS search \cite{Sirunyan:2017yrk} using 12.9 fb${}^{-1}$ integrated luminosity, assuming $pp\to S_1 S_1$ followed by $S_1\to c \tau$ decay, with the $r_{du}$ dependent branching ratio in \eqref{eq:S1:Brs}. We have assumed the $S_1$ best fit mass relation~\eqref{eq:zu23} to $R(\dds)$ data to derive these bounds. 

The orange shaded region in Fig. \ref{fig:S1bound} shows the 95\% CL constraint from the recast of the 13 TeV ATLAS $pp\to \tau\tau$ search at $36{}^{-1}$fb integrated luminosity \cite{Aaboud:2017sjh}, performed in Ref. \cite{Mandal:2018kau}. The bounds in Fig. 3 (left) in Ref. \cite{Mandal:2018kau} can be reinterpreted in terms of the $S_1$ model coupling to a right-handed neutrino by making the replacement $\lambda_{23}^L\to \big[(z_u^{23})^2+(z_Q^{23})^2\big]^{1/2}$.

The combined set of constraints indicates that the $S_1$ leptoquark can be consistent with the $R(D^{(*)})$ anomaly for $m_{S_1}$ as low as 1000 GeV, and with perturbative couplings (the required values of $z_u^{23}$ are shown by dashed blue lines in Fig. \ref{fig:S1bound}).

\subsection{Scalar leptoquark $\tilde R_2$}
The scalar leptoquark $\tilde R_2\sim(3,2)_{1/6}$ has the following interaction Lagrangian, 
\beq
{\cal L}\supset \alpha_{Ld} \big(\bar L_L d_R\big) \epsilon \tilde R_2^\dagger+ \alpha_{QN} \big(\bar Q_L N_R\big) \tilde R_2 +{\rm h.c.}.
\eeq
Integrating out the $\tilde{R}_2$ generates
\begin{equation}
	\frac{c_{\rm SR}^{(\mu)}}{\rho_{\text{SR}}\leff^2}=4 \frac{c_{\rm T}^{(\mu)}}{\rho_{\text{T}}\leff^2} = \frac{\alpha_{Ld}^{33}\alpha_{QN}^{2}}{2m_{\tilde R_2}^2}.
\end{equation}
The best fit values for the $\tilde{R}_2$ WC in Table \ref{tab:bfp} then imply 
\begin{equation}
	 \label{eq:fit:R2:sol}
	m_{\tilde R_2}\simeq  0.95 \big|\alpha_{Ld}^{33}\alpha_{QN}^{2}\big|^{1/2}  \bigg[\frac{40 \times 10^{-3}}{V_{cb}}\bigg]^{1/2}\,\TeV\,.
\end{equation}

\begin{figure}[t]
\begin{center}
\includegraphics[width=8.0cm]{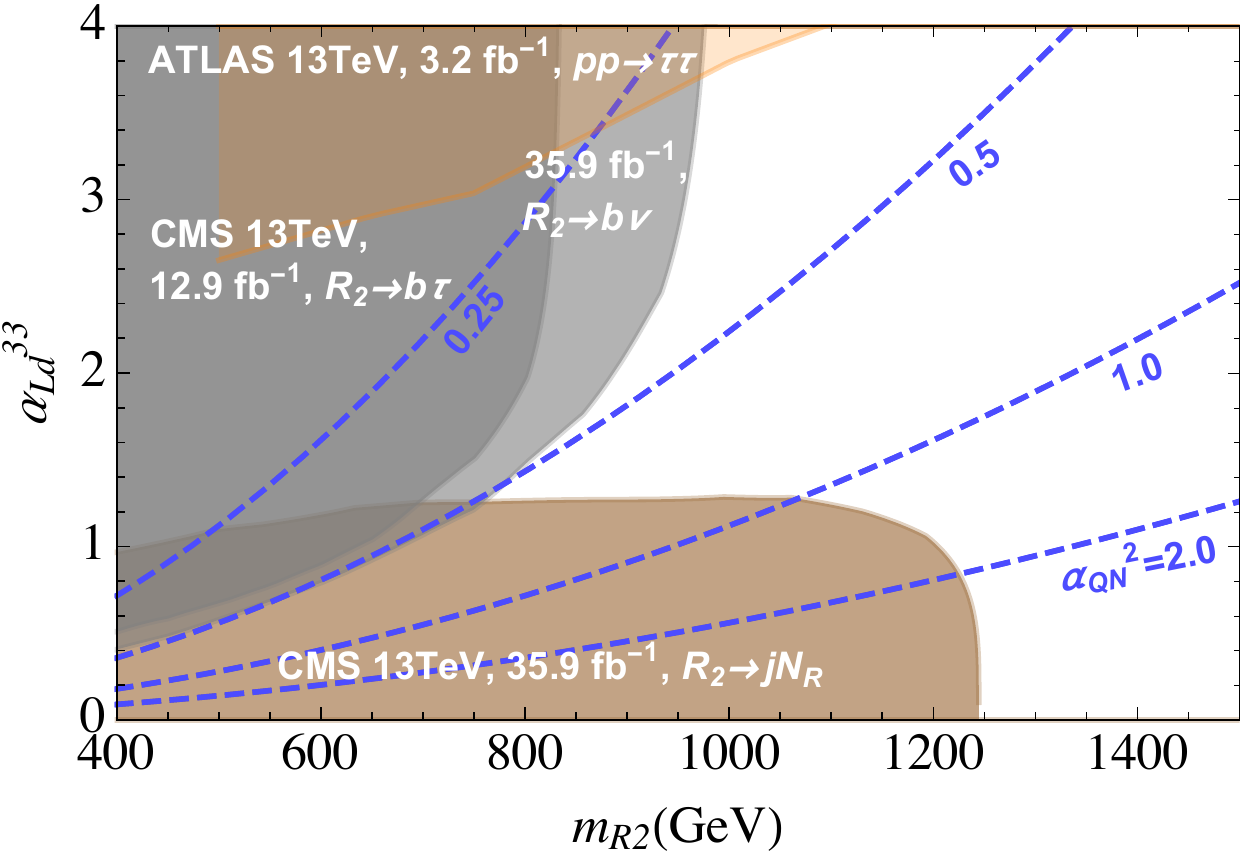}
\end{center}
\caption{The LHC bounds from pair production of $\tilde R_2^{2/3}$ and $\tilde R_2^{-1/3}$ leptoquarks, for the  decay channels $\tilde R_2^{2/3}\to b \bar \tau$ \cite{Sirunyan:2017yrk} (dark grey region), $\tilde R_2^{-1/3}\to b \bar \nu_\tau$ \cite{CMS:2018bhq} (light grey region), $\tilde R_2^{2/3}\to c N_R$, $\tilde R_2^{-1/3}\to s N_R$ \cite{CMS:2018bhq} (brown shaded region), and from $t$-channel exchange in $pp\to \tau\tau$ \cite{Faroughy:2016osc} (orange) as a function of $\tilde R_2$ mass and the coupling constant $\alpha_{Ld}^{33}$. Contours satisfying the $S_1$ best fit mass relation~\eqref{eq:fit:R2:sol} are shown by blue dashed lines, fixing  $\alpha_{QN}^2=0.25, 0.5, 1.0$, and $2.0$.}
\label{fig:R2bound}
\end{figure}

The leptoquark doublet $\tilde R_{2}$ contains two states: the charge $+2/3$ state $\tilde R_2^{2/3}$ and the charge $-1/3$ state $\tilde R_2^{-1/3}$. Keeping only the couplings relevant for the $R(D^{(*)})$ anomaly nonzero, $\alpha_{Ld}^{33},\alpha_{QN}^{2}\ne0$,  the $\tilde R_2$ states have two decay channels
\beq
\frac{\Br[\tilde R_2^{2/3} \to b \bar \tau]}{\Br[\tilde R_2^{2/3} \to c N_R]}=
\frac{\Br[\tilde R_2^{-1/3} \to b \bar \nu_\tau]}{\Br[\tilde R_2^{-1/3} \to s N_R]}=
\Big(\frac{\alpha_{Ld}^{33}}{\alpha_{QN}^{2}}\Big)^2,
\eeq 
where we have neglected differences due to the masses of the final state particles. 

Assuming  $\tilde R_2^{2/3}$ and $\tilde R_2^{-1/3}$ are degenerate, the LHC bounds from leptoquark pair production are shown in Fig.~\ref{fig:R2bound} as a function of $m_{\tilde R_2}$ and the $\alpha_{Ld}^{33}$ coupling. The remaining coupling, $\alpha_{QN}^2$, is set by the $\tilde{R}_2$ best fit mass relation~\eqref{eq:fit:R2:sol}. We show bounds from LHC searches for all four decay channels:  $\tilde R_2^{2/3}\to b \bar \tau$ \cite{Sirunyan:2017yrk} (dark grey region), $\tilde R_2^{-1/3}\to b \bar \nu_\tau$ \cite{CMS:2018bhq} (light grey), and the combined $pp\to \tilde R_2^{2/3} \tilde R_2^{2/3*}$ and $pp\to \tilde R_2^{-1/3} \tilde R_2^{-1/3*}$ cross sections, followed by  $\tilde R_2^{2/3}\to c N_R$ and $\tilde R_2^{-1/3}\to s N_R$ decays, which appear in the detector as 2j+MET \cite{CMS:2018bhq} (brown shaded region). The orange shaded region shows the bounds from $pp\to \tau\tau$ searches \cite{Faroughy:2016osc}, where $\tilde R_2$ can correct the tails of the distributions through the new $t$-channel exchange contribution.
We see that $m_{\tilde R_2}\gtrsim 800$ GeV consistent with the $R(\dds)$ anomaly is allowed, with perturbative couplings, even if no other decay channels are open. 


\section{Sterile Neutrino Phenomenology}
\label{sec:sterile:neutrino}

In this section, we discuss the phenomenology associated with the right-handed (sterile) neutrino $N_R$.  As we will see below, the coupling of $N_R$ to the SM fermions through one of the higher dimension operators in Eq.\,\eqref{eqn:O}, needed to explain $R(\dds)$, carries interesting implications for neutrino masses, cosmology, and collider signatures.  
We will assume that $N_R$ is a Majorana fermion with mass $\lesssim \mathcal{O}(100)$ MeV so that it remains compatible with the measured missing invariant mass spectrum in the $\bddstn$ decay chain. As in Sec.~\ref{sec:simplified}, we do not consider the $\Phi$ model as it is excluded by $B_c \to \tau \nu$ constraints.

\subsection{Neutrino masses}
The effective operators~\eqref{eqn:O} induce a $N_R$--$\nu_L$ Dirac mass at the two loop order via contributions of the form 
\begin{equation}
	m_D \bar{N}_R \nu_L \qquad \sim \qquad \begin{gathered} \includegraphics[width = 6cm]{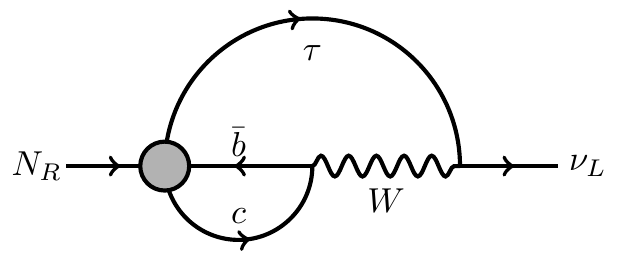} \end{gathered}\,.
\end{equation}
Here, the simplified model mediator has been integrated out, producing an effective four-fermion vertex, shown in gray. Depending on the chiral structure of the simplified model, various mass insertions are mandated on the internal quark and lepton lines. In particular, the $\mathcal{O}_{\text{VR}}$ operator requires three mass insertions, while the scalar and tensor operators require only one. The corresponding Dirac masses can be estimated as
\begin{subequations}
\label{eqn:numass}
\begin{align}
	W': && m_D &\sim \frac{c_{\text{VR}}}{\leff^2}  \frac{g_2^2}{2}\frac{V_{cb}}{(16\pi^2)^2}m_b m_c m_\tau \sim c_{\text{VR}} 10^{-3}\,\text{eV},
	\\
	\tilde{R}_2: && m_D & \sim  c_{\text{SR}}m_b  \frac{g_2^2}{2}\frac{V_{cb}}{(16\pi^2)^2}  \sim c_{\text{SR}} 10^{2}\,\text{eV},
	\\
	U_1: && m_D & \sim  \bigg[c_{\text{SL}}  m_c+\frac{c_{\text{VR}}}{\leff^2}m_b m_c m_\tau \bigg] \frac{g_2^2}{2} \frac{ V_{cb}}{(16\pi^2)^2}
	\sim (c_{\text{SL}}10^{2} + c_{\text{VR}} 10^{-3})\,\text{eV},
	\\
	S_1: && m_D & \sim \bigg[c_{\text{SR}} m_b +\frac{c_{\text{VR}}}{\leff^2} m_b m_c m_\tau \bigg] \frac{g_2^2}{2} \frac{ V_{cb}}{(16\pi^2)^2} 
	\sim (c_{\text{SR}}10^{2} + c_{\text{VR}} 10^{-3})\,\text{eV}\,.
\end{align}
\end{subequations}
In the above estimates, we have ignored $\mathcal{O}(1)$ prefactors and loop integral factors apart from those implied by na\"\i ve dimensional analysis. Note that for diagrams with a single mass insertion, the Wilson coefficients $c_{\rm SL}$, $c_{\rm SR}$ appear without the $1/\leff^2$ prefactor. In such cases, strictly speaking, it is the couplings of the mediators rather than the Wilson coefficients that should appear in the estimates. However, since the collider constraints
require mediators to be heavy, with mass  approximately equal to $\leff$, it is a reasonable approximation to use the Wilson coefficients everywhere in the above estimates.  

Furthermore, for $\tilde R_2$, $U_1$, and $S_1$ mediators, which couple to the left-handed $\tau_L$, there are additional two loop contributions to the neutrino mass matrix arising from
 the $SU(2)_L$ related operators involving $\nu_L$. A representative diagram is shown in Fig.\,\ref{fig:second2loop}. While such diagrams contain similar mass insertions and WC scalings as the corresponding $c_{\rm SL, SR}$ terms in Eqs. \eqref{eqn:numass}, they are GIM suppressed and thus expected to produce only subleading corrections to the Dirac mass estimates in Eqs. \eqref{eqn:numass}. 

\begin{figure}[t]
\begin{center}
\includegraphics[width=7.0cm]{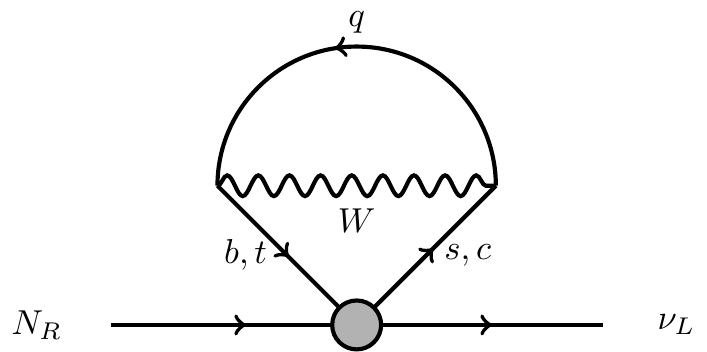}
\end{center}
\caption{Dirac mass contribution by virtue of SU(2) counterparts of the four-Fermi operators that give rise to the $R(\dds)$ enhancements. These diagrams are GIM suppressed and give subdominant contributions to the Dirac mass.}
\label{fig:second2loop}
\end{figure}

Since $N_R$ is assumed to have a Majorana mass $m_{N_R}\lesssim 100$ MeV, the contribution to the SM neutrino masses is $\sim m_D^2/m_{N_R}$, which should not exceed the observed neutrino mass scale $m_\nu\sim 0.1$ eV. From the best fit regions shown in Figs.~\ref{fig:fits} or~\ref{fig:fitsph} (and the best fit values from Table \ref{tab:bfp}), it follows that the $W'$-mediated diagram gives a Dirac mass $m_D\sim10^{-3}$ eV, which is consistent with observed neutrino masses, whereas the $R_2$ mediated digram gives $m_D\sim100$ eV, which is in some tension for $m_{N_R} \lesssim 10$\,keV.  Likewise, the $U_1$ and $S_1$ models produce similarly problematic contributions to the neutrino masses at their best fit points (see Table \ref{tab:bfp}). 
However, from Figs~\ref{fig:fits} and~\ref{fig:fitsph} we also see that the $1\sigma$ CLs of the $U_1$ and $S_1$ models do contain regions with the scalar Wilson coefficients $|c_{\text{SL}, \text{SR}}| \ll 1$, corresponding to small couplings $\alpha_{LQ} \ll 1$ and $z_Q \ll 1$ (cf. Eqs.~\eqref{eqn:U1:Cs} and~\eqref{eqn:S1:Cs}), which remain compatible with observed neutrino masses.

If additional operators are present, neutrino mass contributions can also be generated at one loop. For instance, as discussed in Sec.\,\ref{sec:Fits}, new operators coupling to second generation quark doublets can be introduced to cancel away large contributions to $b\to s\nu\bar \nu$ from the operators in Eq.\,(\ref{eqn:Os}). Such 1-loop neutrino mass contributions scale as $m\sim\frac{1}{16\pi^2}m_f$ and, depending on whether the new operators couple to $\nu\nu$ or $\nu N_R$, contribute to the Majorana or Dirac mass terms for the neutrinos. Unless suppressed by small couplings in the diagram, such mass contributions are generally several orders of magnitude larger than what is allowed by the observed neutrino mass scale $m_\nu\sim 0.1$ eV, and would need to be cancelled by fine-tuned values of bare neutrino masses.

Additional Dirac mass contributions beyond the diagrams considered above could worsen or improve the outlook. For instance, if the mediators also couple to other quarks, in particular the top quark, the corresponding two loop diagrams with a top quark mass insertion would lead to unacceptably large contributions to neutrino masses. On the other hand, additional Dirac mass terms that interfere destructively with the two loop contributions here could restore consistency in otherwise problematic regions of parameter space, albeit at the cost of some fine-tuning of parameters. 

\subsection{Sterile Neutrino Decay}

The two loop diagrams considered above also give rise to the decay process $N_R\to \nu\gamma$ via the emission of a photon from one of the internal propagator lines (a representative diagram is shown in Fig.\,\ref{fig:NRdecay} (left)). The approximate $N_R\to \nu\gamma$ decay rates\footnote{The mass insertion required by the helicity flip for the emission of a photon can occur on an internal fermion line, and does not incur the cost of a mass suppression on an external fermion leg, in contrast to $f_1\to f_2\gamma$ diagrams via an $SU(2)_L$ electroweak loop.} for the simplified models, along with the corresponding decay lifetime estimates, are listed in Table \ref{tab:lifetimes} (for related calculations, see Ref.~\cite{Lavoura:2003xp,Wong:1992qa,Bezrukov:2009th,Aparici:2012vx}). Note that for a given mediator and sterile neutrino mass $m_{N_R}$, the decay rate is completely fixed by the Wilson coefficients consistent with the $R(\dds)$ anomaly. 

\begin{figure}[t]
\begin{center}
\includegraphics[width=5.0cm]{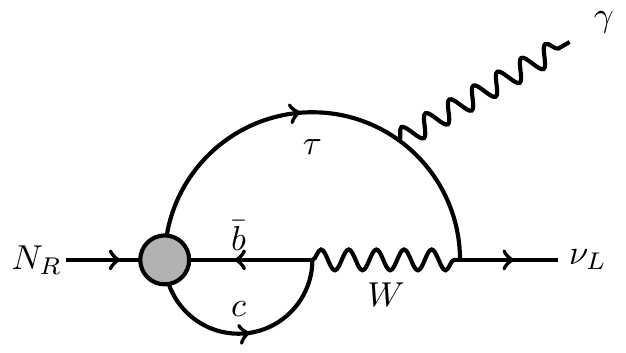}~~~~\includegraphics[width=7.0cm]{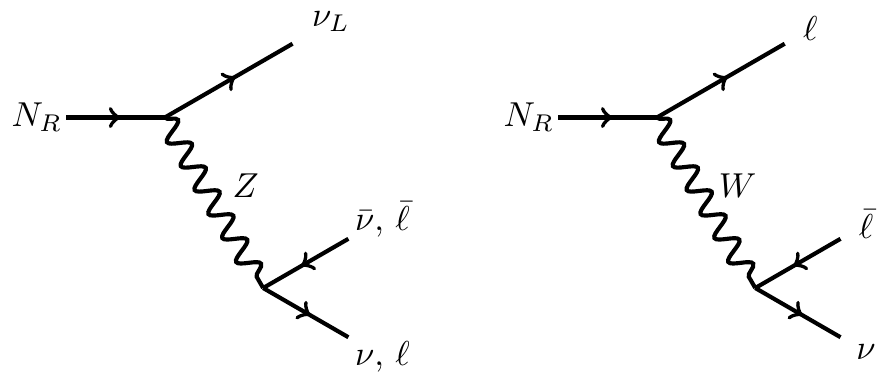}
\end{center}
\caption{Sterile neutrino decay modes induced by the NP couplings (left) and by tree level sterile-active mixing (centre, right).}
\label{fig:NRdecay}
\end{figure}

\begin{table}[t]
\begin{center}
\renewcommand{\arraystretch}{1.4}
\newcolumntype{C}{ >{\centering\arraybackslash $} m{1.25cm} <{$}}
\newcolumntype{E}{ >{\centering\arraybackslash $} m{6cm} <{$}}
\newcolumntype{D}{ >{\centering\arraybackslash $} m{6cm} <{$}}
\begin{tabular}{CED}
	\hline\hline
	\text{Model} & \Gamma_{N_R\to \nu\gamma} & \text{lifetime (s)} \\
	\hline 
	W' 			&  \frac{c_{\text{VR}}^2}{\leff^4}\frac{\alpha}{32\, \pi^8}\, V_{cb}^2\, G_F^2\, m_\tau^2\, m_b^2\, m_c^2\, m_{N_R}^3 				& c_{\text{VR}}^{-2}\,10^{24}\, \left(m_{N_R}/{\text{keV}}\right)^{-3} 		\\
	\tilde{R}_2 	& c_{\text{SR}}^2\frac{\alpha}{32\, \pi^8}\, V_{cb}^2\, G_F^2\, m_b^2\, m_{N_R}^3		& c_{\text{SR}}^{-2}\,10^{13}\, \left(m_{N_R}/{\text{keV}}\right)^{-3} 		\\
	U_1  & c_{\text{SL}}^2\frac{\alpha}{32\, \pi^8}\, V_{cb}^2\, G_F^2\, m_c^2\, m_{N_R}^3		& c_{\text{SL}}^{-2}\,10^{14}\, \left(m_{N_R}/{\text{keV}}\right)^{-3} 		\\
	S_1  & c_{\text{SR}}^2\frac{\alpha}{32\, \pi^8}\, V_{cb}^2\, G_F^2\, m_b^2\, m_{N_R}^3		& c_{\text{SR}}^{-2}\,10^{13}\, \left(m_{N_R}/{\text{keV}}\right)^{-3} 		\\
	\hline\hline
 \end{tabular}
\end{center}
\caption{ \label{tab:lifetimes} Approximate $N_R\to \nu\gamma$ decay rates (middle column) and lifetimes (final column) for the mediators listed in the first column. For $U_1 (S_1)$, we only show the contribution from the $c_{\text{SL}}(c_{\text{SR}})$ operators, which are expected to dominate; if these coefficients vanish, the decay rates and lifetimes get contributions from $c_{\text{VR}}$ of the same form as that for the $W'$ operator. } 
\end{table}

For appreciable mixing between $N_R$ and the SM neutrinos, the leading tree-level decay is into three SM neutrinos (Fig.\,\ref{fig:NRdecay} center) and, if kinematically accessible, into charged leptons (Fig.\,\ref{fig:NRdecay} right). The $N_R \to 3\nu$ decay rate is
\begin{equation}
\Gamma_{N_R\to 3\nu}
	 \simeq\frac{G_F^2}{192\, \pi^3}\, m_{N_R}^5 \sin^2\theta \simeq 10^{-48} \bigg(\frac{m_{N_R}}{\text{keV}}\bigg)^5 \bigg(\frac{\sin^2\theta}{10^{-4}}\bigg)\, \GeV,
\end{equation}
where $\theta$ is the mixing angle between $N_R$ and the SM neutrino. The $N_R\to 3\nu$ decay width is in general subdominant to the $N_R \to\nu\gamma$ decay width induced by the $R(\dds)$ anomaly. For a direct comparison, one can rewrite the $N_R\to \nu\gamma$ decay rate in Table\,\ref{tab:lifetimes} in terms of the Dirac mass from Eq.\,\ref{eqn:numass}, then convert to the mixing angle via sin $\theta \approx m_D/m_N$. For instance, for $S_1$ this gives $\Gamma(N\to\nu \gamma) \sim 32\, \alpha\,\sin^2\theta \,m_N^5\, G_F^2  / \pi^4/g^4$. Thus
\begin{equation}
\frac{\Gamma(N\to\nu \gamma)_{S_1}}{\Gamma(N_R\to 3\nu)_{S_1}}\approx\frac{32\times192\,\alpha}{\pi\,g^4}\sim10^3.
\end{equation}

\subsection{Sterile Neutrino Cosmology}

The above estimates imply that the sterile neutrino $N_R$ can be fairly long-lived. The interactions with SM fermions mandated by consistency with the $R(\dds)$ anomaly also lead to copious production of $N_R$ in the early Universe. The cosmological aspects of the sterile neutrino therefore require careful treatment. 

The interactions with SM fermions thermalize the $N_R$ population with the SM bath at high temperatures. These interactions are active until the temperature drops below the masses of the SM fermions involved in these interactions, i.e., around the GeV scale. Since we have assumed $m_{N_R}\lesssim\,100$ MeV, the $N_R$ abundance is not Boltzmann suppressed, and $N_R$ survives as an additional relativistic neutrino species in the early Universe. It then becomes crucial to determine the fate of this $N_R$ population.

For the $\tilde{R}_2,\,U_1,$ and $S_1$ mediated models, it follows from Table~\ref{tab:lifetimes} that the $N_R$ lifetime is $\sim 10^{14} (m_{N_R}/\text{keV})^{-3}$\,s. For $m_{N_R} \sim \mathcal{O}(\text{eV--keV})$, this implies a late decay of the $N_R$ population into the $\gamma\nu$ channel, which injects an unacceptable amount of photons into the diffuse photon background. The exception are masses close to the upper limit of the range we consider,  $m_{N_R}\lesssim100$ MeV, for which the lifetime is reduced to $\lesssim 1$\,s. The decays then occur before big bang nucleosynthesis (BBN) and do not leave any visible imprints. 

In contrast, for the $W'$ mediated case (or for $U_1,\,S_1$ in the parts of the Wilson coefficient $1\sigma$ CL regions where $c_{SL},\,c_{SR}$ are vanishingly small), the lifetime is much longer because of the additional mass insertions in the decay diagrams, and a lifetime $\lesssim 1$s cannot be achieved for any realistic choices of parameters. However, for $m_{N_R} \lesssim 100$ keV, the sterile neutrino has a lifetime greater than the age of the Universe and could in principle form a component of dark matter or dark radiation. 

The dark matter and dark radiation possibilities of $N_R$ in the $W'$ model have been extensively discussed in Ref.~\cite{Greljo:2018ogz}. In contrast to traditionally studied frameworks of sterile neutrino dark matter, where the relic abundance is produced via freeze-in mechanisms (see, e.g., \cite{Dodelson:1993je,Shi:1998km,Shakya:2015xnx,Shakya:2016oxf,Roland:2014vba,Shakya:2018qzg}), the $W'$ model involves the sterile neutrino freezing out as a relativistic species, leading to too large of a relic abundance for masses greater than $\mathcal{O}($keV). This can be fixed with appropriate entropy dilution from, for instance, late decays of GeV scale sterile neutrinos \cite{Scherrer:1984fd,Asaka:2006ek,Bezrukov:2009th,Bezrukov:2009th}, which also makes the dark matter colder, improving compatibility with warm dark matter constraints. The $\gamma$-ray bounds from various observations \cite{Essig:2013goa} rule out dark matter lifetimes of $\mathcal{O} (10^{26-28})$\,s in the keV-MeV window, ruling out the case that $N_R$ constitutes all of dark matter. This leaves us with the possibility that $N_R$ may constitutes a small fraction -- at the sub-percent level -- of dark matter. Future $\gamma$-ray observations will probe this possibility and could discover a line signal from the $N_R\to \gamma \nu$ decay. For masses  $m_{N_R} \lesssim$ keV, $N_R$ can act as dark radiation and contribute to the effective number of relativistic degrees of freedom $\Delta N_{\text{eff}}\approx\mathcal{O}(0.1)$ at BBN and/or CMB decoupling, which could be detected with future instruments such as CMB-S4 \cite{Abazajian:2016yjj}. Lifetimes shorter than the age of the Universe, however, are incompatible with current observational constraints. 

\subsection{Displaced Decays at Direct Searches and Colliders}

As discussed in the previous section, in the $\tilde{R}_2,\,U_1,$ and $S_1$ models, cosmology favors the regime $m_{N_R}\sim100$ MeV, with a lifetime $\lesssim 1$s. Since the dominant decay channel is $N_R\to \nu\gamma$, this would give rise to displaced decays into a photon+MET. Such displaced signals could provide an interesting, but challenging, target for proposed detectors such as SHiP~\cite{Anelli:2015pba}, MATHUSLA~\cite{Chou:2016lxi}, FASER~\cite{Feng:2017uoz}, and CODEX-b~\cite{Gligorov:2017nwh}. Displaced decays can also occur in the $W'$ UV completion of Refs.~\cite{Asadi:2018wea,Greljo:2018ogz}, where, as discussed earlier, GeV scale sterile neutrinos with lifetimes $\lesssim 1$s might be needed to entropy dilute problematic overabundances of the $N_R$; these can also lead to several other observable signals at various direct and cosmological probes (see, e.g., the discussion in \cite{Drewes:2015iva}).

\section{Conclusions}

\label{sec:conclusions}

We have performed an EFT study of the lowest dimension electroweak operators that can account for the $R(\dds)$ anomalies, assuming they arise because of incoherent contributions from semitauonic decays involving a right-handed sterile neutrino $N_R$. These dimension-six operators can arise from a tree-level mediator exchange in five possible simplified models. We examined the fits and constraints for each simplified model. While all five models have $1\sigma$ fit regions consistent with the $R(\dds)$ data, the case of the scalar doublet mediator is conservatively in tension with constraints from Br$[B_c \to \tau \nu]$, while the experimental bounds on $b\to s\nu\bar \nu$ rates are in tension with the predicted rates from the scalar leptoquark $\tilde R_2$ .

The fit regions of the remaining three simplified models imply sizable semileptonic branching ratios for the tree-level mediators. We find that each model already faces fairly stringent collider constraints.  The searches for the $W'$ mediator in the $W'\to \tau \nu$ channel exclude the model for perturbative couplings, where the calculations are reliable, with the possible exception of very light $W'$ masses (see Fig.~\ref{fig:W'bound} and surrounding discussion). The two leptoquark models are consistent with LHC search results provided the mediator masses are $\mathcal{O}(\TeV)$, while their couplings may still remain in the perturbative regime. 
Our analysis indicates promising paths to future discovery of the tree-level mediators at the LHC, with couplings and masses consistent with the fit to the $R(\dds)$ data. The vector leptoquark $U_1^\mu$ can best be probed at the LHC with simultaneous fits to the three decays $U_1\to cN_R, U_1\to b\tau$ and $U_1\to t\nu_\tau$. Likewise, the scalar leptoquark $S_1$ can be probed via $S_1\to b N_R$ and  $S_1\to c \tau$ decays. Since the mediators cannot be arbitrarily heavy if the couplings are to remain perturbative, prospects of detecting them at the LHC are quite encouraging.  

We have also discussed the phenomenology associated with the sterile neutrino $N_R$. In simplified models involving $\tilde{R}_2,\,U_1,$ and $S_1$, constraints from contributions to neutrino masses as well as cosmology indicate a preference for $m_{N_R}\sim10$\,--\,$100$\,MeV with a decay lifetime $\lesssim 1$s in the dominant channel $N_R\to \nu\gamma$. This opens up the potential for detecting displaced decays of $N_R$ at various detectors. It also implies potentially measurable distortions of the kinematical distributions in semileptonic $B$ meson decays due to the heavy sterile neutrino in the final state. For the $W'$ simplified model, the predicted contribution to neutrino masses is much smaller and poses no constraints on the model. The predicted decay lifetime of $N_R$ is correspondingly much longer than the age of the Universe. Consequently, a significant relic abundance of $N_R$ is likely present in the universe, which can contribute to dark radiation and give measurable deviations to the effective number of relativistic degrees of freedom $\Delta N_{\text{eff}}\approx\mathcal{O}(0.1)$ at BBN and/or CMB decoupling for $m_{N_R} \lesssim$ keV, or constitute a small fraction of dark matter for $N_R$ in the keV-MeV mass range with possible gamma ray signals at future probes.

The interpretation of the $R(\dds)$ anomaly in terms of new physics coupling the SM fermions to a right-handed sterile neutrino is therefore an exciting possibility with testable predictions in multiple directions, spanning kinematic distributions of the measured $B$ meson decays, searches for heavy TeV scale particles at the LHC, displaced decay signals at various detectors, as well as astrophysical and cosmological signatures.

\acknowledgments We thank Damir Be\v cirevi\' c, Florian Bernlochner, Admir Greljo, Svjetlana Fajfer, Nejc Ko\v snik for helpful discussions and communications about their work.  JZ acknowledges support in part by the DOE grant DE-SC0011784. BS was
partially supported by the NSF CAREER grant PHY1654502.
This work was performed in part at the Aspen Center for Physics, which is supported by National Science Foundation grant PHY-1066293. DR thanks Florian Bernlochner, Stephan Duell, Zoltan Ligeti and Michele Papucci for their ongoing collaboration in the development of \texttt{Hammer}, which was used for part of the analysis in this work. The work of DR was supported in part by NSF grant PHY-1720252.
 
\appendix

\section{Differential distributions}
\label{app:dd}
In this appendix we collect the predictions for several normalized differential distributions for $\Bbar \to (D^* \to D\pi)(\tau \to \ell \bar\nu_\ell \nt)\bar\nu$ and $\Bbar \to D(\tau \to \ell \bar\nu_\ell \nt)\bar\nu$ decay chains, shown in the left and right columns in Figs.~\ref{fig:1DWphistos}--\ref{fig:1DS1histos}, respectively. In each plot, the SM predictions (blue dashed curves) are compared with the predictions for the particular simplified model (grey bands), obtained by varying the relevant Wilson coefficients over the $2\sigma$ regions in Fig.~\ref{fig:fits}. In each of the figures the first row shows the normalized distribution $(1/\Gamma)(d\Gamma/dE_D)$, where $E_D$ is the energy of the outgoing $D$ meson in the $B$ meson rest frame. The second row contains the $(1/\Gamma)(d\Gamma/dE_\ell)$ distribution, with $E_\ell$ the energy of the final state charged lepton, while the third row shows the $(1/\Gamma)(d\Gamma/dm^2_{\rm miss})$ distribution, with $m^2_{\rm miss}$ the combined invariant mass of the system of three final state neutrinos. The final row in each Figure shows the $(1/\Gamma)(d\Gamma/d\cos\theta_{D\ell})$ normalized distribution, where $\theta_{D\ell}$ is the angle between the three momenta of the $D$ meson and the charged lepton, $\ell$, in the rest frame of the $B$ meson. 

The comparison between the SM predictions (blue dashed curves) and the predictions for the $W'$ simplified model (grey bands) is shown in Fig.~\ref{fig:1DWphistos}. The differences between the two predictions are small, below about 10\% for $(1/\Gamma)(d\Gamma/dE_\ell)$ and well below this for the other distributions. Similarly small corrections from NP to the shapes of distributions are found for the $\tilde R_2$ model, Fig.~\ref{fig:1DR2histos}.  In this case the largest deviation is found for the $(1/\Gamma)(d\Gamma/dE_D)$ distribution for the $\Bbar \to D^* \to D\pi$ decay (Fig.~\ref{fig:1DR2histos}, first row, right panel) and is at the level of about ${\mathcal O}(20\%)$. The deviations are potentially sizable for the $U_1$ and $S_1$ models for at least some of the distributions, see Figs.~\ref{fig:1DU1histos} and~\ref{fig:1DS1histos}, respectively.

\begin{figure}[t]
\centering{
\includegraphics[width = 0.47\linewidth]{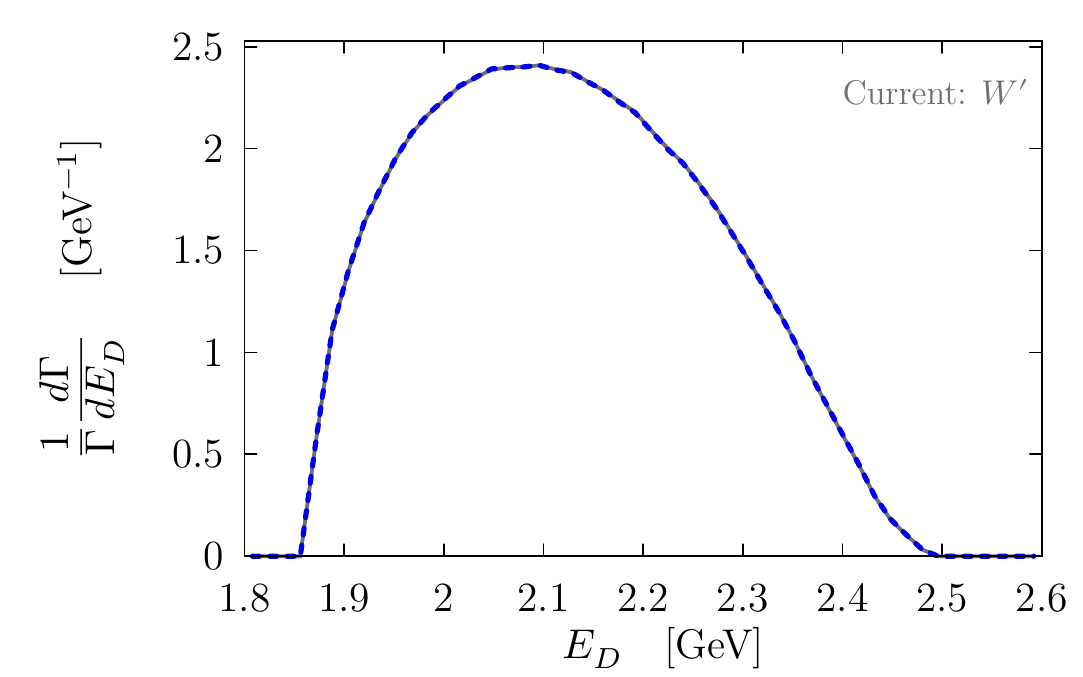}\hfill
\includegraphics[width = 0.47\linewidth]{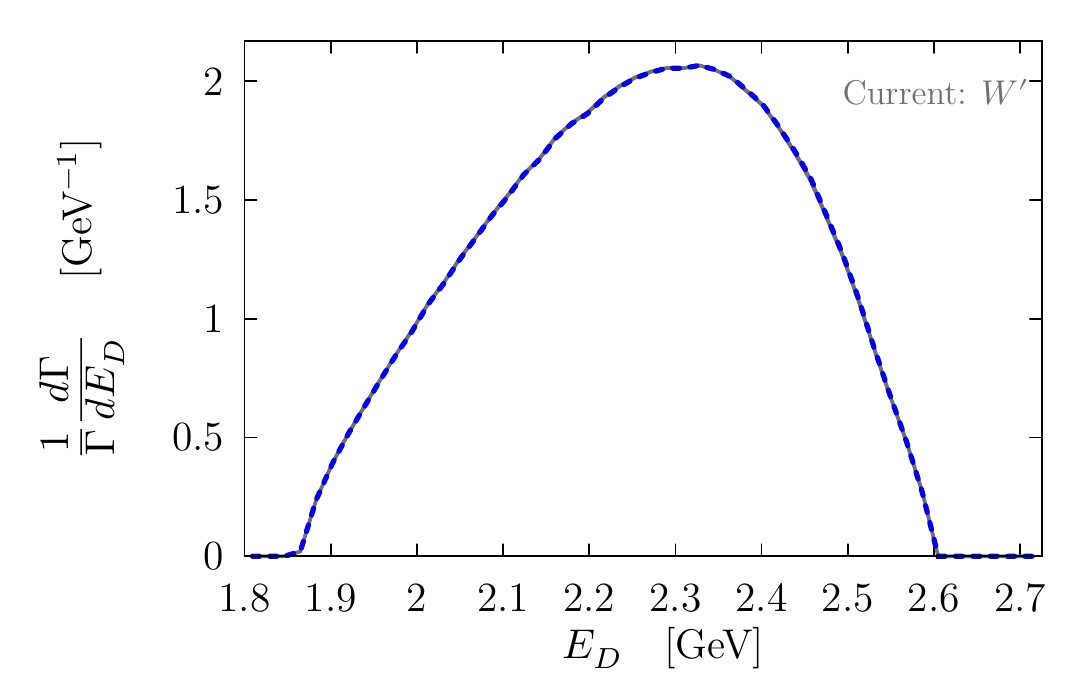} \\
\includegraphics[width = 0.47\linewidth]{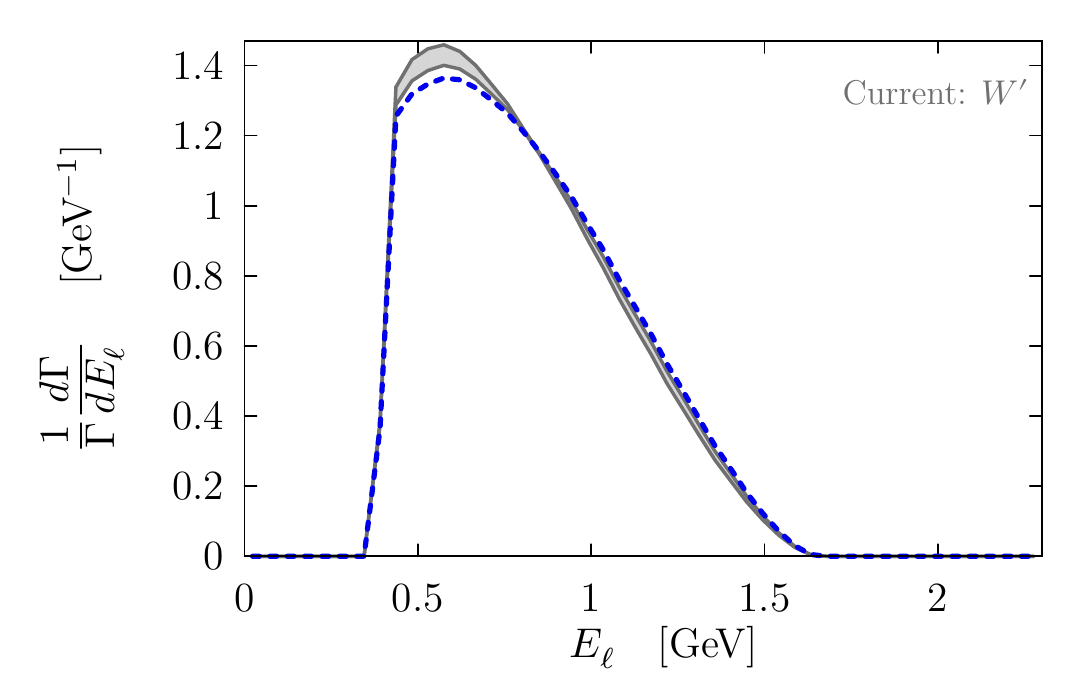}\hfill
\includegraphics[width = 0.47\linewidth]{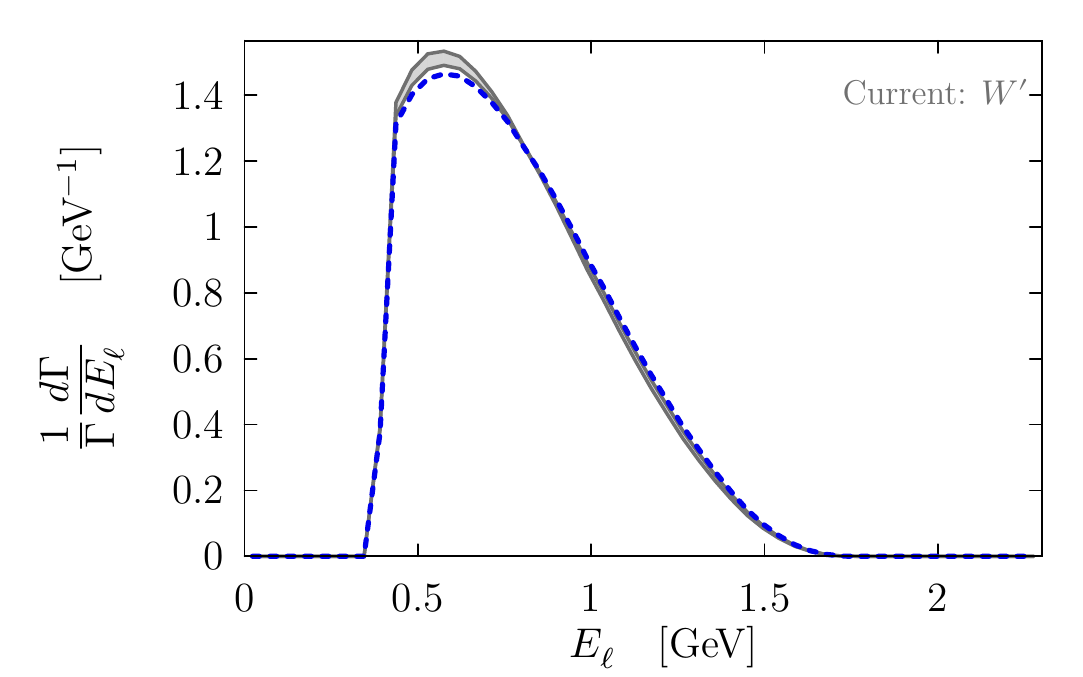} \\
\includegraphics[width = 0.47\linewidth]{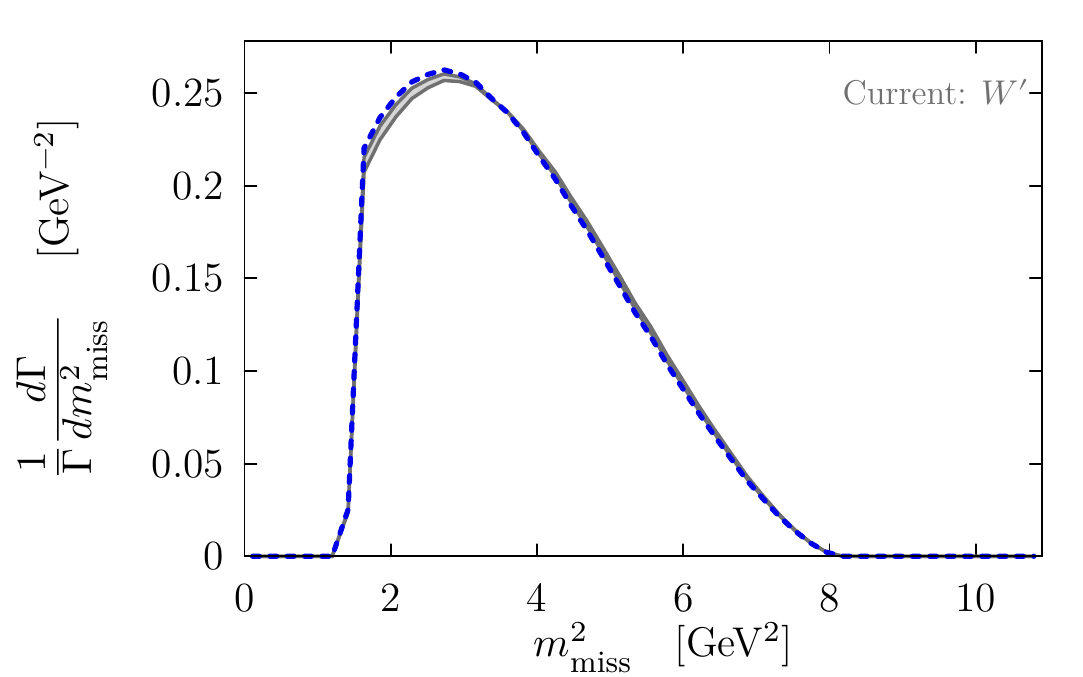} \hfill 
\includegraphics[width = 0.47\linewidth]{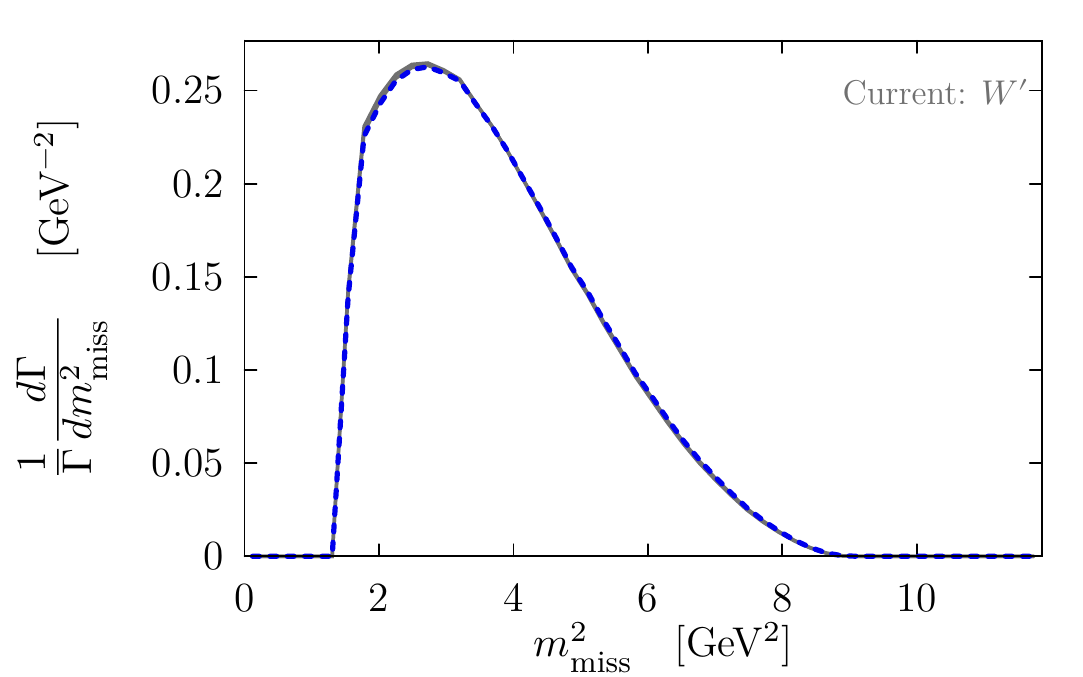}\\
\includegraphics[width = 0.47\linewidth]{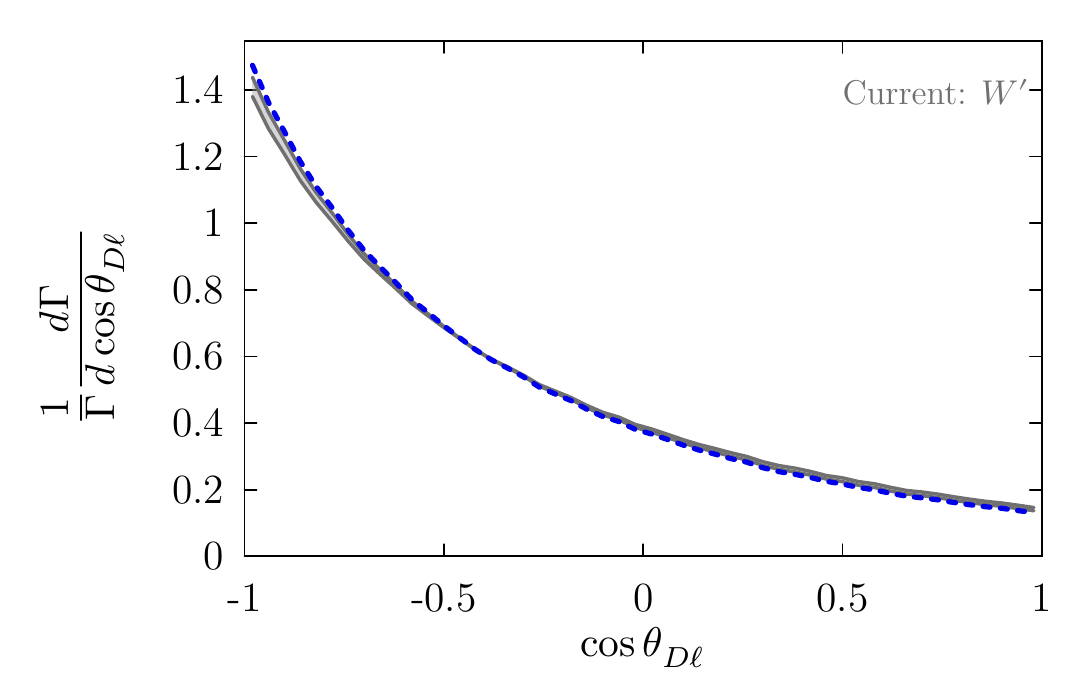} \hfill
\includegraphics[width = 0.47\linewidth]{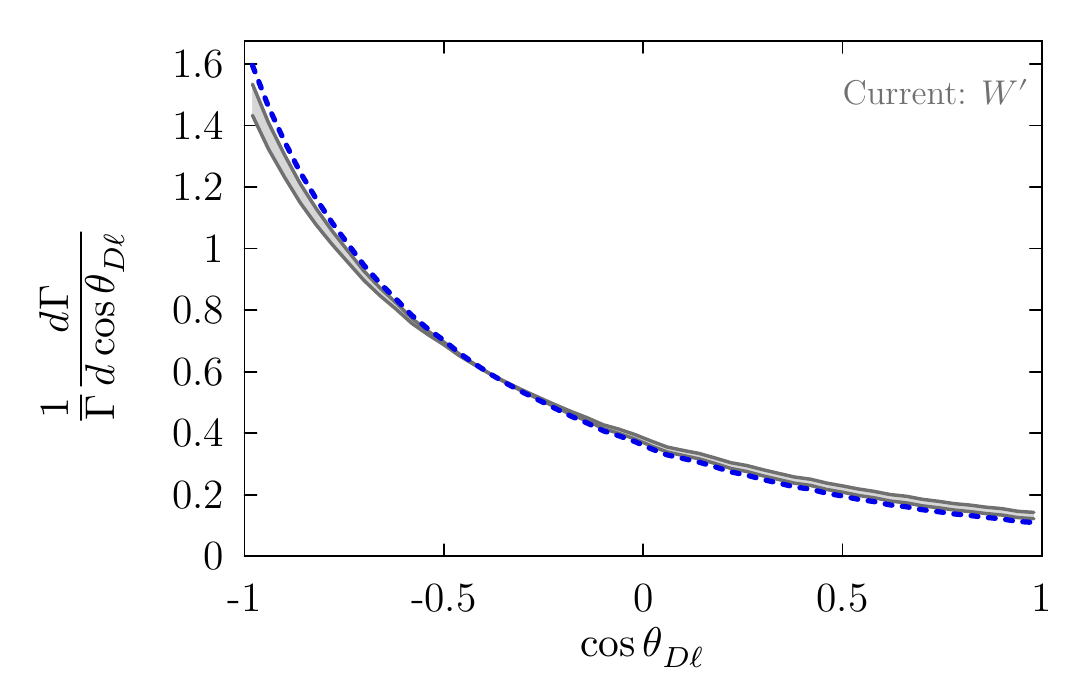} \\
}
\caption{Gray bands show kinematic distributions for $\Bbar \to (D^* \to D\pi)(\tau \to \ell \bar\nu_\ell \nt)\bar\nu$ (left) and $\Bbar \to D(\tau \to \ell \bar\nu_\ell \nt)\bar\nu$ (right) in the $B$ rest frame for the $W'$ simplified model in Table \ref{tab:mediators}, with the Wilson coefficient $c_{\rm VR}$ ranging over $2\sigma$ best fit regions in Fig. \ref{fig:fits}, and applying the phase space cuts~\eqref{eqn:PScut}. The blue dashed curves show the SM prediction.}
\label{fig:1DWphistos}
\end{figure}

\begin{figure}[t]
\centering{
\includegraphics[width = 0.47\linewidth]{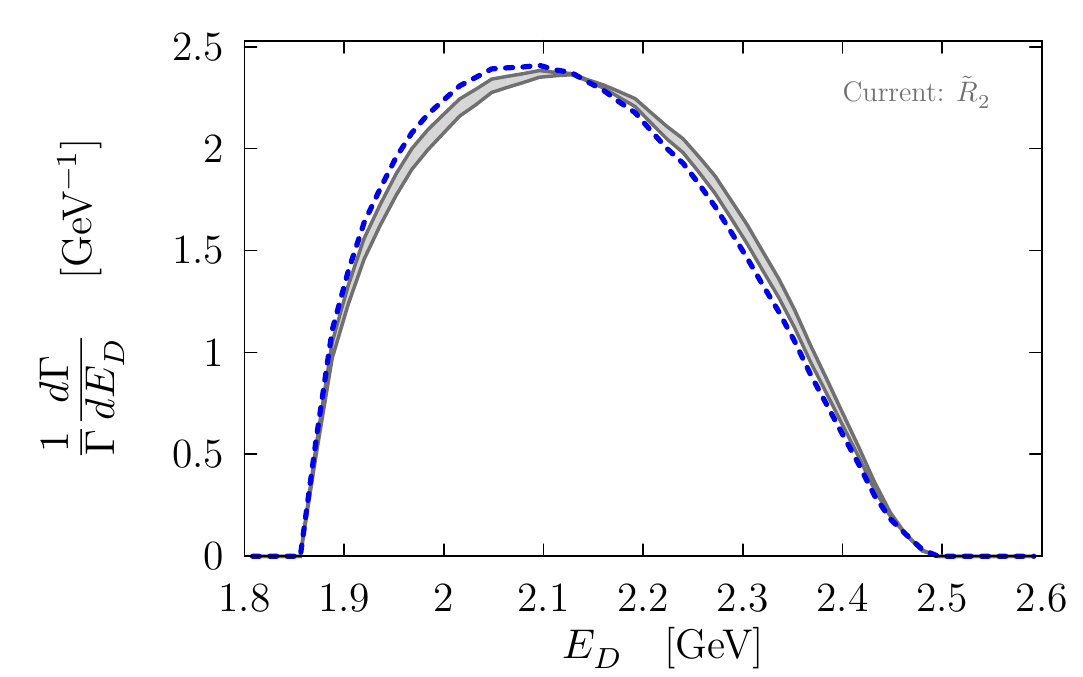}\hfill
\includegraphics[width = 0.47\linewidth]{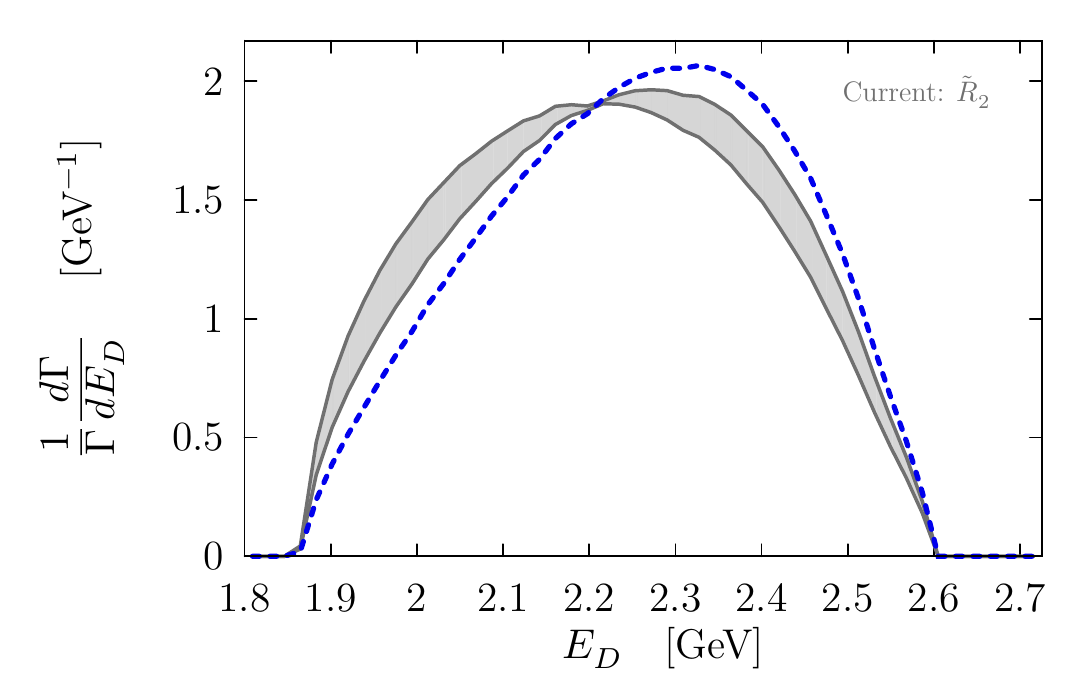} \\
\includegraphics[width = 0.47\linewidth]{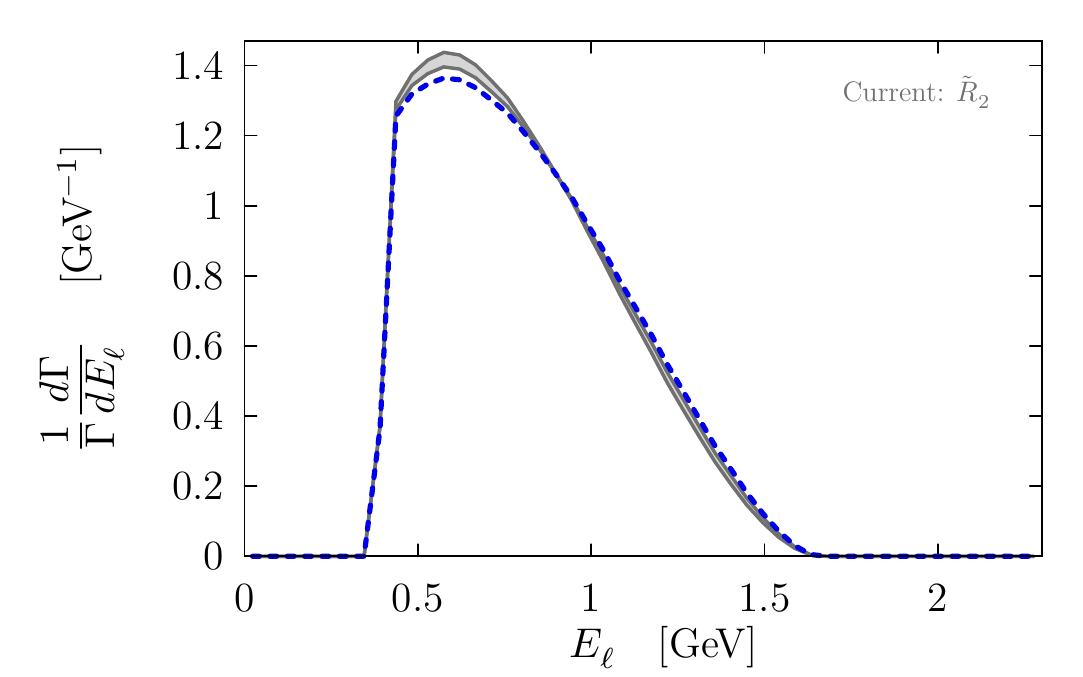}\hfill
\includegraphics[width = 0.47\linewidth]{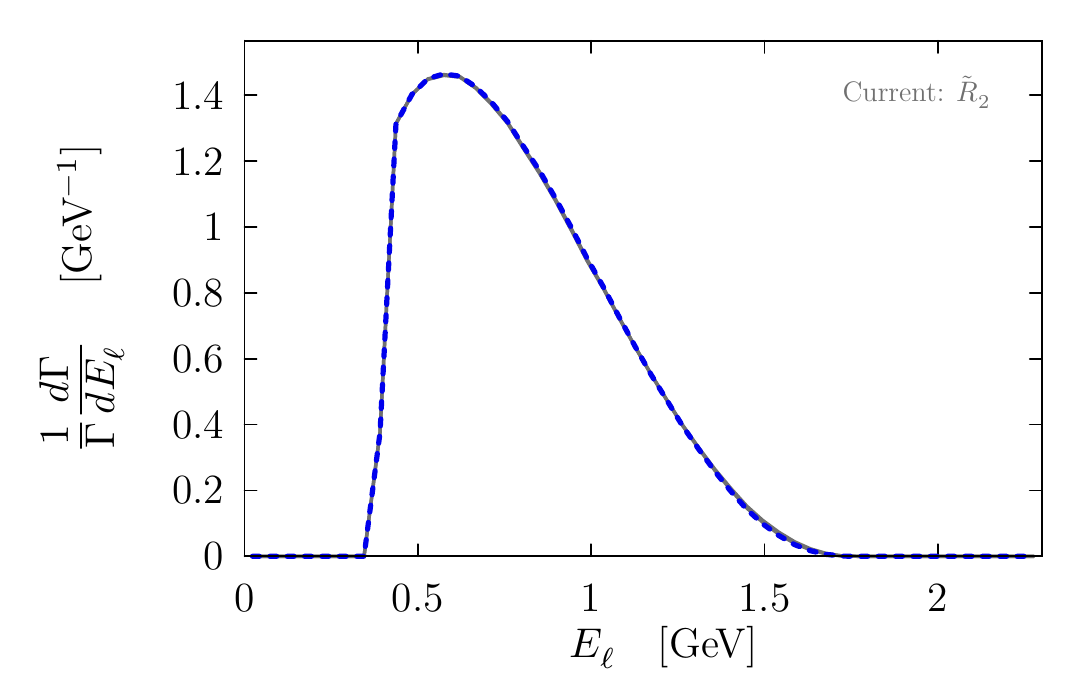} \\
\includegraphics[width = 0.47\linewidth]{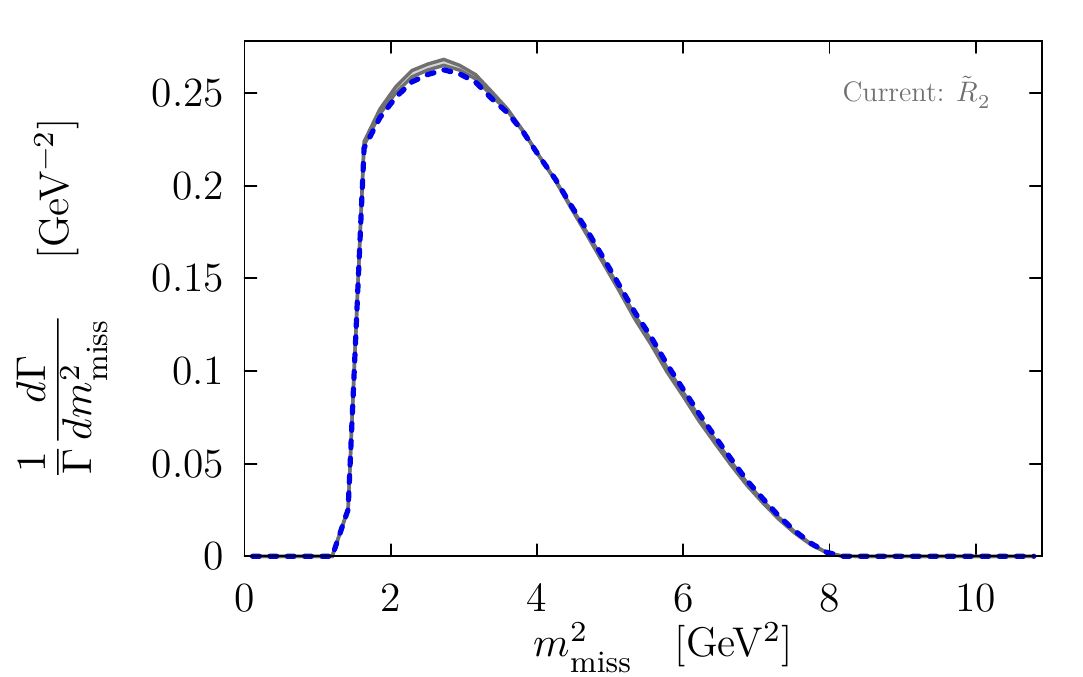} \hfill 
\includegraphics[width = 0.47\linewidth]{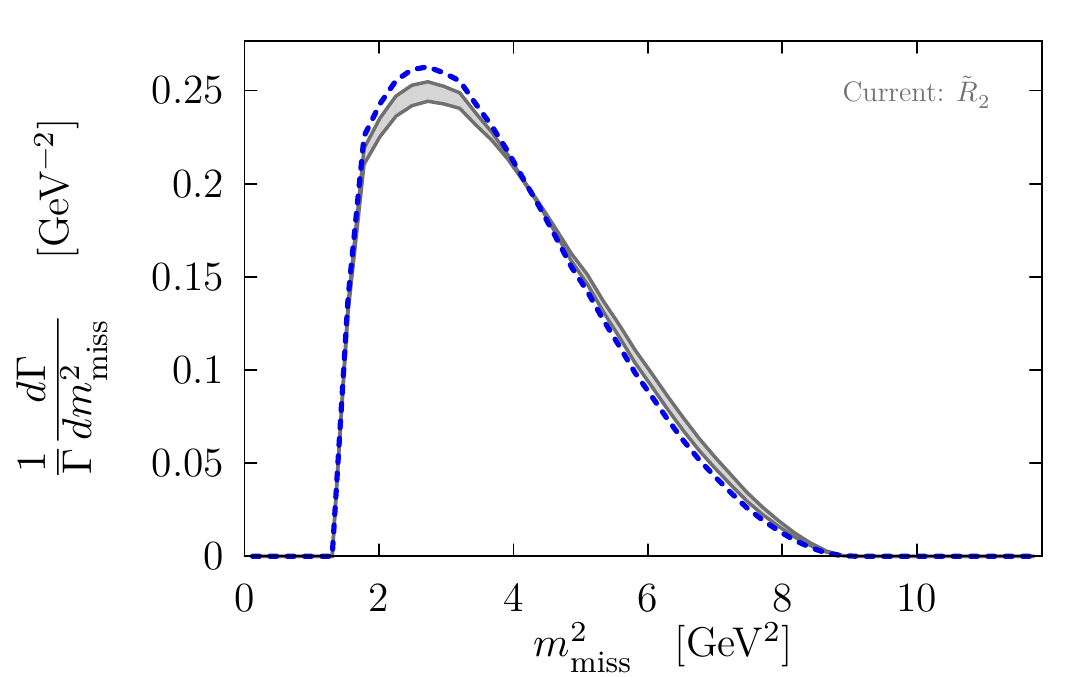}\\
\includegraphics[width = 0.47\linewidth]{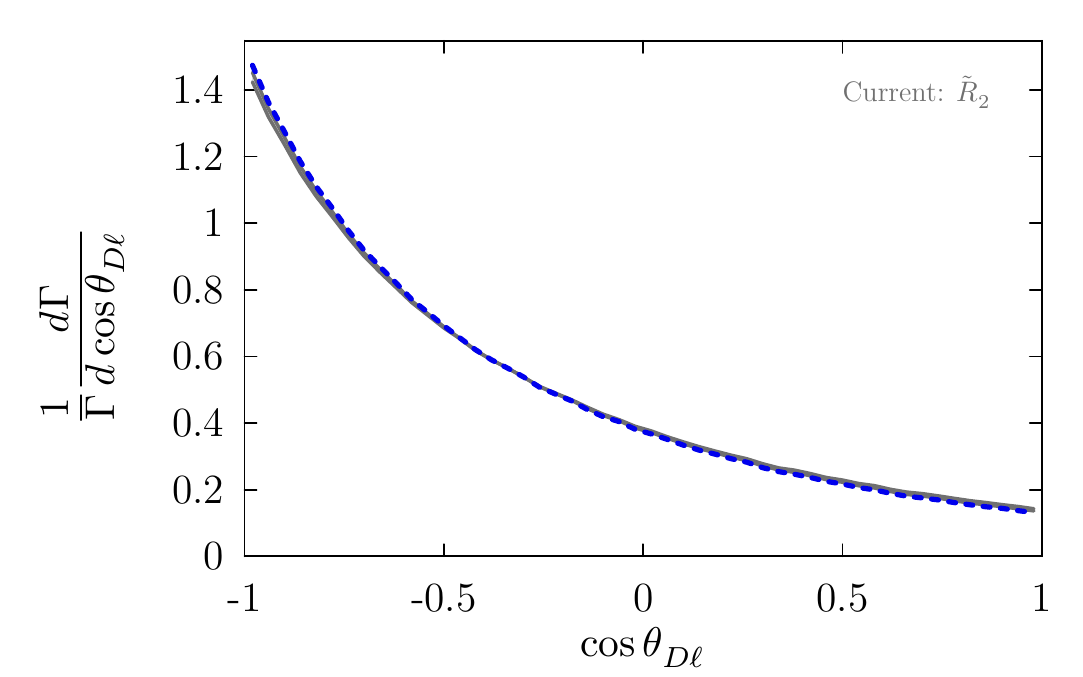} \hfill
\includegraphics[width = 0.47\linewidth]{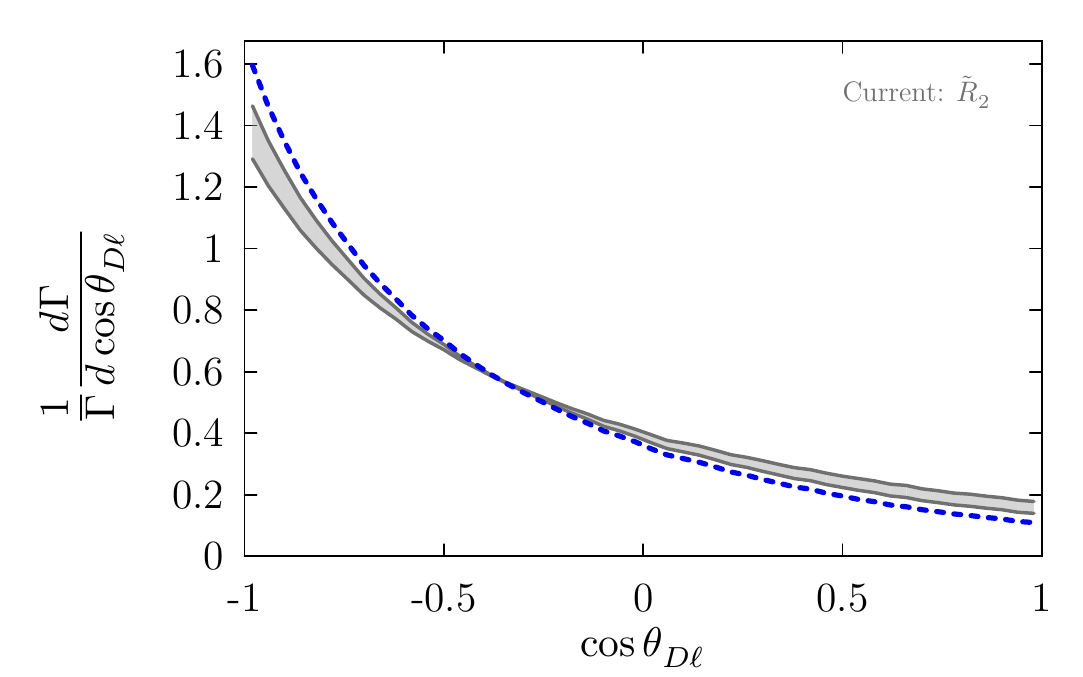} \\
}
\caption{Gray bands show kinematic distributions for $\Bbar \to (D^* \to D\pi)(\tau \to \ell \bar\nu_\ell \nt)\bar\nu$ (left) and $\Bbar \to D(\tau \to \ell \bar\nu_\ell \nt)\bar\nu$ (right) in the $B$ rest frame for the $\tilde{R}_2$ simplified model in Table \ref{tab:mediators}, with the Wilson coefficients $c_{\rm SR}=4c_{\rm T}$ ranging over $2\sigma$ best fit regions in Fig.~\ref{fig:fits}, and applying the phase space cuts~\eqref{eqn:PScut}. The blue dashed curves show the SM prediction.}
\label{fig:1DR2histos}
\end{figure}

\begin{figure}[t]
\centering{
\includegraphics[width = 0.47\linewidth]{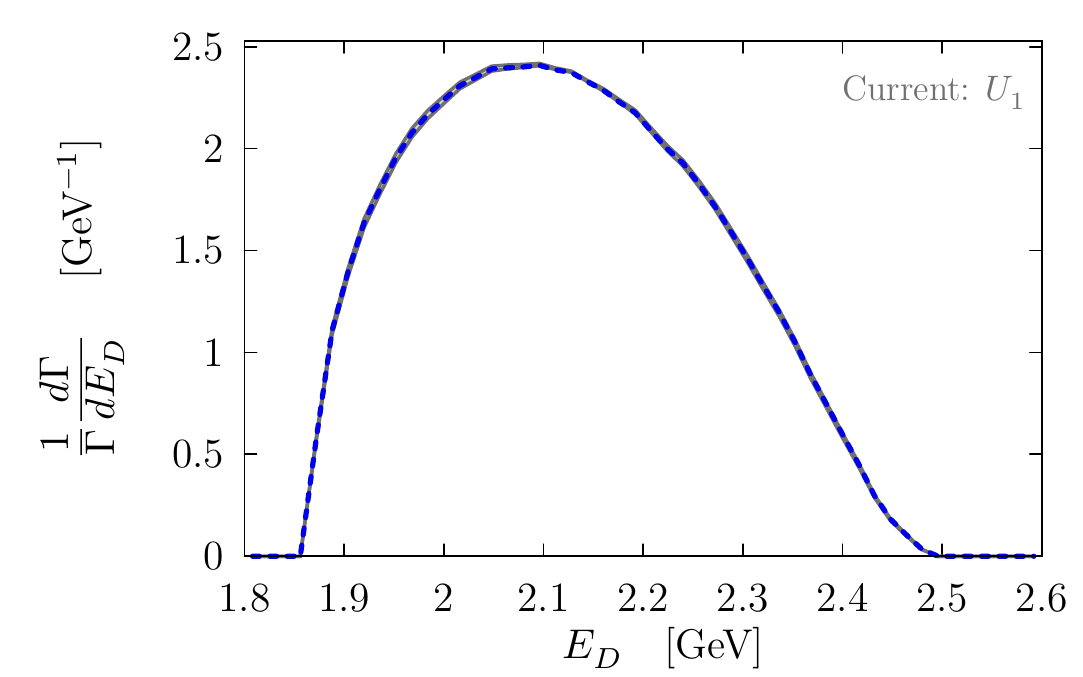}\hfill
\includegraphics[width = 0.47\linewidth]{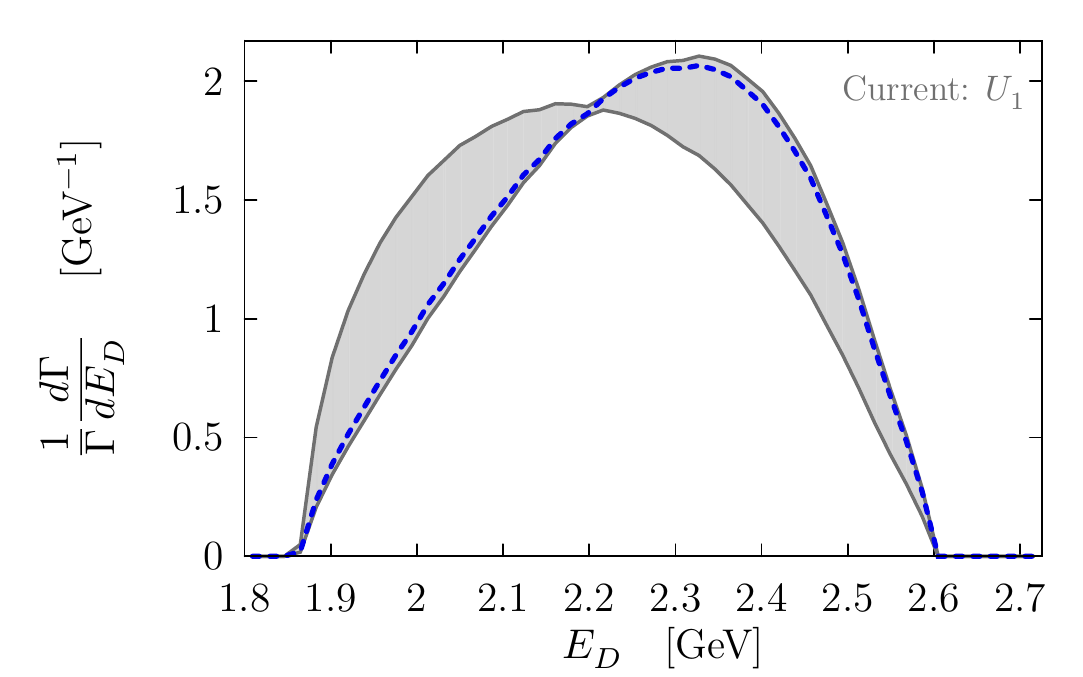} \\
\includegraphics[width = 0.47\linewidth]{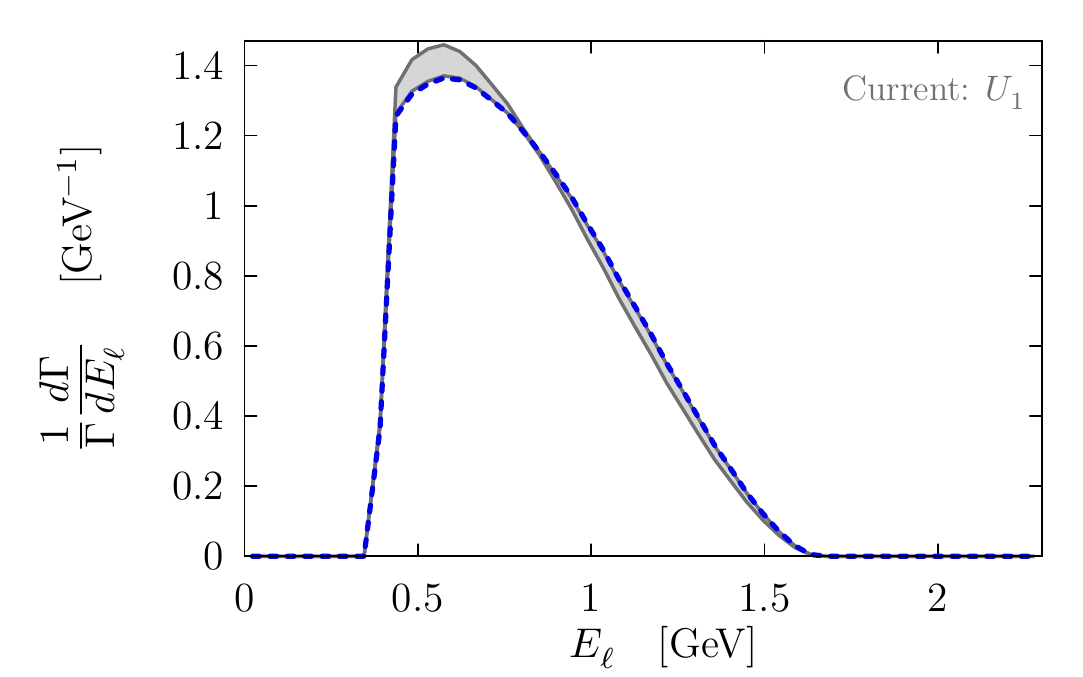}\hfill
\includegraphics[width = 0.47\linewidth]{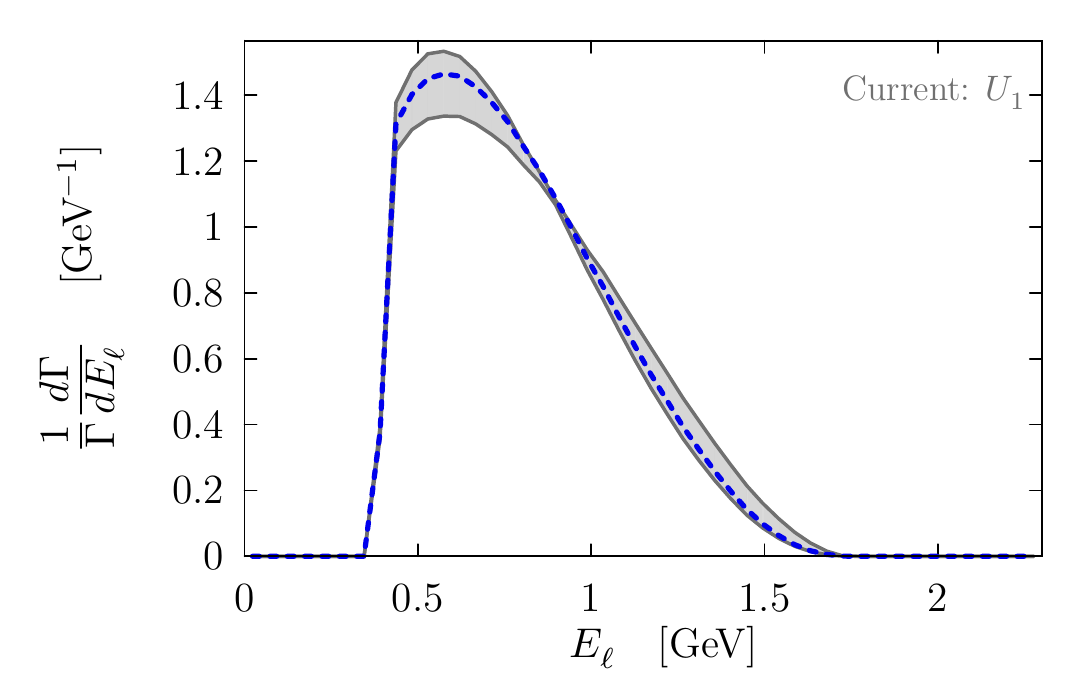} \\
\includegraphics[width = 0.47\linewidth]{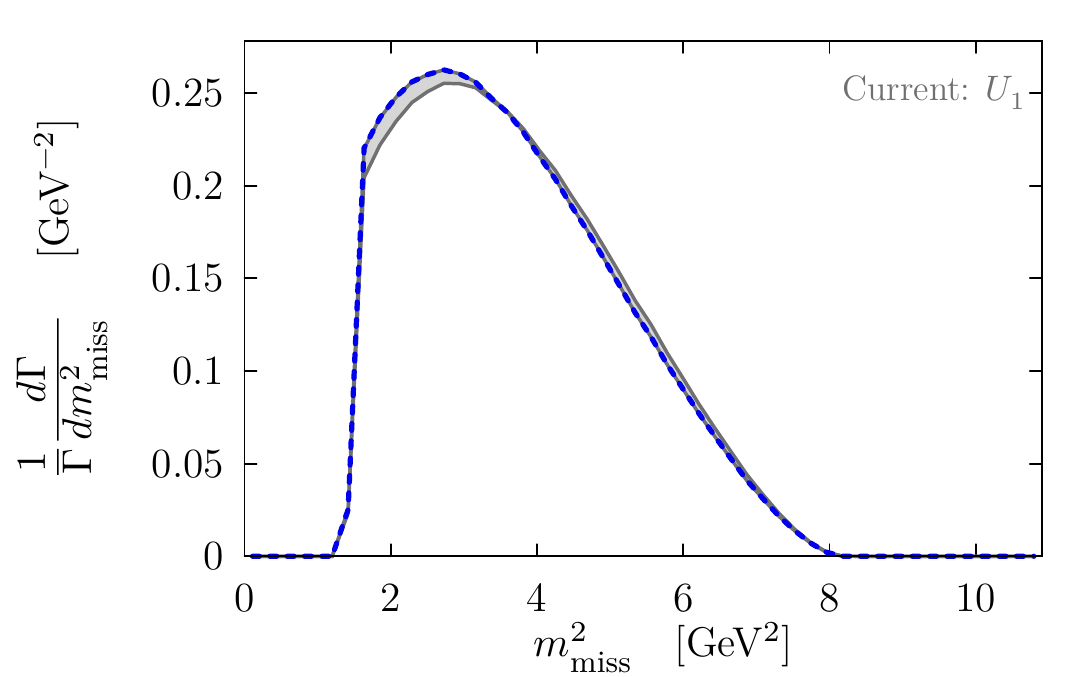} \hfill 
\includegraphics[width = 0.47\linewidth]{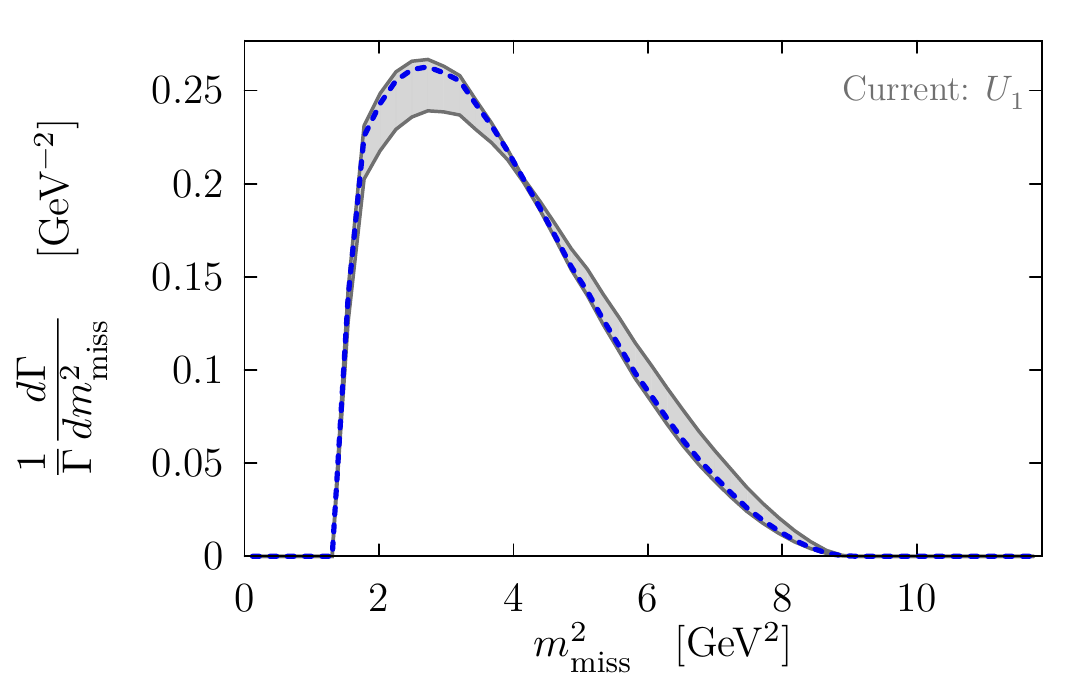}\\
\includegraphics[width = 0.47\linewidth]{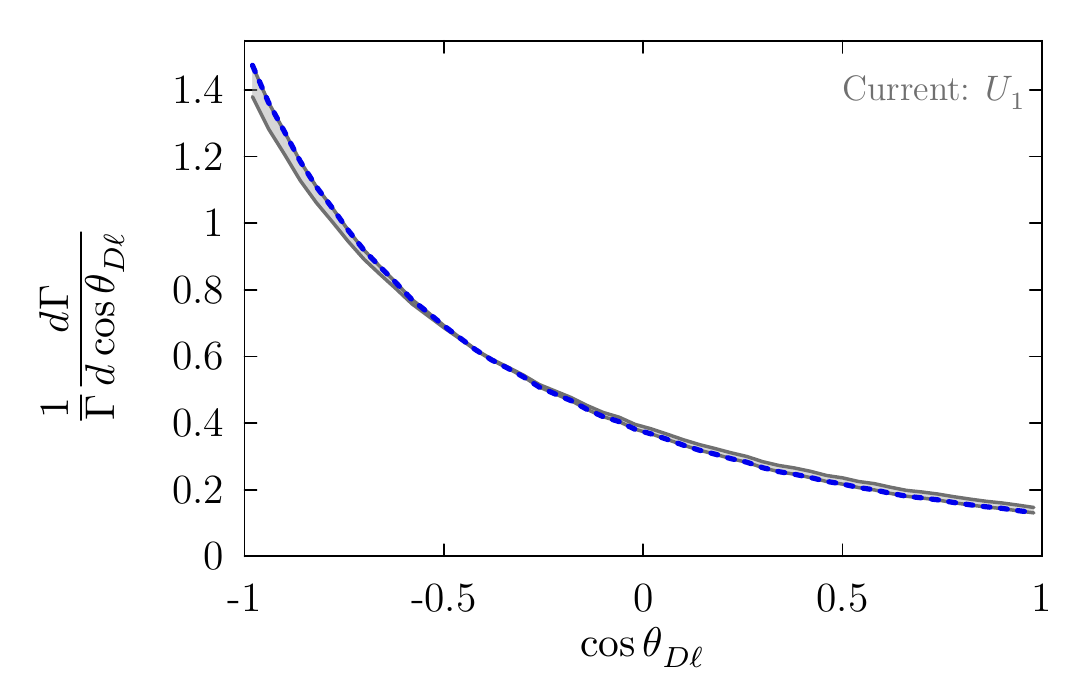} \hfill
\includegraphics[width = 0.47\linewidth]{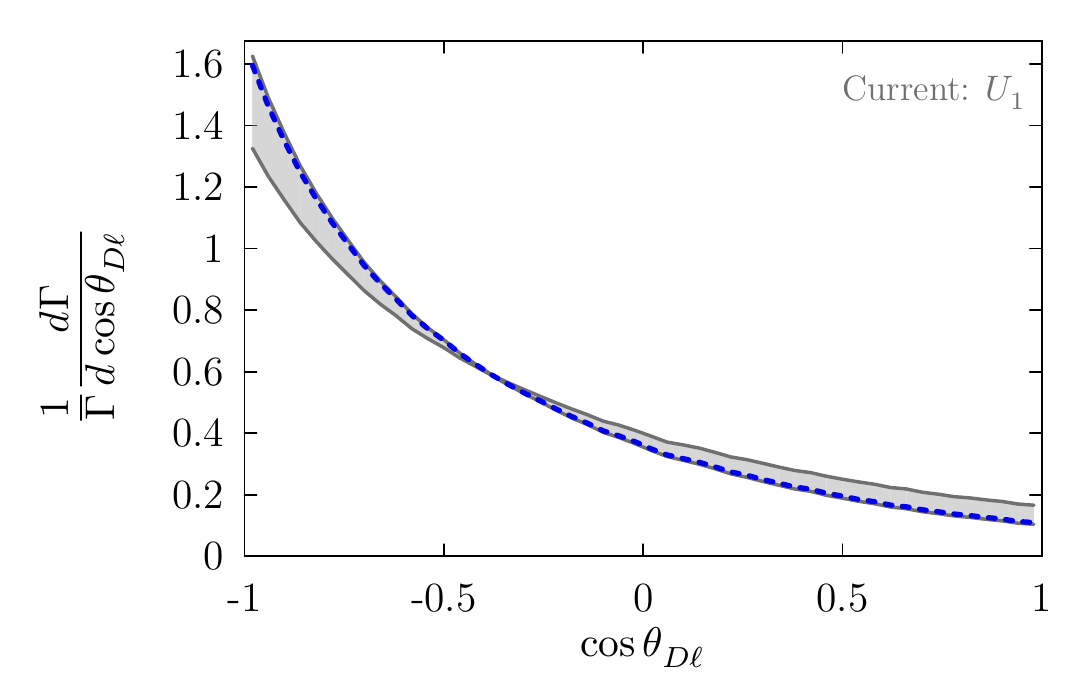} \\
}
\caption{Gray bands show kinematic distributions for $\Bbar \to (D^* \to D\pi)(\tau \to \ell \bar\nu_\ell \nt)\bar\nu$ (left) and $\Bbar \to D(\tau \to \ell \bar\nu_\ell \nt)\bar\nu$ (right) in the $B$ rest frame for the $U_1$ simplified model in Table \ref{tab:mediators}, with the Wilson coefficients $c_{\rm SL}, c_{\rm VR}$ ranging over $2\sigma$ best fit regions in Fig.~\ref{fig:fits}, and applying the phase space cuts~\eqref{eqn:PScut}. The blue dashed curves show the SM prediction.}
\label{fig:1DU1histos}
\end{figure}

\begin{figure}[t]
\centering{
\includegraphics[width = 0.47\linewidth]{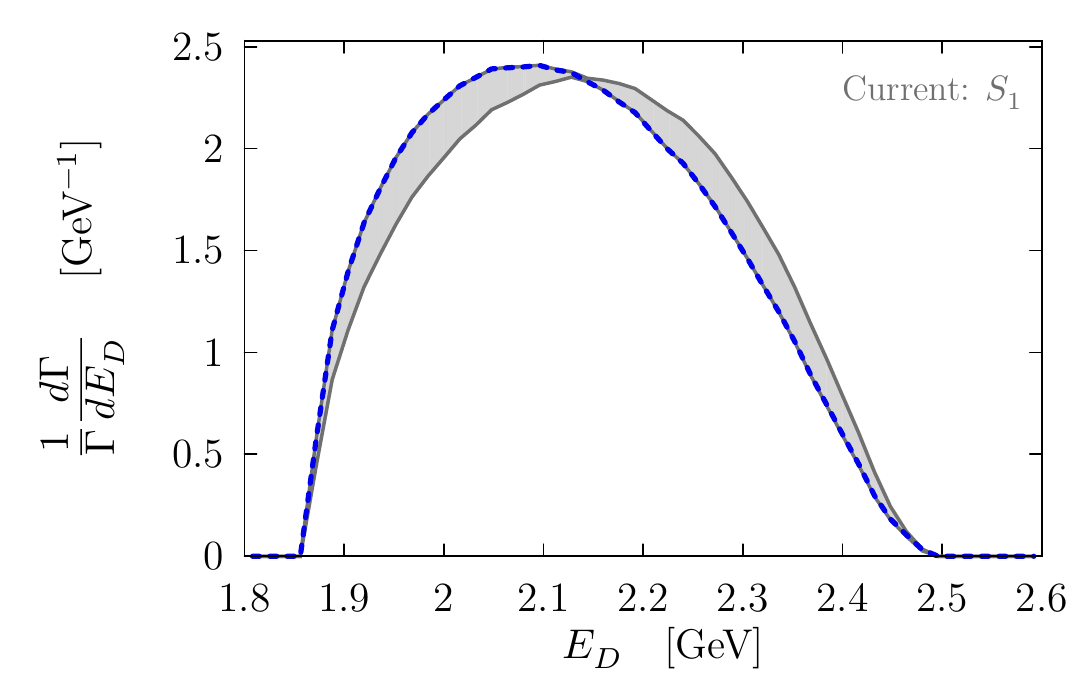}\hfill
\includegraphics[width = 0.47\linewidth]{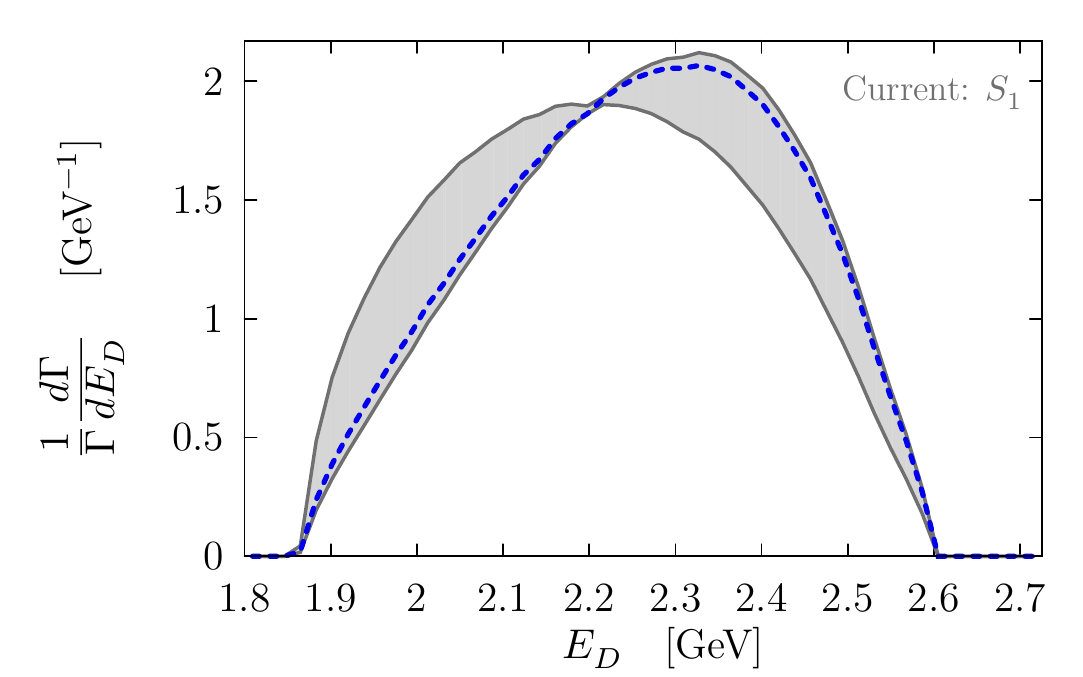} \\
\includegraphics[width = 0.47\linewidth]{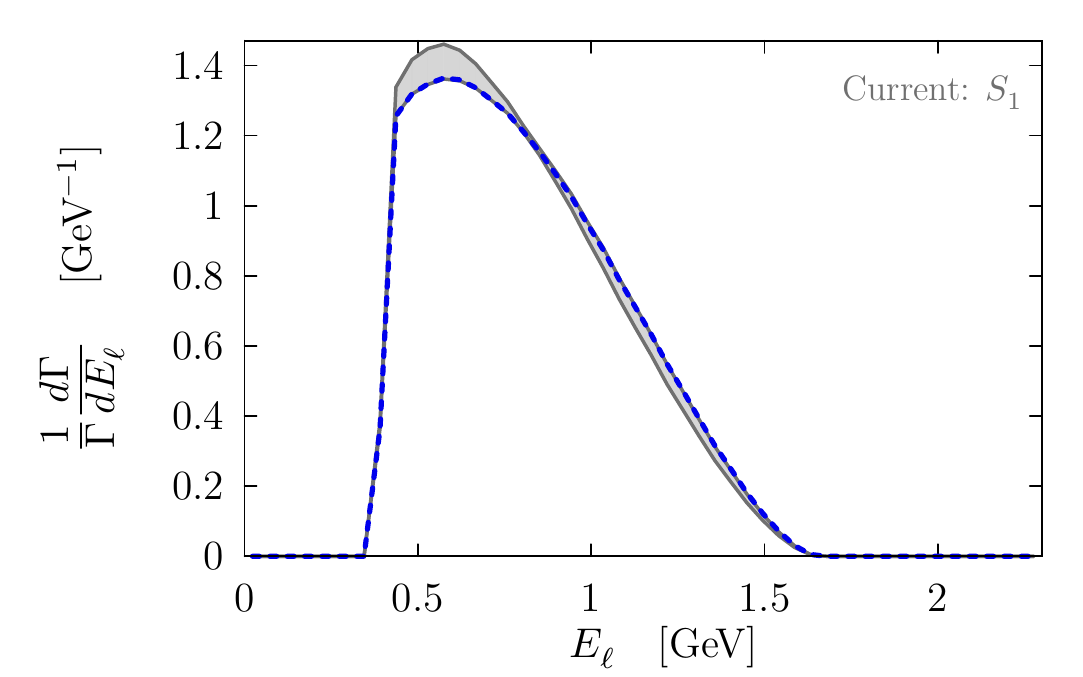}\hfill
\includegraphics[width = 0.47\linewidth]{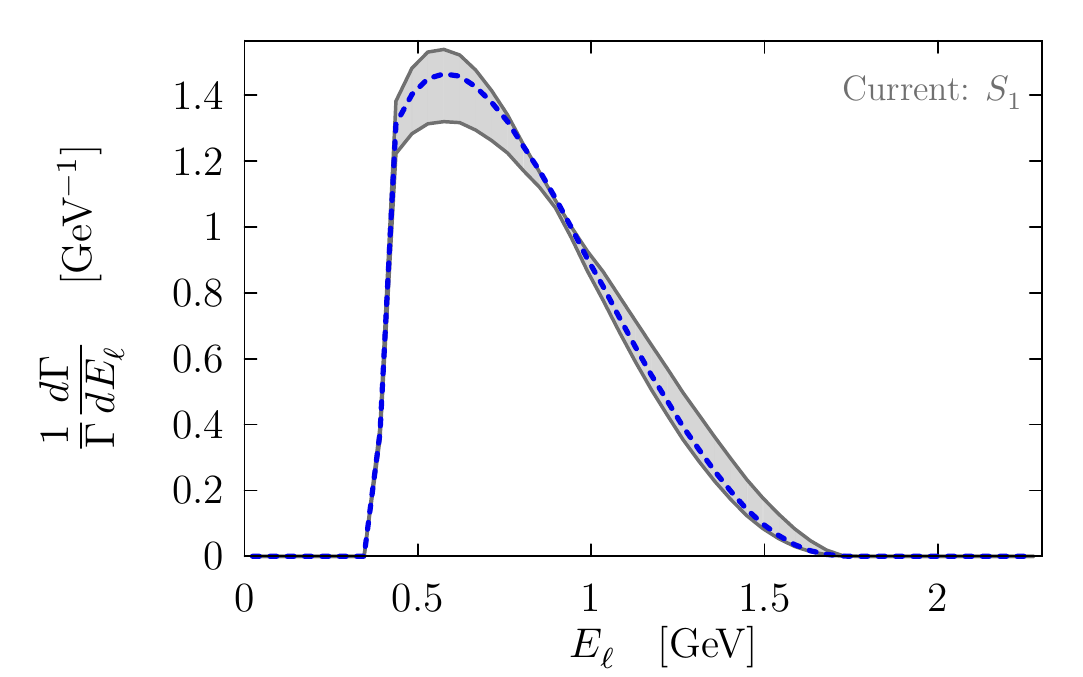} \\
\includegraphics[width = 0.47\linewidth]{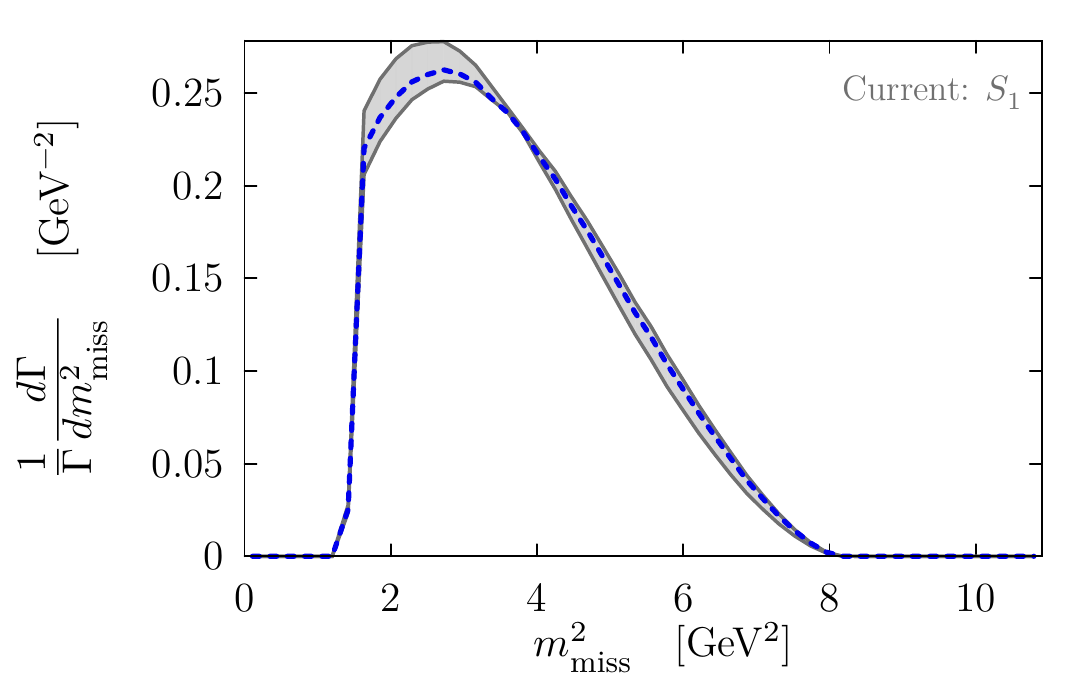} \hfill 
\includegraphics[width = 0.47\linewidth]{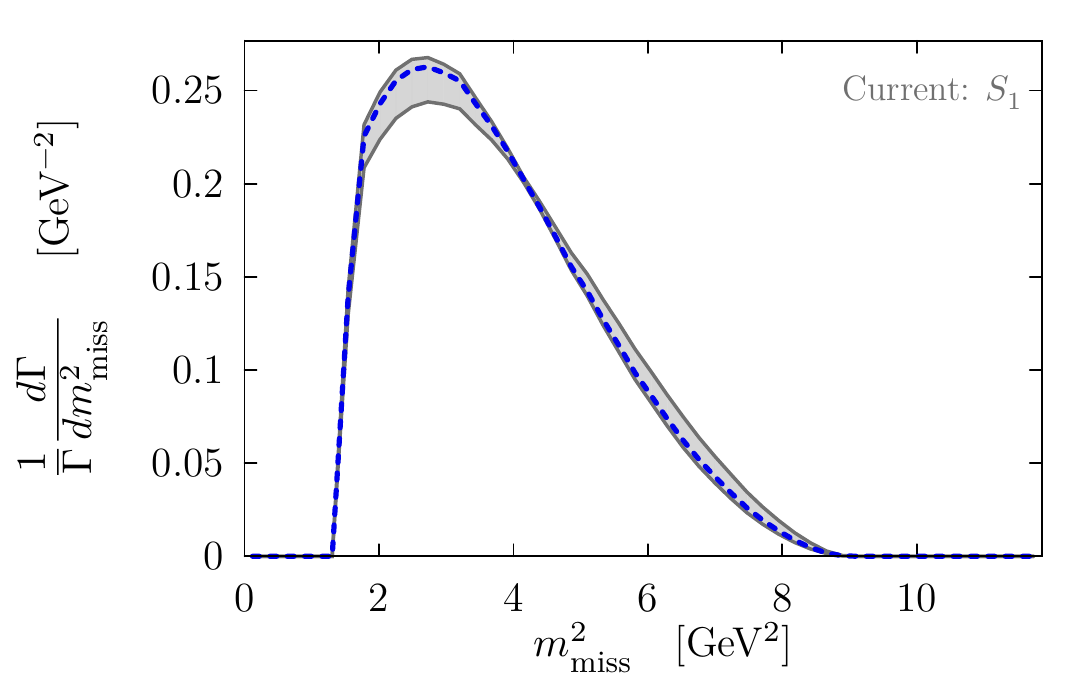}\\
\includegraphics[width = 0.47\linewidth]{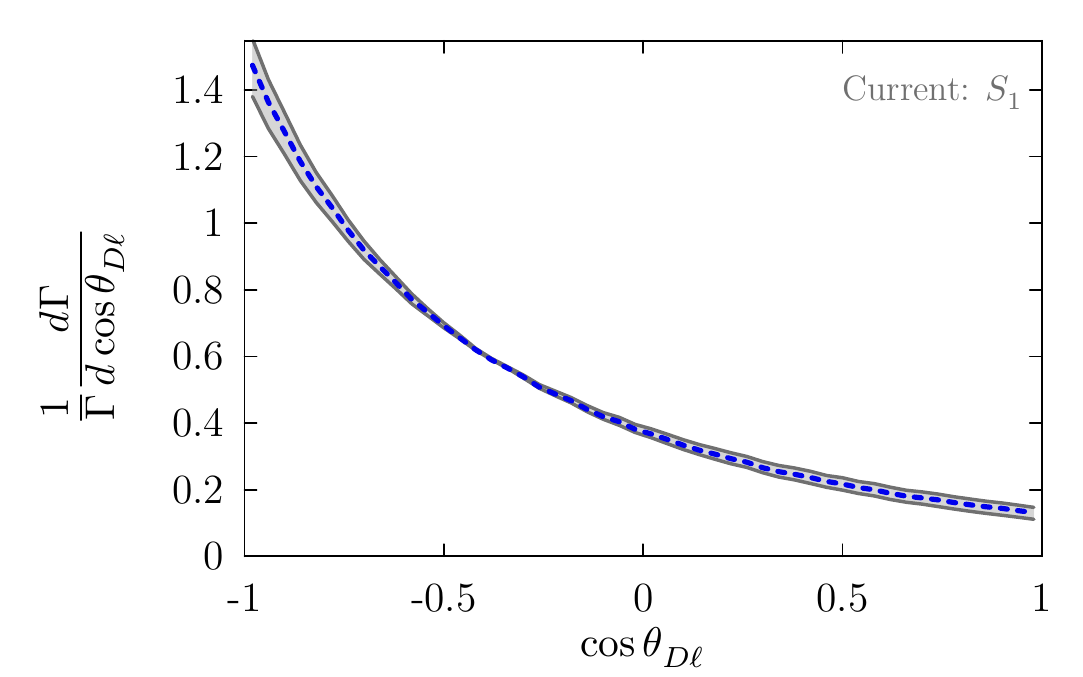} \hfill
\includegraphics[width = 0.47\linewidth]{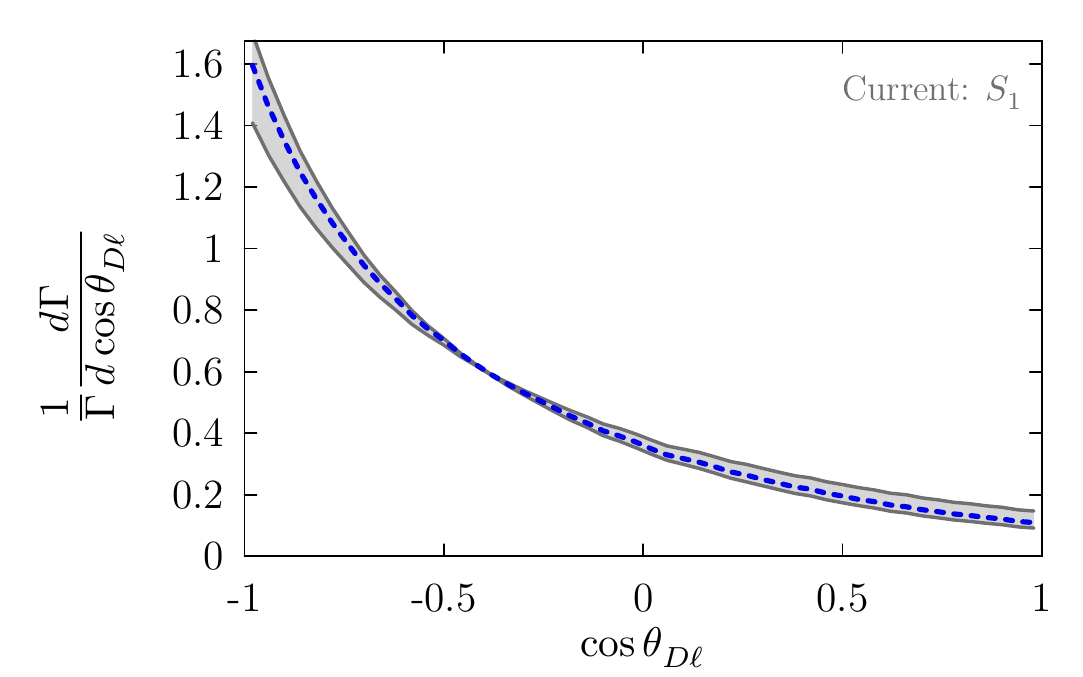} \\
}
\caption{Gray bands show kinematic distributions for $\Bbar \to (D^* \to D\pi)(\tau \to \ell \bar\nu_\ell \nt)\bar\nu$ (left) and $\Bbar \to D(\tau \to \ell \bar\nu_\ell \nt)\bar\nu$ (right) in the $B$ rest frame for the $S_1$ simplified model in Table \ref{tab:mediators}, with the Wilson coefficients $c_{\rm VR}, c_{\rm SR}=-4 c_T$ ranging over $2\sigma$ best fit regions in Fig.~\ref{fig:fits}, and applying the phase space cuts~\eqref{eqn:PScut}. The blue dashed curves show the SM prediction.}
\label{fig:1DS1histos}
\end{figure}

\clearpage

\bibliographystyle{h-physrev}
\bibliography{RDbiblio}

\begin{thebibliography}{10}

\bibitem{Lees:2012xj}
BaBar Collaboration, J.~P. Lees {\em et~al.},
\newblock Phys. Rev. Lett. {\bf 109}, 101802 (2012), 1205.5442.

\bibitem{Lees:2013uzd}
BaBar Collaboration, J.~P. Lees {\em et~al.},
\newblock Phys. Rev. {\bf D88}, 072012 (2013), 1303.0571.

\bibitem{Huschle:2015rga}
Belle Collaboration, M.~Huschle {\em et~al.},
\newblock Phys. Rev. {\bf D92}, 072014 (2015), 1507.03233.

\bibitem{Abdesselam:2016cgx}
Belle Collaboration, A.~Abdesselam {\em et~al.},
\newblock (2016), 1603.06711.

\bibitem{Abdesselam:2016xqt}
Belle Collaboration, A.~Abdesselam {\em et~al.},
\newblock (2016), 1608.06391.

\bibitem{Aaij:2015yra}
LHCb Collaboration, R.~Aaij {\em et~al.},
\newblock Phys. Rev. Lett. {\bf 115}, 111803 (2015), 1506.08614,
\newblock [Addendum: Phys. Rev. Lett. 115, no.15, 159901 (2015)].

\bibitem{HFAG}
Heavy Flavor Averaging Group, Y.~Amhis {\em et~al.},
\newblock (2016), 1612.07233,
\newblock and updates at \url{http://www.slac.stanford.edu/xorg/hfag/}.

\bibitem{Bernlochner:2017jka}
F.~U. Bernlochner, Z.~Ligeti, M.~Papucci, and D.~J. Robinson,
\newblock Phys. Rev. {\bf D95}, 115008 (2017), 1703.05330.

\bibitem{Bigi:2017jbd}
D.~Bigi, P.~Gambino, and S.~Schacht,
\newblock JHEP {\bf 11}, 061 (2017), 1707.09509.

\bibitem{Jaiswal:2017rve}
S.~Jaiswal, S.~Nandi, and S.~K. Patra,
\newblock JHEP {\bf 12}, 060 (2017), 1707.09977.

\bibitem{Alok:2016qyh}
A.~K. Alok, D.~Kumar, S.~Kumbhakar, and S.~U. Sankar,
\newblock Phys. Rev. {\bf D95}, 115038 (2017), 1606.03164.

\bibitem{Bhattacharya:2016zcw}
S.~Bhattacharya, S.~Nandi, and S.~K. Patra,
\newblock Phys. Rev. {\bf D95}, 075012 (2017), 1611.04605.

\bibitem{Faroughy:2016osc}
D.~A. Faroughy, A.~Greljo, and J.~F. Kamenik,
\newblock Phys. Lett. {\bf B764}, 126 (2017), 1609.07138.

\bibitem{Feruglio:2016gvd}
F.~Feruglio, P.~Paradisi, and A.~Pattori,
\newblock Phys. Rev. Lett. {\bf 118}, 011801 (2017), 1606.00524.

\bibitem{Feruglio:2017rjo}
F.~Feruglio, P.~Paradisi, and A.~Pattori,
\newblock JHEP {\bf 09}, 061 (2017), 1705.00929.

\bibitem{Asadi:2018wea}
P.~Asadi, M.~R. Buckley, and D.~Shih,
\newblock (2018), 1804.04135.

\bibitem{Greljo:2018ogz}
A.~Greljo, D.~J. Robinson, B.~Shakya, and J.~Zupan,
\newblock (2018), 1804.04642.

\bibitem{He:2012zp}
X.-G. He and G.~Valencia,
\newblock Phys. Rev. {\bf D87}, 014014 (2013), 1211.0348.

\bibitem{He:2017bft}
X.-G. He and G.~Valencia,
\newblock Phys. Lett. {\bf B779}, 52 (2018), 1711.09525.

\bibitem{Fajfer:2012jt}
S.~Fajfer, J.~F. Kamenik, I.~Nisandzic, and J.~Zupan,
\newblock Phys. Rev. Lett. {\bf 109}, 161801 (2012), 1206.1872.

\bibitem{Becirevic:2016yqi}
D.~Becirevic, S.~Fajfer, N.~Kosnik, and O.~Sumensari,
\newblock Phys. Rev. {\bf D94}, 115021 (2016), 1608.08501.

\bibitem{Cvetic:2017gkt}
G.~Cvetic, F.~Halzen, C.~S. Kim, and S.~Oh,
\newblock Chin. Phys. {\bf C41}, 113102 (2017), 1702.04335.

\bibitem{Li:2016vvp}
X.-Q. Li, Y.-D. Yang, and X.~Zhang,
\newblock JHEP {\bf 08}, 054 (2016), 1605.09308.

\bibitem{Alonso:2016oyd}
R.~Alonso, B.~Grinstein, and J.~Martin~Camalich,
\newblock Phys. Rev. Lett. {\bf 118}, 081802 (2017), 1611.06676.

\bibitem{Celis:2016azn}
A.~Celis, M.~Jung, X.-Q. Li, and A.~Pich,
\newblock Phys. Lett. {\bf B771}, 168 (2017), 1612.07757.

\bibitem{Buchalla:1995vs}
G.~Buchalla, A.~J. Buras, and M.~E. Lautenbacher,
\newblock Rev. Mod. Phys. {\bf 68}, 1125 (1996), hep-ph/9512380.

\bibitem{Freytsis:2015qca}
M.~Freytsis, Z.~Ligeti, and J.~T. Ruderman,
\newblock Phys. Rev. {\bf D92}, 054018 (2015), 1506.08896.

\bibitem{Dorsner:2013tla}
JHEP {\bf 11}, 084 (2013), 1306.6493.

\bibitem{Dorsner:2016wpm}
I.~Dorsner, S.~Fajfer, A.~Greljo, J.~F. Kamenik, and N.~Kosnik,
\newblock Phys. Rept. {\bf 641}, 1 (2016), 1603.04993.

\bibitem{Ligeti:2016npd}
Z.~Ligeti, M.~Papucci, and D.~J. Robinson,
\newblock JHEP {\bf 01}, 083 (2017), 1610.02045.

\bibitem{Colquhoun:2015oha}
HPQCD, B.~Colquhoun {\em et~al.},
\newblock Phys. Rev. {\bf D91}, 114509 (2015), 1503.05762.

\bibitem{PDG}
Particle Data Group, C.~Patrignani {\em et~al.},
\newblock Chin. Phys. {\bf C40}, 100001 (2016).

\bibitem{Kamenik:2009kc}
J.~F. Kamenik and C.~Smith,
\newblock Phys. Lett. {\bf B680}, 471 (2009), 0908.1174.

\bibitem{Kamenik:2011vy}
J.~F. Kamenik and C.~Smith,
\newblock JHEP {\bf 03}, 090 (2012), 1111.6402.

\bibitem{Tanabashi:2018oca}
ParticleDataGroup, M.~Tanabashi {\em et~al.},
\newblock Phys. Rev. {\bf D98}, 030001 (2018).

\bibitem{Buras:2014fpa}
A.~J. Buras, J.~Girrbach-Noe, C.~Niehoff, and D.~M. Straub,
\newblock JHEP {\bf 02}, 184 (2015), 1409.4557.

\bibitem{Hammer_paper}
F.~Bernlochner, S.~Duell, Z.~Ligeti, M.~Papucci, and D.~J. Robinson,
\newblock In preparation  (2018).

\bibitem{Aaboud:2018vgh}
ATLAS, M.~Aaboud {\em et~al.},
\newblock Phys. Rev. Lett. {\bf 120}, 161802 (2018), 1801.06992.

\bibitem{Sirunyan:2018lbg}
CMS, A.~M. Sirunyan {\em et~al.},
\newblock Submitted to: Phys. Lett.  (2018), 1807.11421.

\bibitem{Khachatryan:2016jww}
CMS, V.~Khachatryan {\em et~al.},
\newblock Phys. Lett. {\bf B770}, 278 (2017), 1612.09274.

\bibitem{CMS:2016ppa}
CMS, C.~Collaboration,
\newblock CERN Report No. CMS-PAS-EXO-16-006, 2016 (unpublished).

\bibitem{DiLuzio:2017chi}
L.~Di~Luzio and M.~Nardecchia,
\newblock Eur. Phys. J. {\bf C77}, 536 (2017), 1706.01868.

\bibitem{Greljo:2018tzh}
(2018), 1811.07920.

\bibitem{Sirunyan:2017nvi}
CMS, A.~M. Sirunyan {\em et~al.},
\newblock (2017), 1710.00159.

\bibitem{Khachatryan:2016ecr}
CMS, V.~Khachatryan {\em et~al.},
\newblock Phys. Rev. Lett. {\bf 117}, 031802 (2016), 1604.08907.

\bibitem{Sirunyan:2016iap}
CMS, A.~M. Sirunyan {\em et~al.},
\newblock Phys. Lett. {\bf B769}, 520 (2017), 1611.03568,
\newblock [Erratum: Phys. Lett.B772,882(2017)].

\bibitem{Aad:2011aj}
ATLAS, G.~Aad {\em et~al.},
\newblock New J. Phys. {\bf 13}, 053044 (2011), 1103.3864.

\bibitem{Abe:1997hm}
CDF, F.~Abe {\em et~al.},
\newblock Phys. Rev. {\bf D55}, R5263 (1997), hep-ex/9702004.

\bibitem{Knapen:2015hia}
S.~Knapen and D.~J. Robinson,
\newblock Phys. Rev. Lett. {\bf 115}, 161803 (2015), 1507.00009.

\bibitem{Dorsner:2018ynv}
I.~Dor\v~sner and A.~Greljo,
\newblock (2018), 1801.07641.

\bibitem{CMS:2018bhq}
CMS, C.~Collaboration,
\newblock  Report No. CMS-PAS-SUS-18-001, 2018 (unpublished).

\bibitem{Sirunyan:2017yrk}
CMS, A.~M. Sirunyan {\em et~al.},
\newblock JHEP {\bf 07}, 121 (2017), 1703.03995.

\bibitem{Aaboud:2016cre}
ATLAS, M.~Aaboud {\em et~al.},
\newblock Eur. Phys. J. {\bf C76}, 585 (2016), 1608.00890.

\bibitem{Aaboud:2017sjh}
ATLAS, M.~Aaboud {\em et~al.},
\newblock JHEP {\bf 01}, 055 (2018), 1709.07242.

\bibitem{Mandal:2018kau}
T.~Mandal, S.~Mitra, and S.~Raz,
\newblock (2018), 1811.03561.

\bibitem{Lavoura:2003xp}
L.~Lavoura,
\newblock Eur. Phys. J. {\bf C29}, 191 (2003), hep-ph/0302221.

\bibitem{Wong:1992qa}
G.-G. Wong,
\newblock Phys. Rev. {\bf D46}, 3987 (1992).

\bibitem{Bezrukov:2009th}
F.~Bezrukov, H.~Hettmansperger, and M.~Lindner,
\newblock Phys. Rev. {\bf D81}, 085032 (2010), 0912.4415.

\bibitem{Aparici:2012vx}
A.~Aparici, J.~Herrero-Garcia, N.~Rius, and A.~Santamaria,
\newblock JHEP {\bf 07}, 030 (2012), 1204.1021.

\bibitem{Dodelson:1993je}
S.~Dodelson and L.~M. Widrow,
\newblock Phys. Rev. Lett. {\bf 72}, 17 (1994), hep-ph/9303287.

\bibitem{Shi:1998km}
X.-D. Shi and G.~M. Fuller,
\newblock Phys. Rev. Lett. {\bf 82}, 2832 (1999), astro-ph/9810076.

\bibitem{Shakya:2015xnx}
B.~Shakya,
\newblock Mod. Phys. Lett. {\bf A31}, 1630005 (2016), 1512.02751.

\bibitem{Shakya:2016oxf}
B.~Shakya and J.~D. Wells,
\newblock Phys. Rev. {\bf D96}, 031702 (2017), 1611.01517.

\bibitem{Roland:2014vba}
S.~B. Roland, B.~Shakya, and J.~D. Wells,
\newblock Phys. Rev. {\bf D92}, 113009 (2015), 1412.4791.

\bibitem{Shakya:2018qzg}
B.~Shakya and J.~D. Wells,
\newblock (2018), 1801.02640.

\bibitem{Scherrer:1984fd}
R.~J. Scherrer and M.~S. Turner,
\newblock Phys. Rev. {\bf D31}, 681 (1985).

\bibitem{Asaka:2006ek}
T.~Asaka, M.~Shaposhnikov, and A.~Kusenko,
\newblock Phys. Lett. {\bf B638}, 401 (2006), hep-ph/0602150.

\bibitem{Essig:2013goa}
R.~Essig, E.~Kuflik, S.~D. McDermott, T.~Volansky, and K.~M. Zurek,
\newblock JHEP {\bf 11}, 193 (2013), 1309.4091.

\bibitem{Abazajian:2016yjj}
CMB-S4, K.~N. Abazajian {\em et~al.},
\newblock (2016), 1610.02743.

\bibitem{Anelli:2015pba}
SHiP, M.~Anelli {\em et~al.},
\newblock (2015), 1504.04956.

\bibitem{Chou:2016lxi}
J.~P. Chou, D.~Curtin, and H.~J. Lubatti,
\newblock Phys. Lett. {\bf B767}, 29 (2017), 1606.06298.

\bibitem{Feng:2017uoz}
J.~Feng, I.~Galon, F.~Kling, and S.~Trojanowski,
\newblock Phys. Rev. {\bf D97}, 035001 (2018), 1708.09389.

\bibitem{Gligorov:2017nwh}
V.~V. Gligorov, S.~Knapen, M.~Papucci, and D.~J. Robinson,
\newblock Phys. Rev. {\bf D97}, 015023 (2018), 1708.09395.

\bibitem{Drewes:2015iva}
Nucl. Phys. {\bf B921}, 250 (2017), 1502.00477.

\end{thebibliography}

\end{document}